\documentclass[aps, prd, twocolumn, preprintnumbers, groupedaddress, nofootinbib, amssymb, notitlepage, eqsecnum]{revtex4-2}
\usepackage{bm}
\usepackage{amsmath,amsthm,amssymb}
\usepackage{amsfonts}
\usepackage{hyperref}
\usepackage{graphicx}

\allowdisplaybreaks[1]

\newcommand{\rd}{{\rm d}}
\newcommand{\be}{\begin{equation}}
\newcommand{\ee}{\end{equation}}
\newcommand{\ba}{\begin{eqnarray}}
\newcommand{\ea}{\end{eqnarray}}
\newcommand{\Mpl}{M_{\rm Pl}}
\newcommand{\cL}{{\cal L}}

\begin{document}

\preprint{YITP-24-167, WUCG-24-10}

\title{Nonsingular black holes and spherically symmetric objects \\
in nonlinear electrodynamics with a scalar field}

\author{Antonio De Felice$^{a}$}
\email{antonio.defelice@yukawa.kyoto-u.ac.jp}  
\author{Shinji Tsujikawa$^{b}$}
\email{tsujikawa@waseda.jp} 

\affiliation{$^{a}$Center for Gravitational Physics and Quantum Information, Yukawa Institute for Theoretical Physics, Kyoto University, 606-8502, Kyoto, Japan\\
$^{b}$Department of Physics, Waseda University, 
3-4-1 Okubo, Shinjuku, Tokyo 169-8555, Japan}
 
\date{\today}

\begin{abstract}

In general relativity with vector and scalar fields given by 
the Lagrangian ${\cal L}(F,\phi,X)$, where $F$ is a Maxwell 
term and $X$ is a kinetic term of the scalar field $\phi$, 
we study the linear stability of static 
and spherically symmetric objects without 
curvature singularities at their centers.
We show that the background solutions  
are generally described by either purely electrically or 
magnetically charged objects with a nontrivial 
scalar-field profile. In theories with the Lagrangian 
$\tilde{{\cal L}}(F)+K(\phi, X)$, which 
correspond to nonlinear electrodynamics 
with a k-essence scalar field, angular Laplacian instabilities induced by vector-field perturbations exclude all 
the regular spherically symmetric solutions 
including nonsingular black holes. 
In theories described by the Lagrangian ${\cal L}=X+\mu(\phi)F^n$, 
where $\mu$ is a function of $\phi$ and $n$ is a constant, the absence of angular Laplacian instabilities of spherically symmetric objects requires that $n>1/2$, under which nonsingular 
black holes with apparent horizons 
are not present. 
However, for some particular ranges of $n$, there are horizonless compact objects with neither ghosts nor Laplacian instabilities in the 
small-scale limit.
In theories given by ${\cal L}=X \kappa (F)$, where $\kappa$ is a function of $F$, regular spherically symmetric objects are prone to Laplacian instabilities either around the center or at spatial infinity. 
Thus, in our theoretical framework, we do not find any example of linearly stable nonsingular black holes.

\end{abstract}


\maketitle

\section{Introduction}
\label{Intro}

General Relativity (GR) is a fundamental pillar for describing gravitational interactions in both strong and weak field regimes.
The vacuum solution to the Einstein equation on a static and spherically symmetric (SSS) background is described by a Schwarzschild line metric that contains a mass $M$ of the source. 
In Einstein-Maxwell theory with the electromagnetic Lagrangian $F=-F_{\mu \nu}F^{\mu \nu}/4$, where $F_{\mu \nu}=\partial_{\mu}A_{\nu}
-\partial_{\nu}A_{\mu}$ is the Maxwell tensor with a vector field $A_{\mu}$, the resulting solution is given by a Reissner-Nordstr\"om (RN) metric with electric or magnetic charges. 
For both Schwarzschild and RN black holes (BHs), 
there are singularities at the origin ($r=0$) with divergent curvature quantities. This divergent 
property at $r=0$ also persists for rotating BHs present in the framework of GR. 

In GR, Penrose's singularity 
theorem \cite{Penrose:1964wq} establishes that BH singularities at the origin can arise as a natural consequence of gravitational collapse. The validity of this theorem, however, hinges on several assumptions regarding the structure of spacetime and the properties of matter. Among these is the requirement of global hyperbolicity of spacetime. Violating this condition can potentially lead to the existence of nonsingular BHs.

A notable example of such a solution was first introduced by Bardeen \cite{Bardeen:1968}, who proposed a nonsingular BH with metric components that remain finite as $r \to 0$. Since then, various other regular 
BH metrics have been proposed in the literature, which offers alternative frameworks for addressing the singularity problem in BH physics \cite{Dymnikova:1992ux, Dymnikova:2004zc, Hayward:2005gi, Ansoldi:2008jw, Balart:2014cga, Fan:2016hvf, Simpson:2018tsi}.

Even though the nonsingular 
metrics are given apriori in the aforementioned approach, 
it remains to be seen whether they can be realized in some  
concrete theories. For this purpose, we need to take into account additional degrees of freedom (DOFs) beyond those appearing in GR. 
For example, we may consider 
scalar-tensor theories in which 
a new scalar DOF is incorporated into the gravitational action \cite{Fujii:2003pa}. 
In most general scalar-tensor theories with second-order field equations of motion 
(Horndeski theories \cite{Horndeski:1974wa,Deffayet:2011gz,Kobayashi:2011nu,Charmousis:2011bf}), it is known that the existence of SSS asymptotically-flat hairy BH solutions without ghost/Laplacian instabilities is quite limited \cite{Hui:2012qt,Creminelli:2020lxn,Minamitsuji:2022mlv,Minamitsuji:2022vbi}. 
For a radial dependent 
scalar profile $\phi(r)$, we need a coupling between $\phi$ and a Gauss-Bonnet curvature invariant \cite{Kanti:1995vq,Torii:1996yi,Sotiriou:2013qea,Doneva:2017bvd,Silva:2017uqg,Antoniou:2017acq}, but the Gauss-Bonnet term diverges 
at $r=0$. Hence the construction of nonsingular BHs in the context of scalar-tensor theories is generally challenging. 

If we consider vector-tensor theories 
in nonlinear electrodynamics (NED)
given by the Lagrangian ${\cal L}(F)$, 
where ${\cal L}$ is a nonlinear function 
of $F$, it is possible to realize 
nonsingular BHs without curvature 
singularities at $r=0$ \cite{Ayon-Beato:1998hmi, Ayon-Beato:1999kuh, Ayon-Beato:2000mjt, Bronnikov:2000vy, Dymnikova:2004zc,Balart:2014cga,Rodrigues:2018bdc,Maeda:2021jdc}. The NED Lagrangian accommodates 
Euler-Heisenberg theory \cite{Heisenberg:1936nmg} as well as
Born-Infeld theory \cite{Born:1934gh}. 
In such subclasses of NED theories, 
the resulting SSS BH solutions 
possess
curvature singularities at $r=0$ \cite{Yajima:2000kw, Fernando:2003tz, Cai:2004eh, Dey:2004yt}. 
However, there are nonsingular 
electrically or magnetically charged BHs 
for some specific choices of the NED Lagrangian. Thus, the 
vector field 
with nonlinear Lagrangians 
of $F$ allows an interesting possibility 
for realizing regular BHs even 
at the classical level. 

To determine the stability of nonsingular BHs, it is essential to analyze their linear stability by using BH perturbation theory. 
In Refs.~\cite{Moreno:2002gg, Toshmatov:2018tyo, Toshmatov:2018ell, Toshmatov:2019gxg, Nomura:2020tpc}, the authors discussed the BH stability by considering the propagation of dynamical perturbations in the region outside the event horizon. 
Although the conditions for the absence of ghosts and Laplacian instabilities can be satisfied outside the horizon, a recent analysis \cite{DeFelice:2024seu} shows that there is angular Laplacian instability of vector-field perturbations around the regular center.
This instability manifests for both electric and magnetic BHs, leading to the rapid 
enhancement of metric perturbations. Consequently, the nonsingular background metric cannot be maintained as a steady-state solution.
This means that nonsingular SSS BHs cannot be realized in the context of NED with the Lagrangian ${\cal L}(F)$.

Motivated by the no-go result in the context of NED, we extend our analysis to explore whether similar properties persist in more general classical field theories. To this end, we incorporate a scalar field $\phi$ with a kinetic term $X$ into the NED framework, considering a Lagrangian of the form ${\cal L}(F,\phi, X)$. 
For the gravity sector, we consider GR described by the Lagrangian $\Mpl^2 R/2$, where $\Mpl$ is the reduced Planck mass and $R$ is the Ricci scalar.
Applying such theories to the SSS background, we will show that there are no solutions with mixed electric and magnetic charges (as it happens in NED).
Hence we can focus on either electrically or magnetically charged objects, with 
a nontrivial scalar-field profile.

The theories we will study in this paper belong to a subclass of scalar-vector-tensor theories with second-order field equations of motion. Since they respect $U(1)$ gauge symmetry, there are one scalar, two transverse vectors, and two tensor polarizations as the propagating DOFs.
To derive the stability conditions of those five DOFs, we consider linear perturbations in both odd- and even-parity sectors on the SSS background. We expand the corresponding action up to the second order in perturbations by taking into account both electric and magnetic charges.
For the electric case, a similar analysis was performed in Refs.~\cite{Gannouji:2021oqz,Kase:2023kvq} as a subclass of Maxwell-Horndeski theories.
Since the linear stability of magnetic SSS objects has not been addressed yet, 
we will do so in this paper.

After deriving conditions for the absence 
of ghosts and Laplacian instabilities of
the five dynamical DOFs, we will apply them 
to three subclasses of 
${\cal L}(F,\phi,X)$ theories: 
(i) $\tilde{\cal L}(F)+K(\phi,X)$, 
(ii) ${\cal L}=X+\mu(\phi)F^n$, and 
(iii) ${\cal L}=X \kappa (F)$. 
We will show that regular SSS 
objects realized by theories (i) and 
(iii), which include nonsingular BHs, 
are excluded by Laplacian 
instabilities of vector-field 
perturbations around the origin. 
In theories (ii), the absence of 
angular Laplacian instabilities 
requires the condition $n>1/2$, 
under which there are no regular BHs with apparent  horizons. 
Thus, even by extending NED to more general theories with 
the Lagrangian ${\cal L}(F,\phi,X)$, 
we do not find even a single example 
of nonsingular BHs without instabilities.

This shows the general difficulty of constructing BHs without singularities in classical field theories.
In theories (ii), however, we will show the existence of linearly stable SSS 
compact objects without apparent  horizons. 
These regular solutions are present 
for both electric and magnetic configurations. We will clarify the regions 
of $n$ in which the regular 
horizonless compact objects are subject 
to neither ghosts 
nor Laplacian instabilities.

This paper is organized as follows. 
In Sec.~\ref{backsec}, we derive 
the SSS background solutions and discuss 
the properties of them for the electrically 
and magnetically charged cases.
In Sec.~\ref{persec}, we expand the action up to quadratic order in perturbations and obtain conditions under which neither ghosts nor Laplacian instabilities are present for five dynamical DOFs. 
In Sec.~\ref{kessencesec}, we apply the 
linear stability conditions to theories (i) 
mentioned above and show that nonsingular 
SSS objects are prone to angular Laplacian 
instability. 
In Sec.~\ref{muphisec}, we show that 
the absence of angular Laplacian instabilities demands the condition 
$n>1/2$, under which nonsingular 
BHs with apparent  horizons do not exist.
We also clarify the parameter space of 
$n$ in which horizonless regular 
compact objects suffer from neither 
ghosts nor Laplacian instabilities. 
In Sec.~\ref{kappasec}, we show that 
nonsingular SSS objects in theories (iii) 
are excluded by Laplacian instabilities either at the origin or 
at spatial infinity.
Sec.~\ref{consec} is devoted to conclusions.
 
\section{Field equations on the 
SSS background}
\label{backsec}

We consider theories in which the Lagrangian 
${\cal L}$ in the matter sector depends on 
a scalar field $\phi$ and the two scalar products
\be
X\equiv
-\frac{1}{2}\partial_{\mu}\phi 
\partial^{\mu}\phi\,,
\qquad 
F\equiv-\frac{1}{4}
F_{\mu \nu}F^{\mu \nu}\,,
\label{XFdef}
\ee
where 
$F_{\mu \nu}=\partial_{\mu}A_{\nu}-\partial_{\nu}A_{\mu}$ is the field strength of a covector field $A_{\mu}$. 
For the gravity sector, we consider GR described by the Einstein-Hilbert Lagrangian $\Mpl^2R/2$. 
Then, the total action is given by 
\be
{\cal S}=\int \rd^4 x \sqrt{-g} 
\left[ \frac{\Mpl^2}{2}R
+{\cal L}(F, \phi, X) \right]\,,
\label{action}
\ee
where $g$ is a determinant of the metric tensor $g_{\mu \nu}$.
 
We study SSS solutions on the background given by the line element
\be
\rd s^{2} =-f(r) \rd t^{2}
+h^{-1}(r)\rd r^{2} + r^{2} (\rd \theta^{2}+\sin^{2}\theta\, \rd \varphi^{2})\,,
\label{metric}
\ee
where $f$ and $h$ are functions of 
the radial distance $r$. 
For the positivity of $-g=(f/h)r^4 \sin^2 \theta$ in Eq.~(\ref{action}), we require that $f/h$ is positive. For the later convenience, we introduce the $r$-dependent 
function $N(r)$ satisfying 
\be
f(r)=N(r)h(r)\,,
\ee
so that 
\be
N(r)>0\,.
\label{Ncon}
\ee
In the following, we will use the two functions $N(r)$ and $h(r)$ instead of 
$f(r)$ and $h(r)$.

On the SSS background (\ref{metric}), 
we consider a radial dependent 
scalar-field profile $\phi(r)$. 
For the covector field $A_{\mu}$, the presence of $U(1)$ gauge symmetry in theories given by the action (\ref{action}) 
allows us to express the vector-field  components in the form   
\be
A_{\mu} \rd x^\mu=A_0(r)\rd t -q_M \cos \theta\rd\varphi\,,
\label{Amu}
\ee
where $A_0$ is a function of $r$, and $q_M$ is a constant corresponding to a magnetic charge. The scalar products defined in 
Eq.~(\ref{XFdef}) reduce to 
\be
X=-\frac{1}{2}h \phi'^2\,,\qquad 
F=\frac{A_0'^2}{2N}-\frac{q_M^2}{2r^4}\,,
\label{XF}
\ee
where a prime represents the derivative with respect to~$r$.

Varying the action (\ref{action}) with respect to $N$, $h$, $A_0$, and $\phi$, 
we obtain 
\ba
& &
h'-\frac{1-h}{r}-\frac{r}{\Mpl^2 N}
\left( N {\cal L}-A_0'^2 {\cal L}_{,F} \right)=0\,,
\label{back1}\\
& &
\frac{N'}{N}=\frac{r \phi'^2{\cal L}_{,X}}{\Mpl^2}\,,
\label{back2}\\
& &
\left( \frac{r^2 A_0' {\cal L}_{,F}}{\sqrt{N}} 
\right)' = 0\,,
\label{back3}\\
& &
{\cal E}_{\phi} \equiv
\left(  \sqrt{N} h r^2 \phi'{\cal L}_{,X} \right)'
+\sqrt{N}r^2 {\cal L}_{,\phi}=0\,,
\label{back4}
\ea
where the notations like
${\cal L}_{,F} \equiv \partial{\cal L}/\partial F$ are used 
for partial derivatives. 
If ${\cal L}_{,X}=0$, we require that 
$N'=0$ in general, or $N(r)=1$ by fixing 
boundary conditions at spatial infinity.
If $A_0'\neq0$, we can integrate Eq.~(\ref{back3}) to give 
\be
{\cal L}_{,F}=\frac{q_E \sqrt{N}}{r^2A_0'}\,,
\label{back3D}
\ee
where $q_E$ is an integration constant 
corresponding to an electric charge. 
From Eqs.~(\ref{back1}) and (\ref{back3D}), 
we have
\be
{\cal L}=\frac{\Mpl^2}{r^2} \left( rh'+h-1 \right)
+\frac{q_E A_0'}{\sqrt{N}r^2}\,.
\label{Lag}
\ee
Using Eqs.~(\ref{back2}) and (\ref{back4}), 
we can express ${\cal L}_{,X}$ and 
${\cal L}_{,\phi}$, as 
\ba
{\cal L}_{,X} &=& 
\frac{\Mpl^2 N'}{r\phi'^2 N}\,,
\label{LX} \\
{\cal L}_{,\phi} &=& -\frac{1}{\sqrt{N} r^2} \left( 
\frac{\Mpl^2 N' hr}{\sqrt{N} \phi'} \right)'\,,
\label{Lp}
\ea
which are valid for $\phi' \neq 0$.
Taking the $r$ derivative of 
Eq.~(\ref{Lag}), i.e., 
${\cal L}'(r)={\cal L}_{,F}F'+{\cal L}_{,\phi} \phi'+{\cal L}_{,X}X'$, 
and employing Eqs.~(\ref{XF}), (\ref{back3D}), (\ref{LX}), and (\ref{Lp}), we obtain  
\ba
& &
4q_E r^2 A_0'^2-\Mpl^2 r^2 N^{-3/2} [2N^2 (r^2 h''-2h+2) 
\nonumber \\
& &
+r N (2r h N''+3r h' N'+2 h N')-r^2 h N'^2]A_0'
\nonumber \\
& &
+4 N q_E q_M^2/r^2=0\,,
\label{A0con}
\ea
so that $A_0'$ is known algebraically 
in terms of $h$, $N$, and its $r$ derivatives. Interestingly, the 
$\phi$-dependent terms 
completely vanish in Eq.~(\ref{A0con}).

We will consider SSS objects that are regular at $r=0$, including nonsingular BHs with 
apparent  horizons \cite{Bardeen:1968,Dymnikova:1992ux, Dymnikova:2004zc, Hayward:2005gi, Ansoldi:2008jw, Balart:2014cga, Fan:2016hvf, Simpson:2018tsi}. 
In such cases, the Ricci scalar $R$, 
the squared Ricci tensor 
$R_{\mu \nu}R^{\mu \nu}$, and 
the squared Riemann tensor 
$R_{\mu \nu \rho \sigma}
R^{\mu \nu \rho \sigma}$ 
do not diverge at $r=0$. 
This requires that $h$ and $N$ are expanded 
around $r=0$, as \cite{Frolov:2016pav}
\ba
h(r) &=& 1+\sum_{n=2}^{\infty}h_n r^n\,,\label{hr}\\
N(r) &=& N_0+\sum_{n=2}^{\infty}N_n r^n\,,\label{Nr}
\ea
where $h_n$, $N_0$, and $N_n$ are constants. 
Note that $N_0$ is positive due to 
the condition (\ref{Ncon}).
We substitute Eqs.~(\ref{hr})-(\ref{Nr}) 
and their $r$ derivatives into Eq.~(\ref{A0con}). Solving the resulting equation for $A_0'$ 
and expanding it around $r=0$, we find
\be
A_0' =\pm \frac{\sqrt{N_0} 
\sqrt{-(q_E q_M)^2}}{q_E r^2}
+{\cal O}(r^0)\,.
\ee
Then, we have real solutions to $A_0'$ 
only if 
\be
q_E q_M=0\,,
\ee
and hence the dyon BHs with mixed electric and magnetic 
charges are not 
allowed\footnote{Equivalently, we may study the lowest order of a discriminant of the second-order algebraic equation and 
show that it is always negative.}. 
In the following, we will separate the 
discussion into the electrically 
and magnetically charged cases.

\subsection{Electric case}

For $q_E \neq 0$ and $q_M=0$, the nonvanishing 
solution to Eq.~(\ref{A0con}) is given by 
\ba
A_0' &=&\Mpl^2 [2N^2 (r^2 h''-2h+2)
+r N(2r h N''+3r h' N'
\nonumber \\
&&+2 h N')-r^2 h N'^2]/(4N^{3/2}q_E)\,,
\label{rA0}
\ea
which depends on the background metrics $h$ and $N$. 
Using the expansions (\ref{hr}) and (\ref{Nr}) 
around $r=0$, we have
\be
A_0'=\frac{2\Mpl^2 N_2}{\sqrt{N_0}q_E}r^2+
\frac{\Mpl^2 (4N_0 h_3+9N_3)}{2\sqrt{N_0}q_E}r^3+
{\cal O}(r^4)\,,
\label{A0D}
\ee
which is finite at $r=0$.
Substituting Eq.~(\ref{rA0}) into Eqs.~(\ref{XF}) 
and (\ref{Lag}), we know $F$ and ${\cal L}$ 
in terms of $h$, $N$, and their $r$ derivatives.
Under the expansions (\ref{hr}) and (\ref{Nr}) around $r=0$, 
it follows that 
\ba
F &=& \frac{2\Mpl^4 N_2^2}{N_0^2 q_E^2}r^4
+\frac{\Mpl^4 N_2 (4 N_0 h_3 + 9N_3)}
{N_0^2 q_E^2}r^5 \nonumber \\
& &+{\cal O}(r^6)\,,\\
{\cal L} &=& \frac{\Mpl^2 (3 N_0 h_2 + 2N_2)}{N_0}
+\frac{3\Mpl^2 (4 N_0 h_3 + 3N_3)}{2N_0}r 
\nonumber \\
& &+{\cal O}(r^2)\,,
\ea
which are both finite at $r=0$.

\subsection{Magnetic case}

For $q_M \neq 0$ and $q_E=0$, the solution to Eq.~(\ref{A0con}) yields 
\be
A_0'=0\,.
\ee
From Eq.~(\ref{Lag}), we have  
\be
{\cal L}=\frac{\Mpl^2}{r^2} 
\left( rh'+h-1 \right)\,.
\label{Lmag}
\ee
Applying the expansion (\ref{hr}) to 
Eq.~(\ref{Lmag}) gives
\be
{\cal L}=3\Mpl^2 h_2+4\Mpl^2 h_3 r
+{\cal O}(r^2)\,,
\label{Lre}
\ee
which approaches a constant as $r \to 0$. 
We note that the quantity 
$F=-q_M^2/(2r^4)$ diverges at $r=0$, but the form of ${\cal L}(F,\phi,X)$ 
can be designed to have the regular behavior 
(\ref{Lre}) at the origin. 
Indeed, this was already shown 
for nonsingular magnetic BHs present in 
the pure NED described by the 
Lagrangian ${\cal L}(F)$ \cite{Ayon-Beato:2000mjt,Bronnikov:2000vy}.

\section{Perturbations on the SSS 
background} 
\label{persec}

The linear stability of BHs can be analyzed by considering perturbations on the SSS 
background (\ref{metric}) \cite{Regge:1957td, Zerilli:1970se, Moncrief:1974ng, Zerilli:1974ai}.
We write the metric tensor in the form 
$g_{\mu \nu}=\bar{g}_{\mu \nu}+h_{\mu \nu}$, 
where $\bar{g}_{\mu \nu}$ is the background value and $h_{\mu \nu}$ is the metric perturbation. We expand $h_{\mu \nu}$ in terms of the spherical 
harmonics $Y_{lm}(\theta, \varphi)$. 
Without loss of generality, we will focus on 
the mode $m=0$ and express $Y_{l0}$ as $Y_{l}$ in the following. 
We also omit the summation for $l$ for each perturbed variable. 

We choose the four gauge conditions 
$h_{t \theta}=0$, $h_{\theta \theta}=0$, 
$h_{\varphi \varphi}=0$, and 
$h_{\theta \varphi}=0$. 
In this case, the four components 
of $\xi^{\mu}$ under the infinitesimal 
coordinate transformation 
$x^{\mu} \to x^{\mu}+\xi^{\mu}$ are fixed.
Then, the components of $h_{\mu \nu}$ are 
given by \cite{DeFelice:2011ka,Kobayashi:2012kh,Kobayashi:2014wsa,Kase:2023kvq}\footnote{In the version published in Physical Review D {\bf 111}, 064051 (2025),
there is a typo
in the $h_{rr}$ component of Eq.~(\ref{hcom}). 
Here, we correct $f^{-1}(r)$ to $h^{-1}(r)$.} 
\ba
\hspace{-0.1cm}
& &
h_{tt}=f(r) H_0 (t,r) Y_{l}(\theta), \quad 
h_{tr}=H_1 (t,r) Y_{l}(\theta),\quad 
h_{t \theta}=0, 
\nonumber \\
\hspace{-0.1cm}
& &
h_{t \varphi}=-Q(t,r) (\sin \theta) 
Y_{l, \theta} (\theta),
\;\; 
h_{rr}=h^{-1}(r) H_2(t,r) Y_{l}(\theta),\nonumber \\
\hspace{-0.1cm}
& &
h_{r \theta}=h_1 (t,r)Y_{l, \theta}(\theta),
\quad
h_{r \varphi}=-W(t,r) (\sin \theta) Y_{l,\theta} (\theta), 
\nonumber \\
\hspace{-0.1cm}
& &
h_{\theta \theta}=0,\quad 
h_{\varphi \varphi}=0,\quad
h_{\theta \varphi}=0\,,
\label{hcom}
\ea
where $H_0$, $H_1$, $H_2$, $h_1$, $Q$, 
and $W$ are functions of $t$ and $r$. 

We also decompose the scalar and vector fields, as 
\ba
\phi &=& 
\bar{\phi}(r)+\delta \phi (t,r) Y_{l}(\theta),\\
A_{\mu} &=& 
\bar{A}_{\mu}(r)+\delta A_\mu\,,
\label{perma}
\ea
where
\ba
& &
\delta A_t=\delta A_0 (t,r) Y_{l}(\theta),\qquad 
\delta A_r=\delta A_1 (t,r) Y_{l}(\theta),\qquad \nonumber \\
& &
\delta A_\theta=0,\qquad 
\delta A_{\varphi}=-\delta A(t,r) (\sin \theta) 
Y_{l,\theta}(\theta)\,.
\ea
Here, we have set $\delta A_\theta=0$ by 
exploiting the fact that the 
action (\ref{action}) 
respects $U(1)$ gauge invariance.

The odd-parity sector contains three perturbations 
$Q$, $W$, and $\delta A$, while there are 
seven perturbed fields $H_0$, $H_1$, $H_2$, 
$h_1$, $\delta \phi$, $\delta A_0$, 
and $\delta A_1$ in the even-parity sector. 
After integrating out all nondynamical fields, we have one scalar perturbation 
$\delta \phi$, two vector modes arising from 
$\delta A_{\mu}$, and two tensor polarizations arising from the gravity sector. For the electric BH, the linear stability conditions of such five dynamical perturbations were derived 
in Refs.~\cite{Gannouji:2021oqz,Kase:2023kvq} for more general 
Maxwell-Horndeski theories. 
For the magnetic BH, the stability issue 
has not been addressed yet in theories 
given by the action (\ref{action}).
In the following, we will obtain the full 
second-order action of linear perturbations 
and study the stability of SSS objects with electric and magnetic charges, in turn.

\subsection{Second-order action} 

We expand the action (\ref{action}) up 
to quadratic order in perturbations and integrate it with respect to $\theta$ and $\varphi$. 
After the integration by parts, the 
second-order action can be expressed 
in the form 
\be
{\cal S}^{(2)}=\int \rd t \rd r\,
\left( {\cal L}_1+{\cal L}_2 \right)\,, 
\label{calS}
\ee
where
\begin{widetext}
\ba
{\cal L}_1 &=&
a_0 H_0^2 
+ H_0 \left[ a_1 H_2' + L a_2 h_1' 
+(a_3+L a_4)H_2 + L a_5 h_1+L a_6 \delta A 
+a_7 \delta \phi'+a_8 \delta \phi 
\right]+Lb_1 H_1^2 
\nonumber \\
& &
+H_1 ( b_2 \dot{H}_2+L b_3 \dot{h}_1+b_4 \dot{\delta \phi})
+c_0 H_2^2 +L H_2 ( c_1 h_1+c_2 \delta A) 
+c_3 H_2 \delta \phi'+c_4 H_2 \delta \phi
+L (d_0 \dot{h}_1^2+ d_1 h_1^2) \nonumber \\
& &
+L h_1 (d_2 \delta A_0+d_3 \delta A'
+d_4 \delta \phi)
+s_1 ( \delta A_0'-\dot{\delta A_1} )^2
+(s_2 H_0 +s_3 H_2+L s_4 \delta A
+s_5 \delta \phi'+s_6 \delta \phi)
( \delta A_0'-\dot{\delta A_1}) \nonumber \\
& &
+L ( s_7 \delta A_0^2 +s_8 \delta A_1^2)
+u_1 \dot{\delta \phi}^2+u_2 \delta \phi'^2
+\left( L u_3+\tilde{u}_3 \right) \delta \phi^2+u_4 \delta \phi' \delta A
+u_5 \delta \phi\,\delta A\,,
\label{Lag1}\\
{\cal L}_2 &=& L [ p_1 (r\dot{W}-r Q'+2Q)^2+p_2 \delta A 
(r\dot{W}-r Q'+2Q)+p_3 \dot{\delta A^2}+p_4 \delta A'^2
+L p_5 \delta A^2+(L p_6+p_7)W^2 \nonumber \\
& &~~+(L p_8+p_9)Q^2+p_{10} Q \delta A_0
+p_{11}Q h_1+p_{12} W \delta A_1]\,,
\label{Lag2}
\ea
\end{widetext}
where a dot represents the derivative with 
respect to $t$, the coefficients $a_0$ etc 
are given in Appendix~A, and 
\be
L \equiv l(l+1)\,.
\ee
For both electric and magnetic BHs, we have 
\be
s_4=-\frac{q_M A_0'{\cal L}_{,FF}}{\sqrt{N}\,r^2}=0\,,
\ee
whose condition will be used in the following.

In the odd-parity sector, there are two 
dynamical perturbations \cite{Kase:2023kvq,DeFelice:2024seu}
\ba
& &
\chi \equiv r\dot{W}-rQ'+2Q
-\frac{2{\cal L}_{,F}rA_0'}{\Mpl^2}
\delta A\,,\label{chidef}\\
& &
\delta A\,,
\ea
which correspond to the gravitational and 
vector-field perturbations, respectively.

In the even-parity sector, there are three 
dynamical fields given by 
\ba
\hspace{-1cm}
& &\psi \equiv rH_2-L h_1\,,\\
\hspace{-1cm}
& &V \equiv \delta A_0'-\dot{\delta A}_1 
+\frac{s_2 H_0 +s_3 H_2
+s_5 \delta \phi'+s_6 \delta \phi}{2s_1},
\label{Vdef}\\
\hspace{-1cm}
& &\delta \phi\,,
\ea
which correspond to the gravitational, 
vector-field, and scalar-field perturbations, respectively.

Although we have split perturbations into 
even- and odd-parity modes, the way they couple with each other depends on the background. 
For instance, if $q_M$ does not vanish, 
there are coupling terms between $\delta\phi$ and $\delta A$, as $u_4 \neq 0 \neq u_5$. 

To simplify the analysis of the propagating dynamical DOFs, we introduce two auxiliary fields, $V$ and $\chi$, allowing the action to be reformulated as follows: 
\be
\tilde{\cal S}^{(2)}=
\int \rd t \rd r\,
\left( \tilde{{\cal L}}_1
+\tilde{{\cal L}}_2 \right)\,,
\label{calS2}
\ee
where 
\ba
\hspace{-0.7cm}
\tilde{{\cal L}_1}
&=& {\cal L}_1 
-s_1 \biggl[\delta A_0'-\dot{\delta A}_1 
\nonumber \\
& &
\qquad \quad
+\frac{s_2 H_0 +s_3 H_2
+s_5 \delta \phi'+s_6 \delta \phi}{2s_1}-V 
\biggr]^2,\\
\hspace{-0.7cm}
\tilde{{\cal L}}_2 &=&
{\cal L}_2-L p_1 
\left[ r\dot{W}-rQ'+2Q-\frac{2{\cal L}_{,F}rA_0'}
{\Mpl^2}\delta A-\chi \right]^2.\nonumber \\
\ea
Varying the action (\ref{calS2}) with respect to $V$ and $\chi$, we obtain the relations (\ref{Vdef}) and (\ref{chidef}), respectively.
Then, we find that the action (\ref{calS2}) is equivalent to (\ref{calS}).

\subsection{Stability of electric SSS objects} 

We first study the linear stability 
of electric SSS objects ($q_M=0$) by exploiting the action (\ref{calS2}).
This issue was addressed in Refs.~\cite{Gannouji:2021oqz,Kase:2023kvq}
as a special case of Maxwell-Horndeski theories, but we will revisit it to make a comparison with the stability of 
magnetic SSS objects.
For electric objects, $\tilde{{\cal L}}_1$ is 
composed of even-parity perturbations alone,  
while $\tilde{{\cal L}}_2$ contains 
only odd-parity perturbations.
In the following, we will study the case 
$h>0$, but we will also address 
the case $h<0$ at the end of 
this subsection.

\subsubsection{Odd-parity sector}

We first consider perturbations 
in the odd-parity sector. 
Varying the Lagrangian $\tilde{{\cal L}}_2$ 
with respect to $Q$ and $W$, we obtain 
\ba
Q &=& -\frac{h[2r N \chi'+(2N-rN')\chi]}
{2(L-2)N}\,,\\
W &=& -\frac{r \dot{\chi}}{(L-2)N h}\,.
\ea
Substituting these expressions of $Q$ and $W$ 
and their $t$, $r$ derivatives 
into $\tilde{{\cal L}}_2$ and integrating 
it by parts, the second-order Lagrangian 
is expressed in the form 
\be
\tilde{{\cal L}}_2=
\dot{\vec{\Psi}}_{\rm A}^{t}
{\bm K}_{\rm A} \dot{\vec{\Psi}}_{\rm A}
+\vec{\Psi}'^{t}{\bm G}_{\rm A} \vec{\Psi}_{\rm A}'
+\vec{\Psi}^{t}{\bm M}_{\rm A} \vec{\Psi}_{\rm A}\,,
\label{L2odd}
\ee
where ${\bm K}_{\rm A}$, ${\bm G}_{\rm A}$, 
and ${\bm M}_{\rm A}$ are the $2 \times 2$ 
matrices, and 
\be
\vec{\Psi}_{\rm A}^t=\left( \chi, \delta A 
\right)\,.
\ee
The matrix ${\bm K}_{\rm A}$ has only the diagonal 
components $(K_{\rm A})_{11}$ and 
$(K_{\rm A})_{22}$, so that the no-ghost 
conditions are given by 
\ba
(K_{\rm A})_{11} &=& \frac{\Mpl^2 L}
{4 h N^{3/2}(L-2)}>0\,,
\label{NG1odd} \\
(K_{\rm A})_{11} (K_{\rm A})_{22} &=& 
\frac{\Mpl^2 L^2 
{\cal L}_{,F}}{8h^2 N^2 (L-2)}>0\,.
\label{NG2odd} 
\ea
The inequality (\ref{NG1odd}) automatically 
holds for $l \geq 2$, while the other 
inequality (\ref{NG2odd}) is satisfied if
\be
{\cal L}_{,F}>0\,.
\label{LFcon}
\ee

To study the propagation of dynamical perturbations  
along the radial direction, we first vary the 
Lagrangian (\ref{L2odd}) with respect 
to $\chi$ and $\delta A$. 
Then, we assume solutions to the perturbation 
equations in the WKB form 
\be
\vec{\Psi}_{\rm A}^t=
(\vec{\Psi}_0^t)_{\rm A} 
e^{-i (\omega t-kr)}\,,
\ee
where $(\vec{\Psi}_0^t)_{\rm A}=
(\chi_0, \delta A_0)$ is a constant vector, 
$\omega$ is an angular frequency, and 
$k$ is a wavenumber.
This gives the algebraic equation
${\bm U}_{\rm A}(\vec{\Psi}_0)_{\rm A}=0$, 
where ${\bm U}_{\rm A}$ is a $2 \times 2$ matrix.
To allow the existence of nonvanishing solutions to $(\vec{\Psi}_0)_{\rm A}$, we require that the determinant of ${\bm U}_{\rm A}$ vanishes, i.e., 
\be
{\rm det}~{\bm U}_{\rm A}=0\,.
\label{detUA}
\ee
In the regime $h>0$,  
the radial propagation speed 
$c_r=h^{-1/2} {\rm d}r/{\rm d}\tau$ 
in proper time $\tau=\int 
\sqrt{Nh}\,{\rm d}t$ 
can be derived by substituting 
$\omega=h \sqrt{N} c_r k$ into Eq.~(\ref{detUA}).
Taking the large $\omega$ and $k$ limits,  
we obtain the two solutions
\be
c_r^2=1\,,\quad {\rm for} \quad 
\vec{\Psi}_{\rm A}^t=
\left( \chi, \delta A \right)\,,
\label{crA}
\ee
so that the two dynamical fields $\chi$ and 
$\delta A$ propagate with the speed of light 
along the radial direction.

The angular propagation speed measured by 
the proper time $\tau$ is given by 
$c_{\Omega}=r {\rm d}\theta/{\rm d}\tau
=(r/\sqrt{Nh})(\omega/l)$.
Taking the large $\omega$ and $l$ limits 
in Eq.~(\ref{detUA}), we obtain the following 
two solutions
\be
c_{\Omega}^2=1\,,\quad {\rm for} \quad 
\vec{\Psi}_{\rm A}^t=
\left( \chi, \delta A \right)\,,
\label{cOA}
\ee
and hence the angular propagation speeds of both $\chi$ and $\delta A$ are luminal.

\subsubsection{Even-parity sector}

In the even-parity sector, the product 
$H_0^2$ present in ${\cal L}_1$ disappears in 
$\tilde{\cal L}_1$ as a result of introducing 
the Lagrange multiplier $V$. 
Since $\tilde{\cal L}_1$ only contains terms 
linear in $H_0$, varying $\tilde{\cal L}_1$ 
with respect to $H_0$ puts constraints 
on other perturbed fields. 
We use this equation to express 
$h_1$ in terms of $\psi$, $V$, 
$\delta \phi$, and their derivatives. 
We also vary $\tilde{\cal L}_1$ with respect 
to $H_1$, $\delta A_0$, $\delta A_1$ and 
eliminate these fields from $\tilde{\cal L}_1$ 
by using their equations of motion.
Up to boundary terms, we can express 
$\tilde{\cal L}_1$ in the following form 
\be
\tilde{{\cal L}}_1=
\dot{\vec{\Psi}}_{\rm B}^{t}
{\bm K}_{\rm B} \dot{\vec{\Psi}}_{\rm B}
+\vec{\Psi}_{\rm B}'^{t}{\bm G}_{\rm B} 
\vec{\Psi}_{\rm B}'
+\vec{\Psi}_{\rm B}^{t}{\bm M}_{\rm B} 
\vec{\Psi}_{\rm B}
+\vec{\Psi}_{\rm B}^{t}{\bm Q}_{\rm B} 
\vec{\Psi}_{\rm B}'\,,
\label{L1even}
\ee
where ${\bm K}_{\rm B}$, ${\bm G}_{\rm B}$, 
${\bm M}_{\rm B}$ are $3 \times 3$ 
symmetric matrices, ${\bm Q}_{\rm B}$ is 
an antisymmetric matrix, and 
\be
\vec{\Psi}_{\rm B}^t=
\left( \psi, V, \delta \phi \right)\,.
\ee

The positivity of ${\bm K}_{\rm B}$ determines the no-ghost conditions. 
Taking the limit $l \gg 1$, they are given by 
\ba
\hspace{-0.8cm}
& &
({K_{\rm B}})_{22}=
\frac{r^4 ({\cal L}_{,F}
+2F {\cal L}_{,FF})^2}{2h N^{3/2}L {\cal L}_{,F}}>0\,,\\
\hspace{-0.8cm}
& &
({K_{\rm B}})_{22}({K_{\rm B}})_{33}
-({K_{\rm B}})_{23}^2 \nonumber \\
\hspace{-0.8cm}
& &
=
\frac{r^6 ({\cal L}_{,F}
+2F {\cal L}_{,FF})^2 {\cal L}_{,X}}
{4 h^2 N^2 L {\cal L}_{,F}}>0\,,\\
\hspace{-0.8cm}
{\rm det}\,{\bm K}_{\rm B} &=&
\frac{\Mpl^2 r^6 ({\cal L}_{,F}
+2F {\cal L}_{,FF})^2 {\cal L}_{,X}}
{4 h N^{5/2} L^3 {\cal L}_{,F}}>0\,.
\ea
These inequalities are satisfied if 
\be
{\cal L}_{,X}>0,\quad {\rm and} \quad 
{\cal L}_{,F}>0\,,
\label{NG2}
\ee
where the latter is the same as the no-ghost 
condition (\ref{LFcon}) in the odd-parity sector. 

The perturbation equations of motion for 
$\psi$, $V$, and $\delta \phi$ follow by 
varying (\ref{L1even}) with respect to 
these dynamical fields. We substitute the 
WKB-form solution 
\be
\vec{\Psi}_{\rm B}^t=
(\vec{\Psi}_0^t)_{\rm B} 
e^{-i (\omega t-kr)}\,,
\ee
into these equations, 
where $(\vec{\Psi}_0^t)_{\rm B}=
(\psi_0, V_0, \delta \phi_0)$ is 
a constant vector. 
The resulting algebraic equation
${\bm U}_{\rm B}(\vec{\Psi}_0)_{\rm B}=0$ 
has nonvanishing solutions of  $(\vec{\Psi}_0)_{\rm B}$, so long as the determinant of the $3 \times 3$ matrix 
${\bm U}_{\rm B}$ is vanishing, i.e., 
\be
{\rm det}~{\bm U}_{\rm B}=0\,.
\label{detUB}
\ee

The radial propagation speeds $c_r$ 
can be obtained by taking the limit 
$\omega r_h \approx k r_h \gg l \gg 1$
in Eq.~(\ref{detUB}) and substituting 
the relation $\omega=h \sqrt{N}c_r k$ 
into this equation.
Then, we obtain the following three squared 
propagation speeds
\ba 
& &
c_{r,\psi}^2=1\,,\label{crB1}\\
& &
c_{r,V}^2=1\,,\\
& &
c_{r, \delta \phi}^2=
1+\frac{2X[\cL_{,XX}(\cL_{,F}+2F\cL_{,FF})
-2F \cL_{,FX}^2]}
{\cL_{,X}(\cL_{,F}+2F\cL_{,FF})}. 
\label{crdelta}
\nonumber \\
\ea
In theories where ${\cal L}$ is a function of 
$F$ alone, there are two dynamical 
perturbations $\psi$ and $V$ that propagate 
with the speed of light
\cite{DeFelice:2024seu}. 
In current theories, we have an additional 
scalar perturbation $\delta \phi$ whose 
propagation speed is different from 1. 
To avoid the Laplacian instability of 
$\delta \phi$ along the radial direction, 
we require that $c_{r, \delta \phi}^2>0$.

Taking the large multipole limit characterized 
by the condition $l \approx \omega r_h  
\gg k r_h \gg 1$ in Eq.~(\ref{detUB}), 
the resulting squared angular propagation 
speeds are given by 
\ba
& &
c_{\Omega,\psi}^2=1\,,\label{cO1} \\
& &
c_{\Omega, V}^2=\frac{\cL_{,F}}
{\cL_{,F}+2F \cL_{,FF}}\,,
\label{cO2}\\
& &
c_{\Omega,\delta \phi}^2=1\,.
\label{cO3}
\ea
Here, the vector mode is decoupled from 
the other two modes as 
$K_{22} \omega^2+M_{22}L=0$.
While both $\psi$ and $\delta \phi$ have 
luminal propagation speeds, 
$c_{\Omega, V}^2$ is different from 1 
in theories containing nonlinear 
functions of $F$. The angular Laplacian 
instability of $V$ can be avoided 
if $c_{\Omega, V}^2>0$.

The above discussion is valid in the regime 
characterized by $h>0$, but we can perform 
a similar analysis for $h<0$. 
In the latter regime, the time-like and 
space-like properties of metric components 
$f~(=Nh)$ and $h$ are reversed compared to those in the former regime. 
The no-ghost conditions can be derived 
from the matrices ${\bm G}_{\rm A}$ and 
${\bm G}_{\rm B}$ rather than 
${\bm K}_{\rm A}$ and ${\bm K}_{\rm B}$. 
So long as ${\cal L}_{,F}>0$ and 
${\cal L}_{,X}>0$, there are 
no ghosts in either odd-parity 
or even-parity sectors.

The radial and angular propagation speeds  
of odd- and even-parity perturbations
can be derived by exploiting the WKB 
solutions $\vec{\Psi}_{\rm A}^t=
(\vec{\Psi}_0^t)_{\rm A} 
e^{-i (\omega r-kt)}$ and 
$\vec{\Psi}_{\rm B}^t=(\vec{\Psi}_0^t)_{\rm B} 
e^{-i (\omega r-kt)}$. 
On using the relation $\omega=kc_r/(-h \sqrt{N})$
in the limit $\omega r_h \approx k r_h \gg l 
\gg 1$, we obtain the same values of $c_r^2$ 
as those given in 
Eqs.~(\ref{crA}) and Eqs.~(\ref{crB1})-(\ref{crdelta}). 
Taking the other limit $l \approx \omega r_h 
\gg k r_h \gg 1$ with the relation 
$\omega=c_\Omega l/(r \sqrt{-h})$, we find 
that the squared angular propagation speeds 
are the same as those given 
in Eq.~(\ref{cOA}) and Eqs.~(\ref{cO1})-(\ref{cO3}).

In summary, for both $h>0$ and $h<0$, 
the linear stability of electric SSS 
objects is ensured under the 
four conditions ${\cal L}_{,X}>0$, 
${\cal L}_{,F}>0$, $c_{r, \delta \phi}^2>0$, 
and $c_{\Omega, V}^2>0$, 
where 
$c_{r, \delta \phi}^2$ and 
$c_{\Omega,V}^2$ are given, 
respectively, by Eqs.~(\ref{crdelta}) 
and (\ref{cO2}).

\subsection{Stability of magnetic 
SSS objects} 

Let us proceed to the stability of magnetic 
SSS objects, in which case $A_0'$ 
is vanishing.  
We consider the regime $h>0$, but we will 
briefly mention the case $h<0$ at the end 
of this subsection. 
For $q_M \neq 0$ and $q_E=0$, 
the second-order action (\ref{calS2}) is
decomposed into the two sectors described 
by the combinations of perturbations 
\be
\vec{\Psi}_{\rm C}^{t}=\left( \chi, V 
\right)\,,\qquad
\vec{\Psi}_{\rm D}^{t}=\left( \delta A, 
\psi, \delta \phi \right)\,,
\ee
which we call sectors C and D, 
respectively. The vector 
$\vec{\Psi}_{\rm C}^{t}$ is composed 
of the odd-parity gravitational perturbation 
$\chi$ and the even-parity 
vector-field perturbation $V$.
The vector $\vec{\Psi}_{\rm D}^{t}$ 
consists of the odd-parity vector-field 
perturbation $\delta A$, the even-parity 
gravitational perturbation $\psi$, and 
the even-parity scalar-field perturbation 
$\delta \phi$. Since the sectors C and D 
contain both odd- and even-parity modes, 
we deal with the total action 
(\ref{calS2}) at once.

First of all, we derive the field equations 
of motion for $Q$ and $W$ from 
Eq.~(\ref{calS2}). 
They are used to eliminate $Q$, $W$, and 
their derivatives from the second-order action $\tilde{{\cal S}}^{(2)}$. 
Then, we vary the resulting action 
with respect to $H_0$, $H_1$, 
$\delta A_0$, and $\delta A_1$. 
This allows us to solve these 
perturbation equations for $h_1$, 
$H_1$, $\delta A_0$, and $\delta A_1$, 
so that these fields are removed 
from the action. After the integration 
by parts, the final second-order action 
can be expressed in the form 
\be
\tilde{{\cal S}}^{(2)}=\int 
{\rm d}t {\rm d}r \left( 
\tilde{\cal L}_{\rm C}+
\tilde{\cal L}_{\rm D} \right)\,,
\ee
where $\tilde{\cal L}_{\rm C}$ and 
$\tilde{\cal L}_{\rm D}$ are the Lagrangians containing perturbations in the 
sectors C and D, respectively. 
In the following, we will 
address the linear stability of magnetic 
SSS objects for the sectors C and D, 
in turn.

\subsubsection{Sector C}

The Lagrangian $\tilde{{\cal L}}_{\rm C}$ 
is of the following form 
\be
\tilde{{\cal L}}_{\rm C}=
\dot{\vec{\Psi}}_{\rm C}^{t}
{\bm K}_{\rm C} \dot{\vec{\Psi}}_{\rm C}
+\vec{\Psi}_{\rm C}'^{t}{\bm G}_{\rm C} \vec{\Psi}_{\rm C}'
+\vec{\Psi}_{\rm C}^{t}{\bm M}_{\rm C} 
\vec{\Psi}_{\rm C}
+\vec{\Psi}_{\rm C}^{t}{\bm Q}_{\rm C} 
\vec{\Psi}_{\rm C}'\,,
\label{LC}
\ee
where ${\bm K}_{\rm C}$, 
${\bm G}_{\rm C}$, ${\bm M}_{\rm C}$ 
are $2 \times 2$ symmetric matrices, 
while ${\bm Q}_{\rm C}$ is 
antisymmetric. Unlike ${\bm K}_{\rm A}$, 
the kinetic matrix ${\bm K}_{\rm C}$ 
has both diagonal and off-diagonal components. 
Then, the no-ghost conditions 
for perturbations in the 
sector C are given by 
\ba
\hspace{-0.5cm}
(K_{\rm C})_{11} &=& \frac{\Mpl^2 L}
{4h N^{3/2} (L-2)}>0\,,
\label{NGma1} \\
\hspace{-0.5cm}
(K_{\rm C})_{11}(K_{\rm C})_{22}
-(K_{\rm C})_{12}^2 &=& 
\frac{\Mpl^2 r^4 {\cal L}_{,F}}
{8 h^2 N^3 (L-2)}>0\,.
\label{NGma2}
\ea
The first inequality (\ref{NGma1}) is automatically satisfied for $l \geq 2$, 
whereas the second inequality 
(\ref{NGma2}) holds if 
\be
{\cal L}_{,F}>0\,,
\ee
which is the same as the no-ghost condition 
in the sector~A.

We derive the perturbation equations for 
$\chi$ and $V$ from the Lagrangian 
(\ref{LC}) and substitute the WKB 
solution $\vec{\Psi}_{\rm C}^t=
(\vec{\Psi}_0^t)_{\rm C}\,e^{-i (\omega t-kr)}$ into them. 
The resulting equations are expressed in
the form ${\bm U}_{\rm C}(\vec{\Psi}_0)_{\rm C}=0$. 
Taking the limit $\omega r_h \approx 
k r_h \gg l \gg 1$ and using 
the relation $\omega=h \sqrt{N}c_r k$ 
in the determinant equation 
${\rm det}\,{\bm U}_{\rm C}=0$, 
we obtain the two squared radial 
propagation speeds 
\be
c_r^2=1\,,\quad {\rm for} \quad 
\vec{\Psi}_{\rm C}^t=
\left( \chi, V \right)\,, 
\ee
which are both luminal. 
In the other limit 
$l \approx \omega r_h \gg k r_h \gg 1$, 
we substitute the relation 
$\omega=l \sqrt{Nh}\,c_{\Omega}/r$ 
into ${\rm det}\,{\bm U}_{\rm C}=0$. 
This leads to the two squared angular 
propagation speeds 
\be
c_{\Omega}^2=1\,,\quad {\rm for} \quad 
\vec{\Psi}_{\rm C}^t=
\left( \chi, V \right)\,,
\ee
both of which are luminal as well. 

\subsubsection{Sector D}

The Lagrangian in the sector D 
can be expressed in the form 
\be
\tilde{{\cal L}}_{\rm D}=
\dot{\vec{\Psi}}_{\rm D}^{t}
{\bm K}_{\rm D} \dot{\vec{\Psi}}_{\rm D}
+\vec{\Psi}_{\rm D}'^{t}{\bm G}_{\rm D} \vec{\Psi}_{\rm D}'
+\vec{\Psi}_{\rm D}^{t}{\bm M}_{\rm D} 
\vec{\Psi}_{\rm D}
+\vec{\Psi}_{\rm D}^{t}{\bm Q}_{\rm D} 
\vec{\Psi}_{\rm D}'\,,
\label{LD}
\ee
where ${\bm K}_{\rm D}$, 
${\bm G}_{\rm D}$, ${\bm M}_{\rm D}$ 
are $3 \times 3$ symmetric matrices, 
whereas ${\bm Q}_{\rm D}$ is antisymmetric. 
From the kinetic matrix ${\bm K}_{\rm D}$, 
we obtain the following three no-ghost  conditions
\ba
({K_{\rm D}})_{33}&=&
\frac{r^2 {\cal L}_{,X}}
{2h \sqrt{N}}>0\,,\\
({K_{\rm D}})_{22}({K_{\rm D}})_{33}
&-&({K_{\rm D}})_{23}^2 
=\frac{\Mpl^2 r^2 {\cal L}_{,X}}{2 N L^2}>0\,,\\
{\rm det}\,{\bm K}_{\rm D} &=&
\frac{\Mpl^2 r^2{\cal L}_{,F}
{\cal L}_{,X}}
{4 h L N^{3/2}}>0\,.
\ea
These inequalities are satisfied if
\be
{\cal L}_{,X}>0\,,\quad {\rm and} 
\quad {\cal L}_{,F}>0\,,
\ee
which are the same as the 
no-ghost conditions of electric 
SSS objects in the even-parity sector. 

Varying the Lagrangian (\ref{LD}) with 
respect to $\delta A$, $\psi$, 
$\delta \phi$ and using the 
WKB solution $\vec{\Psi}_{\rm D}^t=
(\vec{\Psi}_0^t)_{\rm D}\,e^{-i (\omega t-kr)}$, 
we can write the perturbation equations in the form 
${\bm U}_{\rm D}(\vec{\Psi}_0)_{\rm D}=0$. 
The squared radial propagation speeds 
can be obtained by taking the limit 
$\omega r_h \approx 
k r_h \gg l \gg 1$ in the determinant 
equation ${\rm det}\,{\bm U}_{\rm D}=0$, 
leading to 
\ba 
& &
c_{r,\delta A}^2=1\,,\\
& &
c_{r,\psi}^2=1\,,\\
& &
c_{r, \delta \phi}^2=
1+\frac{2X {\cal L}_{,XX}}
{{\cal L}_{,X}}\,.
\label{crdelta2}
\ea
In theories with ${\cal L}={\cal L}(F)$, 
the dynamical perturbations $\delta A$ 
and $\psi$ propagate with the speed of 
light, as consistent with the result 
in Ref.~\cite{DeFelice:2024seu}. 
In current theories, the additional 
scalar DOF $\delta \phi$ has 
the propagation speed different from 1. 
As we can compare with Eq.~(\ref{crdelta}), 
the electric SSS object has 
a different value of 
$c_{r, \delta \phi}^2$ in comparison 
to Eq.~(\ref{crdelta2}). 
In theories with ${\cal L}_{,FX}=0$, 
they are identical to each other. 

Taking the other limit 
$l \approx \omega r_h \gg k r_h \gg 1$ in the determinant equation ${\rm det}\,{\bm U}_{\rm D}=0$, 
we obtain the following squared angular 
propagation speeds 
\ba
& &
c_{\Omega,\delta A}^2=
1+\frac{2F{\cal L}_{,FF}}
{{\cal L}_{,F}}\,,
\label{cOdelta1}\\
& &
c_{\Omega,\psi}^2=1\,,\\
& &
c_{\Omega, \delta \phi}^2=1\,.
\label{cOdelta2}
\ea
If ${\cal L}$ contains nonlinear functions 
of $F$, the propagation speed of $\delta A$ 
is different from 1.
The expression of 
$c_{\Omega,\delta A}^2$ coincides with 
the one derived in Ref.~\cite{DeFelice:2024seu} for theories with ${\cal L}={\cal L}(F)$. 

We have also studied the case $h<0$ and 
obtained the same no-ghost 
conditions and radial/angular propagation 
speeds as those derived for $h>0$. 
In summary, the linear 
stability of magnetic SSS objects requires that the four conditions 
${\cal L}_{,X}>0$, ${\cal L}_{,F}>0$, 
$c_{r, \delta \phi}^2>0$, and 
$c_{\Omega,\delta A}^2>0$ 
are satisfied, where 
$c_{r, \delta \phi}^2$ and 
$c_{\Omega,\delta A}^2$ are given, 
respectively by Eqs.~(\ref{crdelta2}) and 
(\ref{cOdelta1}).

\section{k-essence theories with NED} 
\label{kessencesec}

As a first example of the possible 
realization of nonsingular SSS objects, 
we will discuss the case of NED in the 
presence of a k-essence scalar field.
The k-essence Lagrangian of the form 
$K(\phi,X)$ was originally introduced 
in the context of inflation and 
dark energy \cite{Armendariz-Picon:1999hyi, Chiba:1999ka,Armendariz-Picon:2000nqq}.
Now, we consider the following Lagrangian 
\be
\cL=\tilde{\cL}(F)+K(\phi, X)\,,
\label{tLK}
\ee
where $\tilde{\cL}$ is a function of $F$ alone, 
and $K$ depends on $\phi$ and $X$.
Since the electromagnetic field is not 
directly coupled to the scalar field, 
we have that $\cL_{,F}=\tilde{\cL}_{,F}$ 
and ${\cal L}_{,FX}=0$. 
We note that nonsingular magnetic 
black-bounce solutions were recently 
studied in Einstein gravity 
with the Lagrangian (\ref{tLK}) \cite{Pereira:2024rtv}.
In the following, we will study  
the stability of electric and magnetic 
SSS objects in turn.

\subsection{Electric case}

For $q_E \neq 0$ and $q_M=0$, the squared 
propagation speeds (\ref{crdelta}) and 
(\ref{cO2}) in the even-parity sector yield
\ba
c_{r, \delta \phi}^2 &=& 
1+\frac{2X K_{,XX}}{K_{,X}}\,,\\
c_{\Omega, V}^2 &=& 
\frac{\tilde{\cL}_{,F}}
{\tilde{\cL}_{,F}+2F \tilde{\cL}_{,FF}}\,,
\label{cO2D}
\ea
whereas all the other dynamical perturbations 
propagate with the speed of light.
We can compute $\tilde{\cL}_{,FF}$ 
by taking the $r$ derivatives of 
$\tilde{\cL}_{,F}=q_E \sqrt{N}/(r^2 A_0')$ 
and $F=A_0'^2/(2N)$, as 
$\tilde{\cL}_{,FF}=\tilde{\cL}_{,F}'(r)/F'(r)$. 
Then, Eq.~(\ref{cO2D}) reduces to 
\be
c_{\Omega, V}^2=\frac{r(A_0'N'-2A_0'' N)}
{4A_0' N}\,.
\ee
Since $A_0'$ is given by Eq.~(\ref{rA0}), 
we can express 
$c_{\Omega, V}^2$ in terms of $h$, $N$, 
and their $r$ derivatives, as 
\begin{widetext}
\ba
c_{\Omega, V}^2 &=& 
-r( 2N^3r^2 h'''
+ 2 h r^2 N^2 N'''+3 r^2 N^2 N' h'' 
+ 5 r^2 h'N^2 N''-4h r^2 N N'N'' 
- 4 h'r^2 NN'^2  +2 hr^2 N'^3  \nonumber \\
& &
+ 4 N^3 rh''+6 h r N^2 N''+ 8h'r N^2 N' 
- 4hrN N'^2 - 4h' N^3 + 2h N^2 N')
\nonumber \\
& &
/[{2N(2N^2r^2h''+2 hr^2 N N''
+3 h'r^2 N N'-h r^2 N'^2
+ 2 hr N N'- 4h N^2 + 4N^2)}]\,.
\label{cOkes}
\ea
\end{widetext}
Using the expansions (\ref{hr}) 
and (\ref{Nr}) of $h$ and $N$ 
around $r=0$, we obtain 
\be
c_{\Omega, V}^2=-1-\frac{4N_0 h_3
+9N_3}{8N_2}r+{\cal O}(r^2)\,.
\label{cOV1}
\ee
Nonsingular BHs studied in the literature
typically have the properties $h_3=0$ and 
$N_3=0$ \cite{Bardeen:1968,Ayon-Beato:1998hmi,Dymnikova:2004zc,Hayward:2005gi}.
In such cases, the expansion of $c_{\Omega, V}^2$ around $r=0$ leads to
\ba
c_{\Omega, V}^2 
&=&
-1+\frac{3N_2^2-N_0 (5h_2 N_2+8N_4)
-5h_4 N_0^2}{2N_0 N_2}r^2
\nonumber \\
&&+{\cal O}(r^3)\,.
\label{cOV2}
\ea
Since the leading-order contributions to 
$c_{\Omega, V}^2$ are negative for both 
the cases (\ref{cOV1}) and (\ref{cOV2}), 
the vector-field perturbation $V$ is subject 
to Laplacian instability in the 
angular direction. 
We note that NED without the scalar field 
corresponds to $N(r)=1$ for all $r$.
In this case, Eq.~(\ref{cOkes}) 
reduces to the value of $c_{\Omega, V}^2$ 
derived in Ref.~\cite{DeFelice:2024seu}.
For $N(r)=1$, the leading-order term of 
$c_{\Omega, V}^2$ is also negative.

Since the vector-field perturbation is 
coupled to the gravitational perturbation, the Laplacian instability of $V$ leads to the enhancement of $\psi$ along the angular direction.
As studied in Ref.~\cite{DeFelice:2024seu}, 
the typical time scale of instability can be 
estimated as $t_{\rm ins} \simeq 
r/(\sqrt{-c_\Omega^2}\,l)$. 
For $l \gg 1$, $t_{\rm ins}$ is 
infinitely small. Due to this rapid growth 
of even-parity perturbations around $r=0$, 
the line element of nonsingular electric 
SSS objects cannot be sustained 
in a steady state. 
We note that the choice of the 
scalar-field Lagrangian $K(\phi, X)$ 
does not affect the discussion given above. 
In other words, no matter how we choose 
the functional forms of $K(\phi, X)$, 
the angular instabilities of $V$ and $\psi$ 
are inevitable for electric SSS objects.

\subsection{Magnetic case}

For the magnetic case, the squared propagation 
speeds (\ref{crdelta2}) and (\ref{cOdelta1}) 
in the sector D reduce, respectively, to 
\ba
c_{r, \delta \phi}^2 &=& 
1+\frac{2X K_{,XX}}{K_{,X}}\,,\\
c_{\Omega, \delta A}^2 &=& 
\frac{\tilde{\cL}_{,F}+2F \tilde{\cL}_{,FF}}
{\tilde{\cL}_{,F}}\,,
\label{cO3d}
\ea
while all the other dynamical perturbations  
have luminal propagation speeds. 
From Eq.~(\ref{Lmag}), we have
\be
\tilde{\cal L}(F)+K(\phi, X)
=\frac{\Mpl^2}{r^2} 
\left( rh'+h-1 \right)\,,
\label{Lmag2}
\ee
with $F=-q_M^2/(2r^4)$. 
Taking the $r$ derivative of Eq.~(\ref{Lmag2}) 
and exploiting Eqs.~(\ref{back2}) and 
(\ref{back4}), we find that 
$\tilde{\cal L}_{,F}$ is written as 
\ba
\tilde{\cal L}_{,F}
&=& \Mpl^2 r^2 [2N^2 (r^2h''-2h+2)
+2 rh N (r N''+N') \nonumber \\
& &+3r^2 h' N'N-r^2h N'^2]/(4q_M^2 N^2)\,.
\ea
Differentiating this equation with respect to 
$r$, we can express $\tilde{\cal L}_{,FF}$ in terms of $N$, $h$, and their $r$ derivatives. 
Then, it follows that $c_{\Omega, \delta A}^2$ 
is completely identical to $c_{\Omega, V}^2$ 
for the electric configuration  
given by Eq.~(\ref{cOkes}).

Using the expansions of $h$ and $N$ 
around $r=0$, the leading-order contribution 
to $c_{\Omega, \delta A}^2$ is $-1$ 
and hence the odd-parity vector-field 
perturbation $\delta A$ is subject to 
angular Laplacian instability.
This leads to the enhancement of the gravitational perturbation $\psi$. 
Then, the nonsingular magnetic SSS object cannot be present as a stable configuration. 

In summary, for theories with the Lagrangian (\ref{tLK}), we have shown that all the nonsingular electric and magnetic SSS objects 
are excluded by angular Laplacian instabilities 
arising from vector-field perturbations. 
This includes nonsingular BHs 
constructed from the Lagrangian 
(\ref{tLK}), which extends our previous 
results found for NED \cite{DeFelice:2024seu}.

\section{Theories with $\cL=X+\mu(\phi) F^n$}
\label{muphisec}

In theories where the Lagrangian ${\cal L}$ 
contains the nonlinear dependence of $F$, 
$\cL_{,FF}$ does not vanish 
in Eq.~(\ref{cO2}) or Eq.~(\ref{cOdelta1}). 
As we showed in Sec.~\ref{kessencesec}, 
this results in the negative values of $c_{\Omega, V}^2$ or 
$c_{\Omega,\delta A}^2$ around $r=0$. 
If we consider theories in which 
${\cal L}$ contains only a linear term 
in $F$, it is possible to avoid the problem 
of angular Laplacian instabilities. 
We also note that the nonlinear 
dependence of $X$ in ${\cal L}$ 
leads to the deviation 
of $c_{r,\delta \phi}^2$ from 1. 
The linear term in $X$ without a direct 
coupling to $F$ results in the value 
$c_{r,\delta \phi}^2=1$ for both 
electric and magnetic cases.

In this section, we consider theories 
given by the Lagrangian 
\be
\cL=X+\mu(\phi) F^n\,,
\label{EMD}
\ee
where $\mu$ is a function of $\phi$, 
and $n$ is an integer. 
Einstein-Maxwell-scalar theories 
correspond to the particular power $n=1$.
A dilaton field in string theory
has an exponential coupling 
$\mu (\phi)=\mu_0 
e^{-\lambda \phi}$ \cite{Gasperini:2002bn}.
In such Einstein-Maxwell-dilaton theories, 
it is known that there is an exact hairy 
BH solution with a singularity at $r=0$ \cite{Gibbons:1987ps, Garfinkle:1990qj}. 
If $\mu$ contains even power-law functions of 
$\phi$, tachyonic instability of the RN branch can give rise to scalarized charged BH solutions with curvature singularities 
at $r=0$ \cite{Herdeiro:2018wub,Fernandes:2019rez,Myung:2018jvi,Blazquez-Salcedo:2020nhs}.
In this work, we would like 
to explore whether 
stable nonsingular BHs and compact objects can be present in Einstein-Maxwell-scalar theories for general power $n$.

The absence of nonlinear terms in $X$ 
leads to the luminal propagation of 
the scalar field, i.e., 
\be
c_{r,\delta \phi}^2=1\,,
\ee
for both electric and magnetic configurations. 
From Eqs.~(\ref{cO2}) and (\ref{cOdelta1}), 
we have 
\ba
c_{\Omega, V}^2 &=&
\frac{1}{2n-1},\qquad 
{\rm for~the~electric~case}\label{eq:cOm_secV_el}\,,\\
c_{\Omega, \delta A}^2 &=&
2n-1,\qquad\,
{\rm for~the~magnetic~case}\,.\label{eq:cOm_secV_mg}
\ea
Then, the angular Laplacian instability 
is absent if 
\be
n>\frac{1}{2}\,.
\label{ncon}
\ee
One of the no-ghost conditions 
${\cal L}_{,X}>0$ is trivially satisfied. 
On the other hand, the other no-ghost 
condition ${\cal L}_{,F}>0$ is given by 
\be
{\cal L}_{,F}=
n \mu (\phi) \left( 
\frac{A_0'^2}{2N}-\frac{q_M^2}{2r^4} 
\right)^{n-1}>0\,.
\label{muphi}
\ee
For integer-odd values of $n$ 
in the range (\ref{ncon}), 
this inequality translates to $\mu (\phi)>0$.
In the following, we will study the two cases: 
(i) $q_E \neq 0$, $q_M=0$, and 
(ii) $q_M \neq 0$, $q_E=0$, in turn. 
In case (ii), if $n$ is not an integer, 
we should deal with $F^n$ in the Lagrangian, as $(F^2)^m$ with $n=2m$, due to the negativity of $F$.

\subsection{Electric case}

For $q_E \neq 0$ and $q_M=0$, 
Eqs.~(\ref{back3D}) and (\ref{LX}) give 
\ba
\mu(\phi) &=& \frac{q_E \sqrt{N}}{n r^2 A_0' 
F^{n-1}}\,,\label{muco}\\
\phi' &=& \Mpl \sqrt{\frac{N'}{rN}}\,,
\label{phico}
\ea
where we have chosen the branch $\phi'>0$ without 
loss of generality. 
For the realization of the solution (\ref{phico}), 
we require that $N'>0$. 
From Eqs.~(\ref{Lag}) and (\ref{EMD}), we have 
\be
-\frac{1}{2}h \phi'^2+\mu(\phi) 
\left( \frac{A_0'^2}{2N} \right)^n=
\frac{\Mpl^2}{r^2} \left( rh'+h-1 \right)
+\frac{q_E A_0'}{\sqrt{N}r^2}\,.
\label{Lagre}
\ee
Substituting Eqs.~(\ref{rA0}), (\ref{muco}), 
(\ref{phico}) into Eq.~(\ref{Lagre}), 
we find that $h(r)$ and $N(r)$ 
are related to each other, as
\ba
& &
2[ (2n-1)r^2 h''+4nr h'+2h-2]N^2
-(2n-1)r^2 h N'^2\nonumber \\
& &
+[3(2n-1)r^2 h'
+2(4n-1)rh]NN'\nonumber \\
& &
+2(2n-1)r^2 h NN''=0\,.
\label{hNeq}
\ea
By using this relation, we can express $A_0'$ in Eq.~\eqref{rA0} 
in a simpler form
\begin{equation}
\frac{q_E A_0'}{\Mpl^2}=
-\frac{n \left[2\left(r h'
+h -1\right) N +r hN' \right]}{\sqrt{N}\, \left(2 n -1\right)}\,.\label{eq:rA0_secV}
\end{equation}

Since $n > 1/2$, the integrated solution to Eq.~(\ref{hNeq}) is expressed in the form 
\ba
\hspace{-1.0cm}
h(r) &=& 
\frac{2\int_0^r r_1^{-\frac{4(n-1)}
{2n-1}} 
\sqrt{N(r_1)} \bigl(\int_{0}^{r_1} \sqrt{N(r_2)}
\rd r_2\bigr)\rd r_1}
{(2n-1) r^{\frac{2}{2n-1}} N(r)} \nonumber \\
\hspace{-1.0cm}
& &+\frac{c_1}{r^{\frac{2}{2n-1}}N(r)}
+\frac{c_2\int_0^r r_1^{-\frac{4(n-1)}
{2n-1}} \sqrt{N(r_1)}\rd r_1
}{(2n-1) r^{\frac{2}{2n-1}} N(r)}\,,
\label{hint0}
\ea
where $c_1$ and $c_2$ are integration 
constants. We recall that $N(r)$ is 
expanded as Eq.~(\ref{Nr}) around $r=0$. 
To avoid the divergence of $h(r)$ at 
$r=0$, we require that $c_1=0=c_2$.  
Then, the solution (\ref{hint0}) 
reduces to
\be
h(r)=\frac{2\int_0^r r_1^{-\frac{4(n-1)}
{2n-1}} \sqrt{N(r_1)} \bigl(\int_{0}^{r_1} \sqrt{N(r_2)}\rd r_2\bigr)\rd r_1}
{(2n-1) r^{\frac{2}{2n-1}}N(r)}.
\label{hint2}
\ee
Under the linear stability condition 
(\ref{ncon}), $h(r)$ is always positive 
at any distance $r \geq 0$. 
Therefore, we do not have the nonsingular
BH configuration with an apparent  horizon.

On using the expansion (\ref{Nr}) of 
$N(r)$ around $r=0$, we find that 
Eq.~(\ref{hint2}) can be expanded as
\be
h(r)=1-\frac{3n-1}{3n} 
\frac{N_2}{N_0} r^2
+{\cal O}(r^3)\,,
\ee
whose dependence is analogous to the standard 
boundary condition of metrics of stars. 
However, we need to make sure that 
the radial derivatives of $A_0$ and $\phi$
are vanishing at $r=0$ for the realization 
of regular SSS objects. 
From Eq.~(\ref{A0D}), the leading-order 
term of $A_0'(r)$ is proportional to $r^2$ and hence $A_0'(r) \to 0$ as $r \to 0$. 
Applying the expansion (\ref{Nr}) to 
Eq.~(\ref{phico}) for $N_2 \neq 0$, 
it follows that 
\be
\frac{\phi'(r)}\Mpl= \sqrt{\frac{2N_2}{N_0}}
+\frac{3}{4N_2}\sqrt{\frac{2N_2}{N_0}}
N_3\,r+{\cal O}(r^2)\,.
\ee
As $r \to 0$, $\phi'(r)$ approaches a nonvanishing constant $\Mpl \sqrt{2N_2/N_0}$. 
This can lead to a cusp-like structure around the origin, whose property
should be incompatible with the SSS background.
To avoid this behavior, we need to impose 
the condition $N_2=0$. Moreover, 
the realization of the regular solution 
$\phi' \propto r$ around $r=0$ requires that 
$N_3=0$. Then, so long as the expansion of 
$N(r)$ around $r=0$ is given by 
\be
N(r)=N_0+N_4 r^4+{\cal O}(r^5)\,,
\label{Nsmall}
\ee
the scalar-field derivative and 
the kinetic term have
the following behavior 
\begin{align}
\phi'(r)&=2\Mpl \sqrt{\frac{N_4}{N_0}}r
+{\cal O}(r^2)\,,\label{phib}\\
X(r)&=-2\Mpl^2\frac{N_4}{N_0}r^2+{\cal O}(r^3)\,.
\end{align}
From Eq.~(\ref{phib}), we require that 
$N_4>0$. We also have the following expansions
\ba
h(r) 
&=& 
1-\frac{20n-8}{20n-5}\,\frac{N_4}{N_0}\,r^4+\mathcal{O}(r^5)\,,
\label{hr=0} \\
A_0'(r) &=& \frac{12\Mpl^2 n N_4}
{q_E \sqrt{N_0} (4n-1)}r^4
+{\cal O}(r^5)\,,
\label{rA0r=0}
\ea
so that $F=\mathcal{O}(r^8)$. 
We notice that $h(r)<1$ under the inequalities 
$n>1/2$, $N_0>0$, and $N_4>0$.
Thus, $N(r)$ is constrained to be of the form (\ref{Nsmall}) to realize the 
regular behavior $\phi'(r) \propto r$ 
around the origin. 

One of the examples for $N(r)$ 
that has the property (\ref{Nsmall}) 
is given by 
\be
N(r)=\left( \frac{r^4+\sqrt{N_0}r_0^4}
{r^4+r_0^4} \right)^2\,,
\label{Nexample}
\ee
where $N_0$ and $r_0$ are positive constants. 
Around $r=0$, this function is expanded as
\be
N(r)=N_0+\frac{2\sqrt{N_0}(1-\sqrt{N_0})}{r_0^4}r^4+{\cal O}(r^8)\,, 
\ee
and hence $N_4=2\sqrt{N_0}(1-\sqrt{N_0})/r_0^4$. 
Since the inequality $N_4>0$ holds,  
$N_0$ should be in the range 
\be
0<N_0<1\,.
\ee
At spatial infinity, the function 
(\ref{Nexample}) approaches 1 with 
a correction of order $r^{-4}$.

Let us pause for a moment to reflect on the reasoning behind the choice (\ref{Nexample}). 
In the standard approach, one typically specifies the function $\mu(\phi)$ and then searches for solutions by imposing boundary conditions that are consistent with the given 
choice of $\mu(\phi)$. 
In contrast, we adopt a different strategy here. We prescribe the form of $N(r)$ 
and determine the corresponding $\phi(r)$ and $\mu(r)$ by integrating Eqs.~\eqref{phico} and \eqref{eq:rA0_secV}, together with the relation 
(\ref{Lagre}).

As $\phi$ evolves from $\phi_0$ (its value at the origin) to $\phi_\infty$ (its value at infinity), we will demonstrate that it is, in principle, possible to reconstruct $\mu(\phi)$ for $\phi_0<\phi<\phi_\infty$, assuming the branch where $\phi'>0$. Once $\mu(\phi)$ is obtained--at least over this range of $\phi$--we can revert to the conventional approach of solving for configurations compatible with the reconstructed $\mu(\phi)$.

In general, the set of solutions corresponding to this $\mu(\phi)$, if it is neither empty nor a singleton, will 
be distinguished by their respective values of mass 
and charge of the object.\footnote{This would mean that, for a fixed $\mu(\phi)$, there would be more families of profiles, e.g., for $N(r)$ and $\phi(r)$.} The variation in these parameters may result in different forms of 
$N(r)$,\footnote{Notice, however, that 
for the RN solution in GR, although the mass and charge vary, the function $N$ is unity over 
the whole manifold. 
For perfect fluids, we instead have $N\neq1$, when 
a non-negligible pressure is present.} which must satisfy the boundary conditions $N'(r\to0) \propto r^3$ at the origin and $N \to 1$ at infinity.\footnote{In addition, we require the 
scalar-field dependence $\phi'(r \to 0) \propto r$ 
as well as all the other boundary conditions that define regular objects with asymptotically flat spacetime.} Exploring this second avenue, checking the properties of solutions for a fixed $\mu(\phi)$  is an interesting prospect. Still, we believe that its investigation lies beyond the scope of this work and should be addressed in a future study.

For $n$ in the range $n>1/2$ 
with the choice (\ref{Nexample}), 
Eq.~(\ref{hint2}) shows that $h(r)$ 
also approaches 1 as $r \to \infty$.
Then, the background metric components satisfy the condition for asymptotic flatness.
At large distances ($r \gg r_0$), the differential Eq.~\eqref{hNeq} 
is approximately given by 
\begin{align}
(2n-1) h'' \simeq  -\frac{4 n}{r}h'
-\frac{2}{r^2}h+\frac{2}{r^2}\,,
\label{ddeqap}
\end{align}
where we have kept the most dominant $r$-dependent contributions 
in the coefficients of $h'$ and $h$. 
For $n>1/2$, we can integrate Eq.~(\ref{ddeqap})
to give
\be
h(r)=1+\frac{c_1}{r}+c_2 r^{-\frac2{2n-1}}\,,
\qquad {\rm for} \quad n \neq \frac{3}{2}\,,
\label{hLso}
\ee
and 
\be
h=1+\frac{c_1}{r}+c_2 \frac{\ln\,r}{r}\,,
\qquad {\rm for} \quad n=\frac{3}{2}\,,
\label{hLso2}
\ee
where $c_1$ and $c_2$ are constants. 
Up to the next-to-leading order terms 
to $h(r)$, we can classify the large-distance 
behavior of $h(r)$, as 
\ba
h(r) &\simeq& 1+\frac{c_1}{r}\,,\qquad {\rm for} \quad 
\frac{1}{2}<n<\frac{3}{2}\,,\label{hL1}\\
h(r) &\simeq& 1+c_2 \frac{\ln r}{r}\,,\qquad {\rm for} \quad n=\frac{3}{2}\,,\label{hL2}\\
h(r) &\simeq& 1+c_2 r^{-\frac{2}{2n-1}}\,,\qquad {\rm for} \quad n>\frac{3}{2}\,.\label{hL3}
\ea
In the last case (\ref{hL3}), the term 
$r^{-\frac{2}{2n-1}}$ decreases slowly 
relative to $r^{-1}$.

The Arnowitt-Deser-Misner (ADM) mass 
of SSS objects is defined by  
\begin{equation}
M(r)\equiv \lim_{r \to \infty}
\frac{r}{2G}\,(1-h)\,.
\label{ADM}
\end{equation}
where $G$ is the gravitational constant.
From Eqs.~(\ref{hL2}) and (\ref{hL3}),  
for $n \geq 3/2$, the quantity $r(1-h)$ 
increases at large distances. 
In this case, we do not have compact SSS objects. 
For $n$ in the range
\be
\frac{1}{2}<n<\frac{3}{2}\,,
\ee
the metric component behaves as Eq.~(\ref{hL1}) and hence 
$M(r)$ approaches a constant value $-c_1/(2G)$. 
This is the case in which nonsingular SSS objects
satisfy the condition for compactness.

Applying Eq.~(\ref{Nexample}) to 
Eq.~(\ref{phico}), the scalar-field derivative 
has the following behavior in the regime $r \gg r_0$:
\be
\phi'(r)=\frac{2\Mpl r_0^2 
\sqrt{2(1-\sqrt{N_0})}}{r^3}
+{\cal O}(r^{-7})\,.
\label{rphiL}
\ee
We note that the dependence 
$\phi'(r) \propto r^{-3}$ is 
an outcome of the particular 
choice (\ref{Nexample}). 
If we consider $N(r)$ with the 
large-distance behavior 
$N(r)=1+{\cal O}(r^{-2})$, 
then we have the dependence 
$\phi'(r) \propto r^{-2}$.
Substituting the solution \eqref{hLso} of $h(r)$ into Eq.~\eqref{eq:rA0_secV}, 
we find 
\begin{align}
\frac{q_E A_0'}{\Mpl^2}
=-\frac{2 c_{2}   
\left(2 n -3\right) n}
{\left(2 n -1\right)^{2}
r^{\frac{2}{2 n -1}}}+\mathcal{O}(r^{-4})\,,
\label{A0inf}
\end{align}
where the term proportional to 
$r^{-{\frac{2}{2n -1}}}$ 
dominates over $r^{-4}$ for $3/4<n<3/2$.
As we will see later in Fig.~\ref{fig5}, 
the allowed values of $n$ that are consistent 
with no-ghost conditions are indeed larger 
than $3/4$.

\begin{figure}[ht]
\begin{center}
\includegraphics[height=3.3in,width=3.5in]{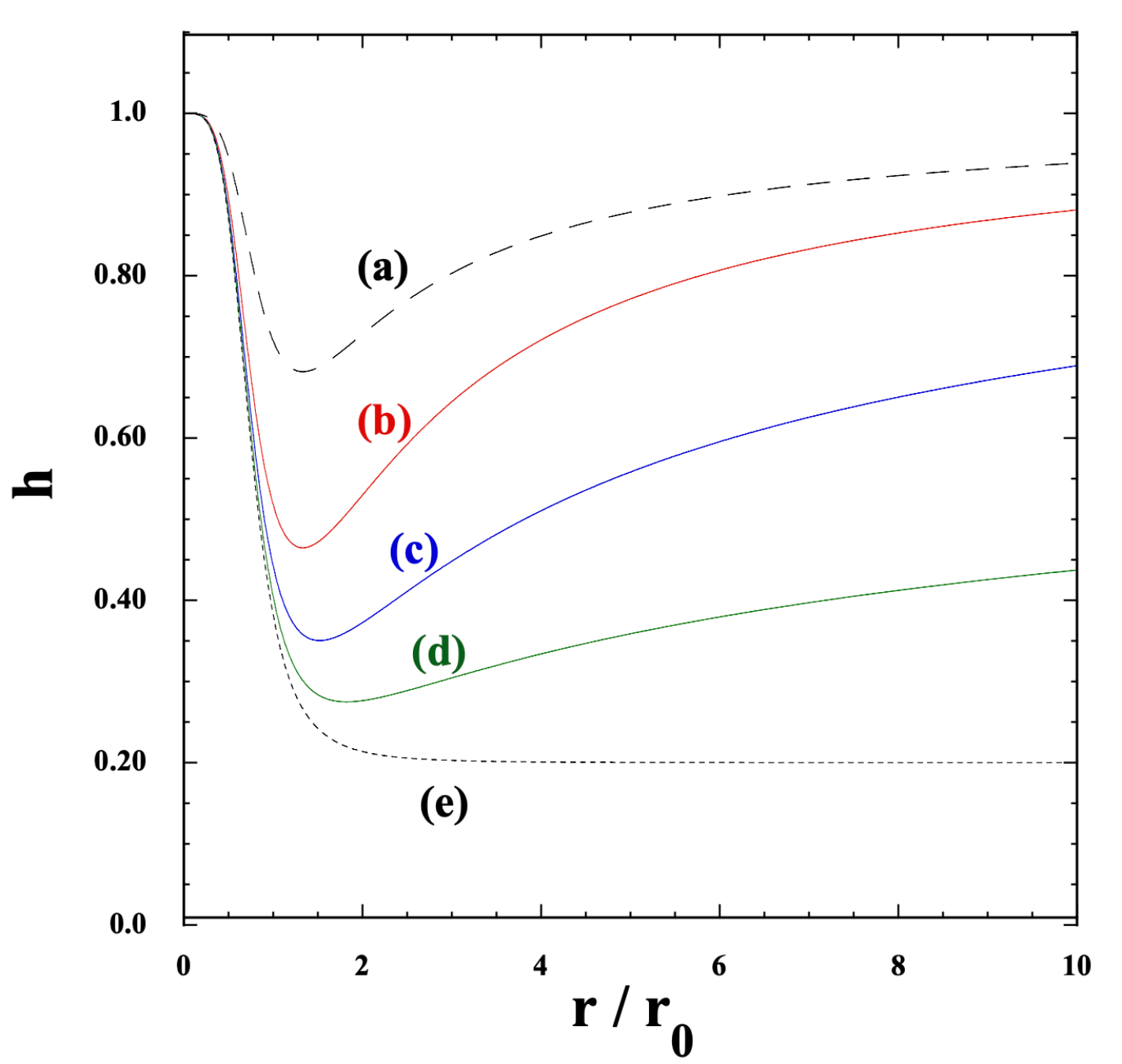}
\end{center}
\caption{
Metric component $h(r)$ versus $r/r_0$ for 
electric SSS objects present in theories 
given by the Lagrangian (\ref{EMD}). 
We choose $N(r)$ of the form 
(\ref{Nexample}) with $N_0=0.2$.  
Each case corresponds to (a) $n=1/2$, 
(b) $n=1$, (c) $n=2$, (d) $n=5$, and 
(e) $n \gg 1$. 
For $n$ in the range $n>1/2$, 
the theoretical lines of $h(r)$ are 
between (a) and (e), so that $h(r)$ is 
always positive at any distance $r$.
\label{fig1}}
\end{figure}

\begin{figure}[ht]
\begin{center}
\includegraphics[height=3.5in,width=3.5in]{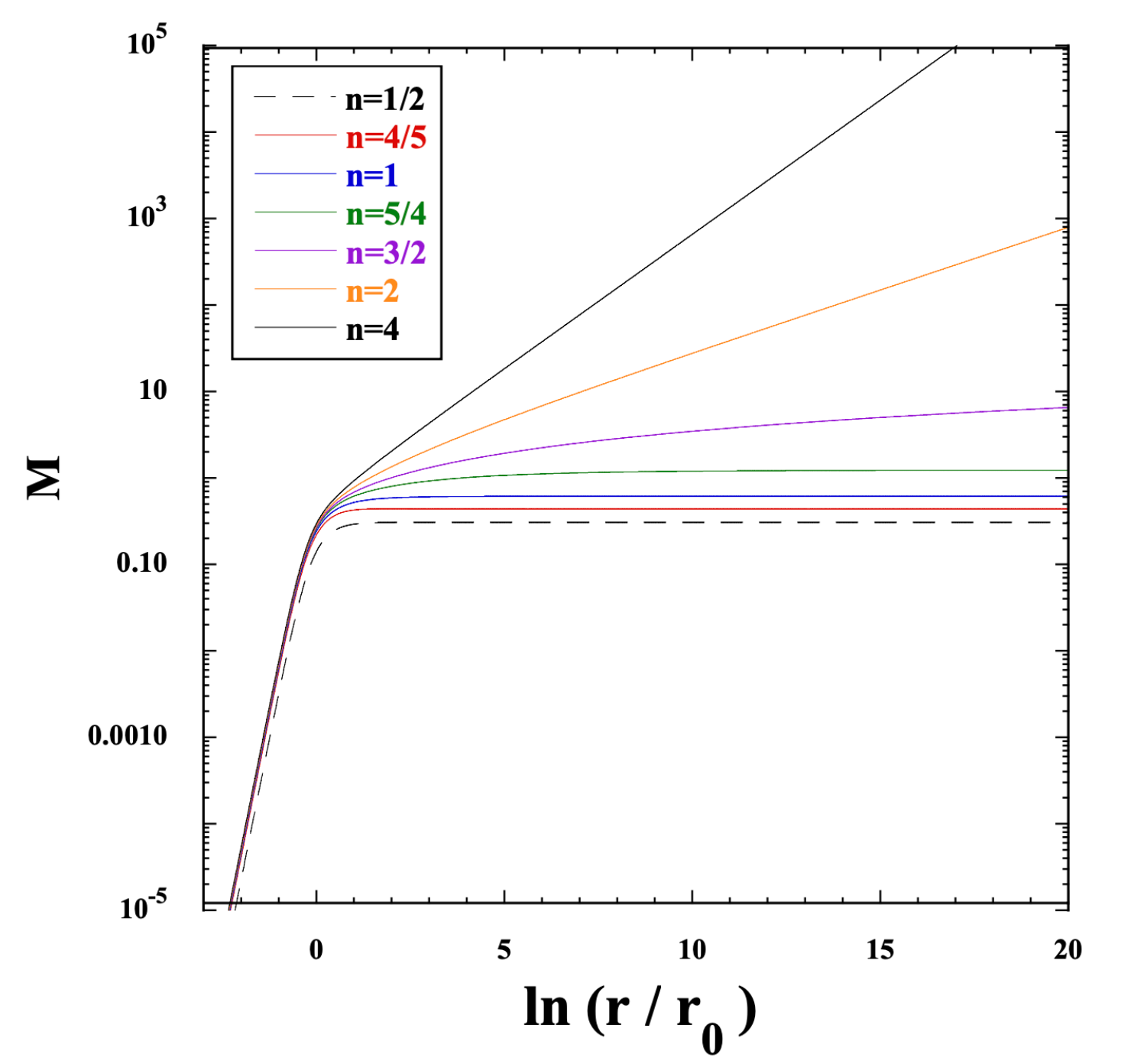}
\end{center}
\caption{ADM mass $M$ (unit of $G=1$) versus 
$r/r_0$ for electric SSS objects present 
in theories given by the Lagrangian (\ref{EMD}). 
We choose $N(r)$ to be  
(\ref{Nexample}) with $N_0=0.2$. 
From bottom to top, each line
corresponds to $n=1/2, 4/5, 1, 5/4, 3/2, 2, 4$. 
For $n<3/2$, $M$ asymptotically approaches 
a constant, while, for $n \geq 3/2$, $M$ grows 
in the regime $r \gg r_0$.
\label{fig2}}
\end{figure}

In the limit $n \to 1/2$, both $h''(r)$ and $N''(r)$ vanish in Eq.~(\ref{hNeq}). 
In this case, the integrated solution to 
Eq.~(\ref{hNeq}) is given by 
\be
h(r)=\frac{1}{r \sqrt{N(r)}} \left( \int_0^{r}
\sqrt{N(r_1)}\,{\rm d}r_1+c_1 \right)\,,
\label{hn1}
\ee
where the constant $c_1$ 
should be 0 to avoid the divergence 
of $h(r)$ at $r=0$. 
Then, Eq.~(\ref{hn1}) reduces to
\be
h_{n\to 1/2}(r)=\frac{1}{r \sqrt{N(r)}} 
\int_0^{r}
\sqrt{N(r_1)}\,{\rm d}r_1 \,,
\label{hn12}
\ee
which is positive at any 
distance $r \geq 0$.

In the limit that $n \gg 1$, 
Eq.~(\ref{hNeq}) yields
\be
h''=\frac{r(hN'^2-3NN'h'-2hNN'')
-4N(Nh'+hN')}{2rN^2}.
\ee
The solution to this equation can be 
written as
\be
h(r)=\frac{c_1}{N(r)}+\frac{c_2}{N(r)}
\int_0^r \frac{\sqrt{N(r_1)}}{r_1^2} 
{\rm d}r_1\,.
\ee
To match this with the expansions (\ref{hr}) and (\ref{Nr}) 
around $r=0$, the integration constants are fixed to be $c_1=N_0$ and $c_2=0$.
Then, we obtain  
\be
h_{n \gg 1}(r)=\frac{N_0}{N(r)}\,,
\label{hngg}
\ee
which is always positive. 
At spatial infinity, $h_{n \gg 1}(r)$ 
approaches $N_0$, so the condition 
for asymptotic flatness is not satisfied 
unless $N_0=1$. 
For finite values of $n$ in the range 
$n>1/2$, however, we have already seen that 
$h(r)$ approaches 1 as $r \to \infty$.

In Fig.~\ref{fig1}, we plot $h$ versus $r/r_0$ 
for $n=1/2, 1, 2, 5$ and $n \gg 1$, with the choice $N_0=0.2$ in Eq.~(\ref{Nexample}). 
In each case, we integrate Eq.~(\ref{hNeq}) 
outward from $r=0$ by using the boundary 
condition (\ref{hr=0}). 
Except for the limit $n \gg 1$, the initial decrease of $h(r)$, which is characterized by 
the solution (\ref{hr=0}), 
changes to its growth toward 
the asymptotic value $h(r) \to 1$.
In the limit $n \gg 1$, i.e., curve (e) in Fig.~\ref{fig1}, $h(r)$ monotonically decreases toward the asymptotic value $N_0$, with $h(r)>0$ for any distance $r$.
Under the linear stability condition $n>1/2$,  
the theoretical curves of $h(r)$ lie 
between (a) and (e) in Fig.~\ref{fig1}. 
In this region, as we showed analytically, 
we have $h(r)>0$ at any distance $r$. 
Thus, so long as the linear stability 
conditions are satisfied, we do not realize 
nonsingular BHs where $h(r)$ becomes negative 
for some range of $r$.

In Fig.~\ref{fig2}, we plot the ADM mass 
(\ref{ADM}) for seven different values 
of $n$, with $N_0=0.2$. 
The analytic estimation (\ref{hL1}) 
of $h(r)$ shows that  
$M$ asymptotically approaches constants
for $1/2<n<3/2$, whose property can be 
confirmed in Fig.~\ref{fig2}. 
For $n=3/2$, the ADM mass exhibits the 
logarithmic growth $M \propto \ln r$ 
due to the property of Eq.~(\ref{hL2}). 
For $n>3/2$, Eq.~(\ref{hL3}) gives the 
analytic dependence 
$M \propto r^{\frac{2n-3}{2n-1}}$. 
In Fig.~\ref{fig2}, we can confirm 
this power-law growth of $M$ for $n>3/2$ 
at large distances. 
Thus, so long as $1/2<n<3/2$,
the electric SSS configuration without 
the apparent horizon can be interpreted as a compact object.

\begin{figure}[ht]
\begin{center}
\includegraphics[height=3.4in,width=3.5in]{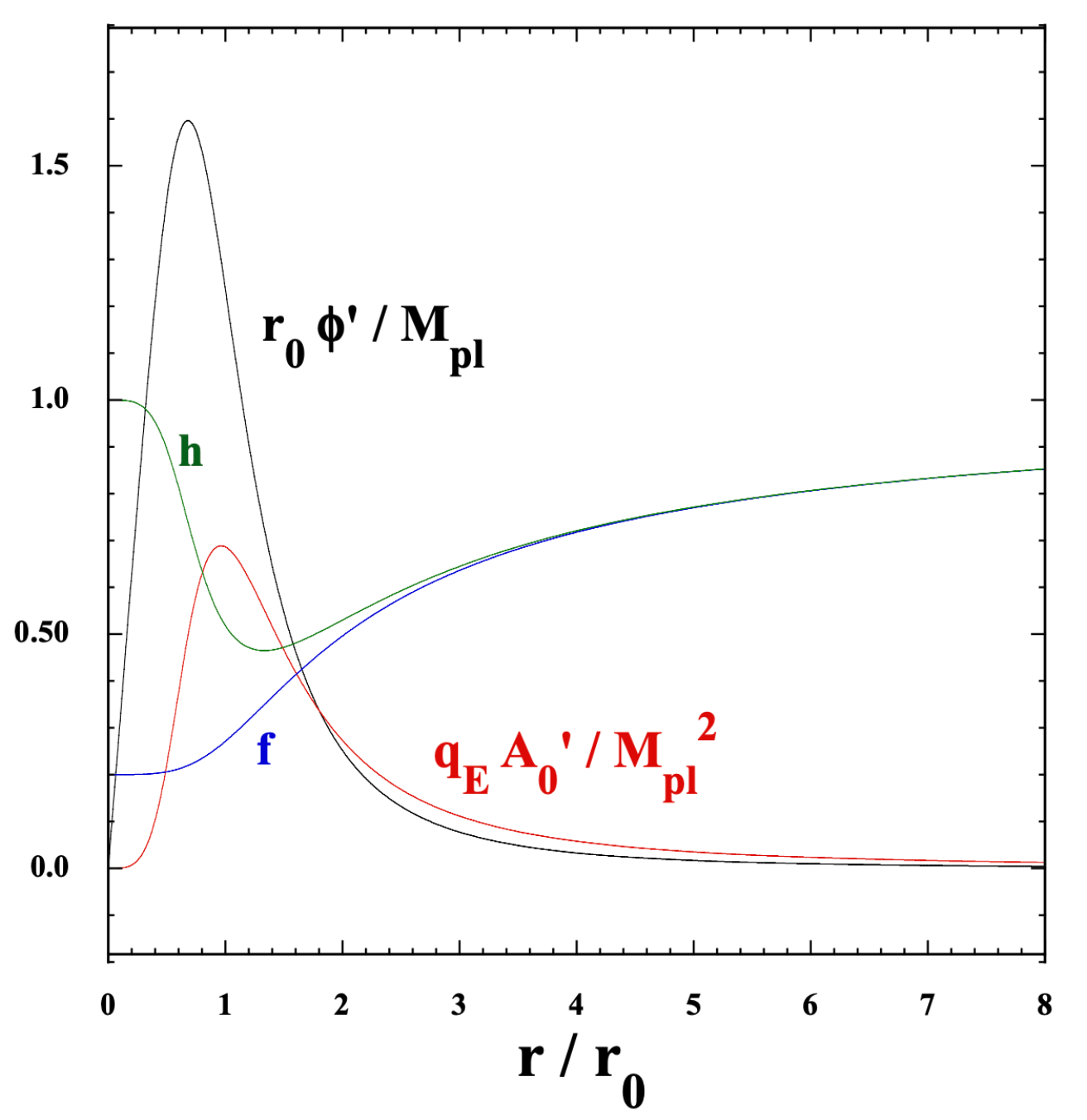}
\end{center}
\caption{We plot $r_0\phi'/\Mpl$, $q_E A_0'/\Mpl^2$, $f$, and $h$ versus $r/r_0$ for the electric BH in theories given by the Lagrangian (\ref{EMD}) with $n=1$. We choose $N(r)$ to be (\ref{Nexample}) 
with $N_0=0.2$. We observe that $\phi'$ and $A_0'$ 
approach 0 as $r \to 0$. 
The two metric components $f$ and $h$ are different around $r=0$, but both converge to 1 
at spatial infinity.
\label{fig3}}
\end{figure}

\begin{figure}[ht]
\begin{center}
\includegraphics[height=3.3in,width=3.5in]{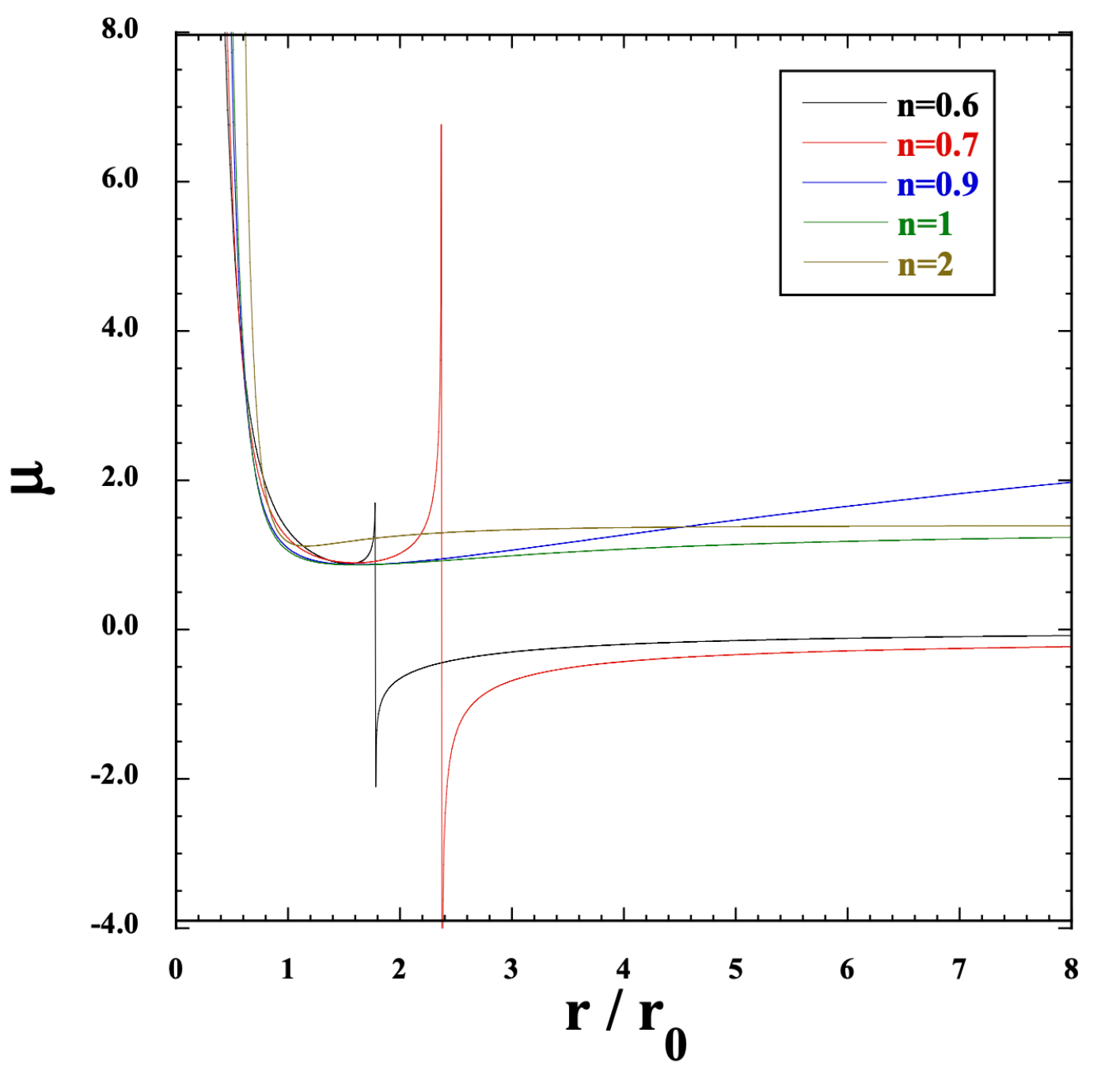}
\end{center}
\caption{The coupling $\mu$ versus $r/r_0$ for the 
electric SSS object with the choice $N_0=0.2$ 
in Eq.~(\ref{Nexample}). 
We choose five different powers: 
$n=0.6, 0.7, 0.9, 1, 2$. 
For $n>0.887$, $\mu$ is positive at any 
distance $r$.
\label{fig4}}
\end{figure}

\begin{figure}[ht]
\begin{center}
\includegraphics[height=3.3in,width=3.5in]{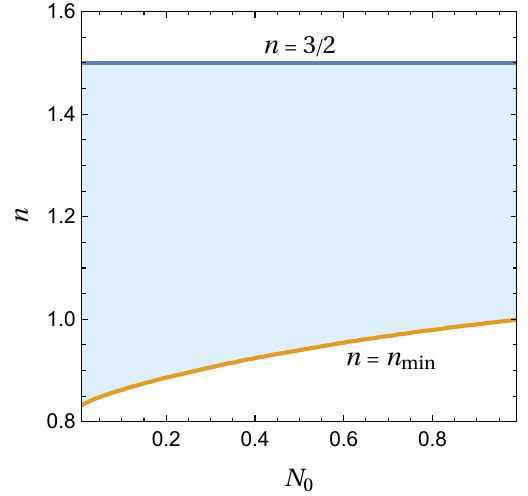}
\end{center}
\caption{The allowed region of parameter space for 
electric compact objects, represented by the 
light-blue shaded area. 
The lower boundary corresponds to the values of 
$n_{\rm min}$, below which the no-ghost condition is violated for a given $N_0$ in the range $0<N_0<1$. 
The upper boundary, represented by $n=3/2$, marks the limit beyond which the solutions no longer 
describe compact objects. 
Thus, the allowed range for $n$ is 
$n_{\rm min}\leq n<3/2$. The curve of 
$n_{\rm min}(N_0)$ is obtained by interpolating over a set of 99 numerically computed values of $n_{\rm min}$ for $N_0$ in the range $0.01 \leq N_0 \leq 0.99$.
\label{fig5}}
\end{figure}

In Fig.~\ref{fig3}, we show $r_0\phi'/\Mpl$, 
$q_E A_0'/\Mpl^2$, $f$, and $h$ as functions of $r/r_0$ for $n=1$ with $N_0=0.2$ in Eq.~(\ref{Nexample}).
As estimated from Eqs.~(\ref{phib}) and (\ref{rA0r=0}), numerical results show the 
properties $\phi' \propto r$ and 
$A_0' \propto r^4$ around $r=0$. 
According to the analytic estimations 
given in Eqs.~(\ref{rphiL}) and (\ref{A0inf}), 
$\phi'$ and $A_0'$ should have the 
large-distance behaviors $\phi' \propto r^{-3}$ 
and $A_0' \propto r^{-2}$ for $n=1$.
Indeed, after reaching the maximum values 
at intermediate distances, we confirm that 
$\phi'$ and $A_0'$ approach these asymptotic 
solutions. As we showed analytically, 
the asymptotic behavior of $A_0'$ is 
different depending on the power $n$.

The metric component $f=Nh$ differs from 
$h$ by the factor $N$, whose difference 
is largest at $r=0$ by the factor $N_0=0.2$. 
Since $N$ grows from $N_0$ to 1 for 
increasing $r$, $f$ approaches $h$ at large 
distances and they finally converge to 
the asymptotic values $f \to 1$ and 
$h \to 1$. The analytic estimation 
(\ref{hLso}) shows that $h(r)$ has the 
large-distance behavior 
$h(r)=1+c_1/r+c_2/r^2$ for $n=1$.
From Fig.~\ref{fig3}, we find that $h(r)<1$ 
and hence $c_1$ is negative. 
As we see in Fig.~\ref{fig2} for $n=1$, 
the ADM mass $M$ approaches a constant positive 
value $-c_1/(2G)$ at spatial infinity.

Let us study the bounds on $n$ derived by the 
no-ghost condition (\ref{muphi}), i.e.,
${\cal L}_{,F}=n \mu (\phi) F^{n-1}>0$, 
where $F=A_0'^2/(2N)$.
Using Eqs.~(\ref{rA0}), (\ref{Lagre}) and (\ref{hNeq}), 
this condition translates to 
\be
{\cal L}_{,F}=\frac{q_E^2 (1-2n)N}
{\Mpl^2 n r^2 [rhN'+2(rh'+h-1)N]}>0\,.
\label{LFel}
\ee
Since we are now considering the case $1/2<n<3/2$ with $F>0$, the no-ghost condition for electric compact objects is equivalent to $\mu>0$, i.e., 
\ba
\hspace{-0.9cm}
\mu &=& 
\frac{2^{n-1}\Mpl^2 [2N(1-h-rh')-rhN']}
{(2n-1)r^2N}  
\nonumber \\
\hspace{-0.9cm}
& &\times
\left[\frac{q_E^2 (2n-1)^2N^2}
{\Mpl^4 n^2
\{2(rh'+h-1)N+rhN'\}^2}
\right]^{n}>0\,.
\label{mue}
\ea
For given values of $N_0$, we study how the no-ghost condition $\mu>0$ depends on the parameter $n$ within the allowed range $1/2<n<3/2$. 
Our numerical analysis reveals the existence 
of a minimum value of $n$ for each $N_0$, 
denoted as $n_{\rm min}$, above which $\mu>0$ 
for all $r$. Conversely, if $n<n_{\rm min}$, $\mu$ becomes negative at some finite distance $r$, 
so these values of $n$ are excluded. 

In Fig.~\ref{fig4}, we plot $\mu$ versus $r/r_0$ 
with the choice $N_0=0.2$ for five different values 
of $n$. In this case, if $n<0.887$, the coupling 
enters the region $\mu<0$ for some values of $r$.
For $n>n_{\rm min}=0.887$, we observe in Fig.~\ref{fig4} that $\mu>0$ at any distance $r$.
For $N_0$ in the range $0<N_0<1$, we numerically obtain the minimum values of $n$ and 
plot $n_{\rm min}$ versus 
$N_0$ in Fig.~\ref{fig5}. 
We find that $n_{\rm min}$ mildly increases 
as a function of $N_0$ and approaches 
1 for $N_0 \to 1$. 
Note that $n_{\rm min}$ is always larger than $1/2$ 
for any values of $N_0$.
So long as $n$ is in the range 
\be
n_{\rm min} \leq n <\frac{3}{2}\,,
\label{nrae}
\ee
there exist horizonless compact objects 
with neither ghosts nor Laplacian instabilities.

The integrated solution to 
Eq.~(\ref{phico}) is given by 
\be
\phi(r)=\phi_\infty-\int_r^{\infty}
\sqrt{\frac{N'}{\tilde{r}N}} \rd 
\tilde{r}\,,
\ee
where $\phi_\infty$ 
is the value of $\phi$ at $r\to\infty$, 
a value that could be matched to 
the cosmological value of the field. 
Since we are considering the branch $\phi'(r)>0$, 
the scalar field increases as a function of $r$. 
We can invert the relation $\phi=\phi(r)$ 
to write $r=\tilde{r} (\phi)$.
Since the coupling (\ref{mue}) depends on $r$, 
i.e., $\mu=\tilde\mu(r)$, we can express it 
in the form $\mu(\phi)=\tilde{\mu}
(\tilde{r}(\phi))$. 
In other words, the functional form of $\mu(\phi)$ 
is determined to realize a desired expression 
of $N(r)$ consistent with the boundary conditions at $r=0$ and at spatial infinity.

For example, let us consider the case $n=1$,  
with $N(r)$ given by Eq.~(\ref{Nexample}).
Around $r=0$, the coupling (\ref{mue}) has the dependence $\mu \propto r^{-6}$. 
In this regime, the scalar field 
behaves as $\phi=\phi_0+\Mpl\sqrt{N_4/N_0}\,r^2
+{\cal O}(r^3)$ from Eq.~(\ref{phib}), where $\phi_0=\phi(r=0)$, so that 
$\mu(\phi) \propto (\phi-\phi_0)^{-3}$. 
Even though $\mu$ is divergent 
as $r \to 0$, the product $\mu (\phi) F$ in the Lagrangian approaches 0 due to the dependence 
$F \propto r^8$. 
At large distances, integrating 
Eq.~(\ref{rphiL}) leads to the 
solution $\phi=\phi_{\infty}
-\Mpl^2 r_0^2 \sqrt{2(1-\sqrt{N_0})} r^{-2}
+{\cal O}(r^{-6})$.
Since the coupling (\ref{mue}) behaves as 
$\mu \to \mu_\infty={\rm constant}$ 
at large distances,\footnote{Using the 
solution \eqref{hLso} for $h$ at 
large distances, we find 
$\mu\simeq  \Mpl^{-4 n +2} n^{-2 n} 2^{-n} c_{2} \left( c_2^2/q_E^2 \right)^{-n} 
\left(2 n -1\right)^{4 n -2}/\left(3-2 n \right)^{2 n -1}$, and hence $c_2$ needs 
to be non-negative.} this translates to 
the $\phi$ dependence $\mu(\phi) 
= \mu(\phi_{\infty})$. 
By fixing $\phi_\infty$ to the cosmological value of $\phi$, one can uniquely determine the functional form of $\mu$, as both $\phi$ and $\mu$ are completely determined as functions of $r$. 

\subsection{Magnetic case}

For $q_M \neq 0$ and $q_E=0$, we have $A_0'=0$ 
and hence Eq.~(\ref{back3}) is automatically satisfied. From Eq.~(\ref{back2}), we have 
\be
\phi'= \Mpl \sqrt{\frac{N'}{rN}}\,,
\label{phid}
\ee
where we have chosen the branch $\phi'>0$. 
The Lagrangian ${\cal L}$ obeys 
the following relation 
\be
-\frac{\Mpl^2 h N'}{2rN}+\mu(\phi) 
\left( -\frac{q_M^2}{2r^4} \right)^n
=\frac{\Mpl^2}{r^2} \left( rh'+h-1 \right)\,.
\label{Lred}
\ee
If $n$ is an integer, the second term in 
Eq.~(\ref{Lred}) reduces to 
$[q_M^2/(2r^4)]^n$ for even $n$ and 
$-[q_M^2/(2r^4)]^n$ for odd $n$.
If $n$ is not an integer, we should 
think of $F^n$ in the Lagrangian 
as $(F^2)^m$, i.e.,\ $n=2m$, so that 
Eq.~(\ref{Lred}) reduces to 
\be
\mu=\Mpl^2 \left( \frac{4r^8}{q_M^4} 
\right)^{n/2} \left( \frac{rh'-h-1}{r^2}
+\frac{hN'}{2rN} \right)\,.
\ee
For both integer and non-integer values of $n$, 
the coupling $\mu(\phi)$ is 
known in terms of $h$, $N$, and 
their $r$ derivatives.

From Eq.~(\ref{back4}), 
we can express $\mu_{,\phi}$, as 
\be
\mu_{,\phi}=-\frac{\Mpl [rhN''
+(2rh'+3h)N']}{2r^{3/2} \sqrt{NN'}}
\left( -\frac{q_M^2}{2r^4} \right)^{-n}\,.
\label{muphi2}
\ee
We take the $r$ derivative of Eq.~(\ref{Lred}) and exploit the relation 
$\mu'=\mu_{,\phi} \phi'$. 
Eliminating the term $\mu_{,\phi}$ on account of Eq.~(\ref{muphi}), it follows that 
\ba
& & 
2(r^2 h''+4nr h'+4nh-2h-4n+2)N^2 
-r^2 h N'^2
\nonumber \\
& &
+(3rh'+4nh)r NN'+2rh N (rN''+N')=0\,.
\label{heqma}
\ea
This equation can be mapped to the differential Eq.~\eqref{hNeq} found 
for the pure electric case by 
the substitution $n \to n/(2n-1)$. 
We use the labels ``e'' and ``m'' for the electric 
and magnetic cases, respectively. 
If $n_{\rm e}>1/2$ for the absence of 
Laplacian instabilities, the mapping 
$n_{\rm m}= n_{\rm e}/(2n_{\rm e}-1)$ implies that 
$n_{\rm e}= n_{\rm m}/(2n_{\rm m}-1)$, and also 
$n_{\rm m}>1/2$. Therefore, 
we have the mirror magnetic solutions 
without Laplacian instabilities identified 
by using the above mapping. The same mapping 
brings Eq.~\eqref{eq:cOm_secV_el} into Eq.~\eqref{eq:cOm_secV_mg} and vice versa.

The integrated solution to Eq.~(\ref{heqma}) can be obtained by replacing the power $n$ to 
$n/(2n-1)$ in Eq.~(\ref{hint0}), so that 
\ba
h(r) &=& 
\frac{2(2n-1) \int_0^r r_1^{4(n-1)}
\sqrt{N(r_1)} \bigl(\int_{0}^{r_1} \sqrt{N(r_2)}
\rd r_2\bigr)\rd r_1}
{r^{2(2n-1)} N(r)} \nonumber \\
& &+\frac{c_1}{r^{2(2n-1)}N(r)} \nonumber \\
& &+\frac{c_2 (2n-1) \int_0^r r_1^{4(n-1)}
\sqrt{N(r_1)}\rd r_1}{r^{2(2n-1)} N(r)}\,.
\label{hintm}
\ea
To avoid the divergence of $h(r)$ at $r=0$ 
for $n>1/2$, we require that $c_1=0=c_2$. 
Then, the resulting solution to $h(r)$ 
is given by 
\be
h(r)= 
\frac{2(2n-1) \int_0^r r_1^{4(n-1)}
\sqrt{N(r_1)} \bigl(\int_{0}^{r_1} \sqrt{N(r_2)}
\rd r_2\bigr)\rd r_1}
{r^{2(2n-1)} N(r)}\,.
\label{hintm2}
\ee
Since $h(r)>0$ at any $r \geq 0$ under the 
Laplacian stability condition $n>1/2$, 
there are no regular magnetic BHs 
with apparent  horizons.

\begin{figure}[ht]
\begin{center}
\includegraphics[height=3.3in,width=3.5in]{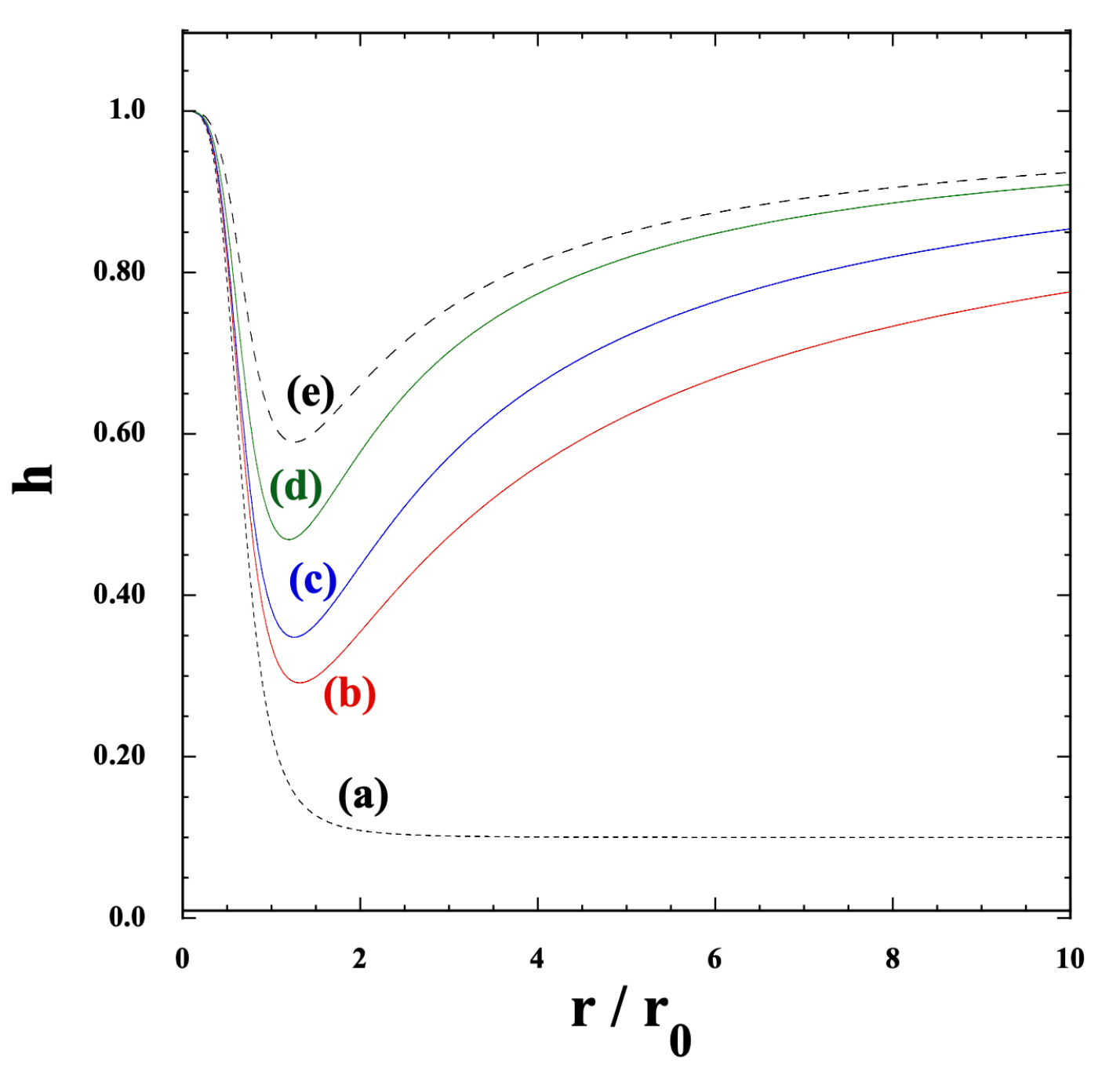}
\end{center}
\caption{
Metric component $h(r)$ versus $r/r_0$ for 
magnetic SSS objects in theories given 
by the Lagrangian (\ref{EMD}). 
We choose $N(r)$ given by Eq.~(\ref{Nexample}) 
with $N_0=0.1$. 
Each case corresponds to (a) $n=1/2$, (b) $n=4/5$, 
(c) $n=1$, (d) $n=2$, and (e) $n \gg 1$. 
The theoretical lines of $h(r)$ are 
positive at any distance $r$.
\label{fig6}
}
\end{figure}

\begin{figure}[ht]
\begin{center}
\includegraphics[height=3.3in,width=3.5in]{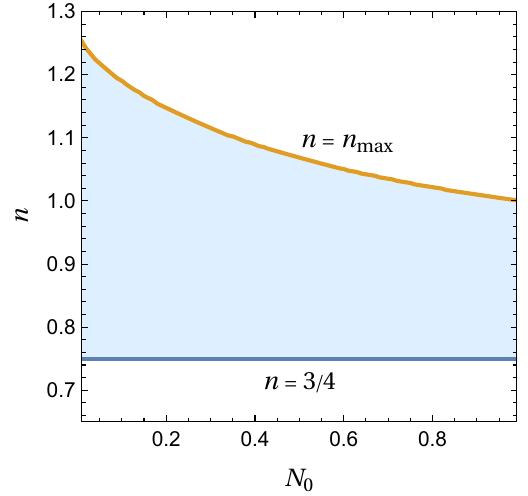}
\end{center}
\caption{This plot depicts the allowed region of parameter space, indicated by the light-blue shaded area. The upper boundary corresponds to the value of $n_{\rm max}$, beyond which the no-ghost condition is violated for a given $N_0$ within the allowed range $0<N_0<1$. The lower boundary, at $n=3/4$, defines the limit below which the solutions no longer describe a compact object. Consequently, the permissible range for $n$ is $3/4<n\leq n_{\rm max}$.
\label{fig7}}
\end{figure}

To realize the behavior 
$\phi'(r) \propto r$ for avoiding the 
cusp-like structure at $r=0$, we require 
that $N(r)$ is expanded as Eq.~(\ref{Nsmall}) 
around $r=0$. Then, the field derivative 
has the desired behavior (\ref{phib}). 
As a possible form of $N(r)$ having 
this property, we can choose the function 
(\ref{Nexample}). For this choice, $\phi'(r)$ 
behaves as Eq.~(\ref{rphiL}) at large distances. 
As we discussed for the electric case, $\phi'(r)$ 
has a maximum at some intermediate distance, 
with the asymptotic behaviors 
$\phi'(r) \propto r$ for $r \ll r_0$ and 
$\phi'(r) \propto r^{-3}$ for $r \gg r_0$. 
We note that $A_0'$ is vanishing 
at any distance $r$ for 
the magnetic configuration.

The metric component $h$ behaves as 
Eq.~(\ref{hr=0}) around $r=0$, where 
$n$ should be replaced with $n/(2n-1)$. 
At large distances, we require that the ADM mass 
$M$ is converging, whose condition corresponds to 
$n<3/2$ for the electric configuration.
Using the electric-magnetic duality, the 
condition for compactness of magnetic SSS 
objects is now given by $n>3/4$. 
For the choice (\ref{Nexample}), the 
large-distance solutions to $h(r)$ 
are given by the dual of 
Eq.~(\ref{hLso}), that is
\be
h(r)=1+\frac{c_1}{r}+\frac{c_2}{r^{4n-2}}
=1+\frac{c_1}{r}+\mathcal{O}(r^{2-4n})\,.
\ee
For the parameter range $n>3/4$, the term of 
order $r^{2-4n}$ is sub-dominant to $c_1/r$.

For $n \to 1/2$, we have the solution 
corresponding to the limit $n \to \infty$ 
in the electric configuration, i.e., 
\be
h_{n \to 1/2}(r)=\frac{N_0}{N(r)}\,.
\label{h12mag}
\ee
For $n \gg 1$, the solution is equivalent 
to that for $n=1/2$ in the electric 
case, i.e., 
\be
h_{n \gg 1}(r)=
\frac{1}{r \sqrt{N(r)}} \int_0^{r}
\sqrt{N(r_1)}\,{\rm d}r_1\,.
\label{hlmag}
\ee

In Fig.~\ref{fig6}, we show $h$ versus $r/r_0$ 
for $n=1/2, 4/5, 1, 2$ and the limit $n \gg 1$, 
with the choice $N_0=0.1$ in Eq.~(\ref{Nexample}).
The theoretical curves are bounded from 
below by the line $h_{n \to 1/2}(r)$. 
We have $h(r)>0$ for all the powers $n$ in 
the range $n>1/2$, showing the absence 
of nonsingular magnetic BHs with 
apparent  horizons. The compactness of 
magnetic SSS objects requires 
that $n>3/4$. 
In such cases, we numerically checked 
the asymptotic behaviors 
$h-1 \propto r^4$ for $r \ll r_0$ and 
$h-1 \propto r^{-1}$ for $r \gg r_0$. 
We note that the radial dependence of $\phi'(r)$ 
is similar to that of the electric case shown 
in Fig.~\ref{fig3}.

Using Eq.~(\ref{Lred}), the no-ghost condition 
${\cal L}_{,F}=n\mu [-q_M^2/(2r^4)]^{n-1}>0$ 
translates to 
\be
{\cal L}_{,F}=\frac{q_E^2 (1-2n)N}
{\Mpl^2 n r^2[r hN'+(rh'+h-1)N]}>0\,.
\label{LFma}
\ee
In comparison to Eq.~(\ref{LFel}) for the electric 
case, there is an electric-magnetic duality in that 
Eq.~(\ref{LFma}) follows from the inverse of 
Eq.~(\ref{LFel}) by the mappings $n \to n/(2n-1)$ 
and $q_E \to q_M$. This means that we can derive 
the no-ghost condition for magnetic SSS objects 
by using the correspondence $n_{\rm e}=
n_{\rm m}/(2n_{\rm m}-1)$. 
Then, in the magnetic configuration, the range 
(\ref{nrae}) obtained for the electric case 
translates to the following region 
\be
\frac{3}{4}<n \leq n_{\rm max}
\equiv
\frac{n_{{\rm e, min}}}
{2 n_{{\rm e, min}}-1}\,,
\label{nram}
\ee
where $n_{{\rm e, min}}$ is the minimum value 
of $n$ given in Eq.~(\ref{nrae}).
In Fig.~\ref{fig7}, we plot the allowed range 
(\ref{nram}) versus $N_0$ for the choice 
(\ref{Nexample}).
We find that $n_{\rm max}$ is larger 
than 1, e.g.\ $n_{\rm max}=1.19$ for $N_0=0.1$. 
As $N_0$ increases in Fig.\ \ref{fig7} from 0.01 to 1, $n_{\rm max}$ decreases from 1.253 to 1. 
In the parameter region shown as a shaded color, 
there are compact magnetic SSS objects with 
neither ghosts nor Laplacian instabilities.

From Eq.~(\ref{Lred}), the coupling $\mu$ 
can be written in terms of $h(r)$, $N(r)$, 
and their $r$ derivatives. 
Integrating Eq.~(\ref{phid}), we can determine 
$\phi$ as a growing function of $r$. 
This allows us to express $\mu$ as a 
function of $\phi$. Since $F$ is negative, 
the no-ghost condition 
$\mathcal{L}_{, F}=n(\mu/F) F^n>0$ 
means that $\mu<0$ for even-integer $n$ 
and $\mu>0$ for odd-integer $n$.
If $n$ is non-integer, we should consider $F^n$ as $(F^2)^m$, so that $n \to 2m$. In this case, we have that $\mathcal{L}_{, F}=n(\mu/F) (F^2)^m$, and the condition $\mathcal{L}_{, F}>0$ 
leads to $\mu/F>0$, i.e., $\mu<0$. 
Thus, depending on the values of $n$, the 
signs of $\mu$ consistent with no-ghost 
conditions are different.

\section{Theories with Lagrangian 
$\mathcal{L}=X \kappa(F)$}
\label{kappasec}

Finally, we consider theories 
given by the Lagrangian
\be
{\cal L}=X \kappa (F)\,,
\label{LagXF}
\ee
where $\kappa$ is a function of $F$. 
The Lagrangian (\ref{LagXF}) respects 
a shift symmetry for the scalar field, 
so that it is invariant under the shift 
$\partial_{\mu} \phi \to 
\partial_{\mu} \phi+c_{\mu}$.
In this case, we have $\mathcal{L}_{,\phi}=0$ 
and hence Eq.~(\ref{Lp}) gives
\be
\phi'=\frac{\Mpl^2 N' hr}{q_s \sqrt{N}}\,,
\label{eq:d_phi_sec6}
\ee
where $q_s$ is a constant. 
From Eq.~(\ref{back2}), we obtain 
\be
\phi'^2=\frac{\Mpl^2 N'}{\kappa r N}\,.
\label{eq:d_phi2_sec6}
\ee
Combining Eq.~(\ref{eq:d_phi2_sec6}) 
with Eq.~(\ref{eq:d_phi_sec6}), 
we can express $\kappa$ in the form 
\be
\kappa=\frac{q_s^2}{\Mpl^2 N' h^2 r^3}\,,
\label{kappa}
\ee
which is a function of $r$. 

\subsection{Electric case}

Let us first study the case $q_E \neq 0$ 
and $q_M=0$. Since ${\cal L}_{,X}=\kappa$, 
${\cal L}_{,XX}=0$, and ${\cal L}_{,FX}=\kappa_{,F}$, the squared propagation speeds (\ref{crdelta}) and (\ref{cO2}) reduce, respectively, to
\ba
c_{r, \delta \phi}^2 
&=&
1-\frac{4F \kappa_{,F}^2}{\kappa 
(\kappa_{,F}+2F \kappa_{,FF})}\,,
\label{crX}\\
c_{\Omega,V}^2 
&=& \frac{\kappa_{,F}}{\kappa_{,F}+2F\kappa_{,FF}}\,.
\label{cOX}
\ea

We first derive the relation between $h(r)$ and $N(r)$ to find the existence of nonsingular BHs. 
Using Eq.~(\ref{eq:d_phi2_sec6}), 
the Lagrangian (\ref{LagXF}) 
is expressed as 
\be
{\cal L}=-\frac{\Mpl^2 hN'}{2rN}\,,
\label{LXre}
\ee
on the SSS background.
Equating Eq.~(\ref{LXre}) with Eq.~(\ref{Lag}) 
and using Eq.~(\ref{rA0}), it follows that 
\ba
& & 
2(r^2 h''+2r h')N^2 -r^2 h N'^2
+(3rh'+2h)r NN'\nonumber \\
& &
+2rh N (rN''+N')=0\,.
\label{heqX}
\ea
This differential equation is the same as 
Eq.~(\ref{heqma}) with $n=1/2$. 
Then, we have the following integrated solution 
\be
h(r)=\frac{c_1}{N(r)}+\frac{c_2}{N(r)}
\int_0^r \frac{\sqrt{N(r_1)}}{r_1^2} 
{\rm d}r_1\,.
\label{hint}
\ee
Imposing regularities at $r=0$ in the forms 
(\ref{hr}) and (\ref{Nr}), we can 
fix integration constants to be
$c_1=N_0$ and $c_2=0$. 
Then, Eq.~(\ref{hint}) reduces to
\be
h(r)=\frac{N_0}{N(r)}\,.
\label{hsoX}
\ee
Since $N(r)>0$ and $N_0>0$, $h(r)$ is positive 
at any distance $r$. 
This means that nonsingular electric BHs 
do not exist in theories given by the 
Lagrangian (\ref{LagXF}).
In particular, by setting asymptotically-flat 
boundary conditions $h \to 1$ and $N \to 1$ 
for $r \to \infty$, we need to impose that $N_0=1$ and hence $h(r)=1/N(r)$. In this case, we have 
$h=N=1$ at both $r=0$ and $r \to \infty$, 
but $h(r)$ and $N(r)$ can differ from 1 at 
intermediate distances.

The above SSS configuration does not correspond to nonsingular BHs, but we study the linear stability of such SSS objects without horizons. From Eqs.~(\ref{kappa}) and (\ref{hsoX}), the coupling $\kappa$ 
can be expressed as 
\be
\kappa=\frac{q_s^2 N^2}
{\Mpl^2 N_0^2 r^3 N'}\,.
\label{kappa2}
\ee
On using this relation with $F=A_0'^2/(2N)$, 
we can compute the quantities $\kappa_{,F}$ 
and $\kappa_{,FF}$ in Eqs.~(\ref{crX}) 
and (\ref{cOX}). 
Then, it follows that 
\begin{widetext}
\ba
c_{r,\delta \phi}^2
&=& \frac{NN'}{2rN'^2-N(rN''+2N')}\,,\\
c_{\Omega,V}^2 
&=& 
\frac{r N_0 N'(rN N''-2r N'^{2}+3 N N')}
{2(2 N_0 N- r N_0 {N'} -2 N^{2})
(rN N'' -2r {N'}^{2}+2 N {N'} )}\,.
\ea
\end{widetext}
Using the expansion (\ref{Nr}) of $N(r)$ 
around $r=0$, we obtain
\ba
c_{r,\delta \phi}^2
&=& -\frac{1}{3}
+\frac{N_3}{6 N_2}\,r+\mathcal{O}(r^2)\,,\\
c_{\Omega,V}^2 
&=& -\frac{1}{3}
-\frac{N_3}{24 N_2}\,r+\mathcal{O}(r^2)\,,
\ea
whose leading-order terms are negative. 
Hence the background SSS solution is prone to Laplacian instability in both radial and angular directions. We stress that this instability arises for the coupling $\kappa$ with the radial dependence given by 
Eq.~(\ref{kappa2}).
In the vicinity of $r=0$, we have 
\ba
\kappa &=& 
\frac{q_s^2}{2\Mpl^2 N_2}r^{-4}
+{\cal O}(r^{-3})\,,
\label{kr0} \\
F &=& \frac{2\Mpl^4 N_2^2}{q_E^2 N_0^2}r^4
+{\cal O}(r^5)\,.
\label{Fr0}
\ea
Then, the leading-order terms of 
$\kappa$ and $F$ have the following 
relation 
\be
\kappa=\frac{\Mpl^2 N_2 q_s^2}
{q_E^2 N_0^2} F^{-1}\,,
\ee
around $r=0$.
The reason why the leading-order terms of 
$c_{r,\delta \phi}^2$ and $c_{\Omega,V}^2$ 
are negative is attributed to the property
$\kappa \propto F^{-1}$.
Indeed, we can obtain the value $-1/3$ 
by substituting the relation 
$\kappa \propto F^{-1}$ into 
Eqs.~(\ref{crX}) and (\ref{cOX}).
Even though the dependence of $\kappa$ on $F$ is different for the distance $r$ away from the origin, the Laplacian instability around $r=0$ is sufficient to exclude the above horizonless solution as a stable SSS configuration. 

\subsection{Magnetic case}

For $q_M \neq 0$ and $q_E=0$, the squared 
radial and angular propagation speeds 
(\ref{crdelta2}) and (\ref{cOdelta1}) 
reduce to
\ba
c_{r, \delta \phi}^2 
&=& 1\,,
\label{crXm}\\
c_{\Omega,\delta A}^2 
&=& \frac{\kappa_{,F}+2F\kappa_{,FF}}{\kappa_{,F}}\,.
\label{cOXm}
\ea
This means that the Laplacian 
instability of $\delta \phi$ along 
the radial direction is absent.

As in the electric case, the background Lagrangian is expressed as Eq.~(\ref{LXre}). 
Combing this with Eq.~(\ref{Lag}) and using 
the property $A_0'=0$, we obtain
\be
2N \left( rh'+h-1 \right)=
-rhN' \,.
\label{eq:h_prime_sec6}
\ee
This differential equation is equivalent 
to the limit $n \gg 1$ in Eq.~(\ref{heqma}). 
Then, the integrated solution to $h(r)$ 
that is regular at $r=0$ is given by 
\be
h(r)=
\frac{1}{r \sqrt{N(r)}} \int_0^{r}
\sqrt{N(r_1)}\,{\rm d}r_1\,,
\label{hXsom}
\ee
which is positive for $r \geq 0$. 
Then, the above solution does not 
correspond to the nonsingular BH 
with an apparent  horizon.

Since the coupling $\kappa$ is given by 
Eq.~(\ref{kappa}), we can express 
$\kappa_{,F}$ and $\kappa_{,FF}$
in terms of $h(r)$, $N(r)$ and its 
$r$ derivatives. 
Then, Eq.~(\ref{cOXm}) yields
\begin{widetext}
\begin{align}
c_{\Omega,\delta A}^2 &=
-(2 r^{2} h^{2} N N''^{2}
-r^{2}h^{2}N N'N'''-r^{2} h^{2} N'^{2}N''
-r h^{2} NN'N''+rh^2 N'^{3}
-h^{2} N N'^{2}+4rh N N' N'' 
-5 r h N'^{3} \nonumber\\
&~~~-2 h N N'^{2}+6 N N'^{2})
/[2 h (r h N'^{2}-r h N N'' 
-h N N' -2 N N')N']\,.
\end{align}
\end{widetext}
Using the expansions (\ref{hr}) and 
(\ref{Nr}) of $h(r)$ and $N(r)$ 
around $r=0$, it follows that 
\be
c_{\Omega,\delta A}^2 = 
1+\frac{9N_3}{16 N_2}\,r
+{\cal O}(r^2)\,.
\ee
Since the leading-order contribution to $c_{\Omega,\delta A}^2$ is positive, the angular Laplacian instability of $\delta A$ is 
absent around $r=0$.
On the other hand, at large distances, 
we exploit the expanded solutions of $N(r)$ and $h(r)$ in the forms
\ba
N(r)=\sum_{n=0}\frac{\tilde{N}_n}{r^n}\,,
\qquad 
h(r)=1+\sum_{n=1}\frac{\tilde{h}_n}{r^n}\,,
\ea
where $\tilde{N}_n$ and $\tilde{h}_n$ are constants. 
Solving Eq.~\eqref{eq:h_prime_sec6} order 
by order, we have $\tilde{N}_1=0$ and $\tilde{h}_2=-\tilde{N}_2/\tilde{N}_0$ 
at lowest order. This leaves $\tilde{h}_1$ 
as a free parameter, related to the ADM 
mass of the SSS object.

Then, the large-distance behavior of $c_{\Omega,\delta A}^2$ is given by 
\begin{widetext}  
\be
c_{\Omega,\delta A}^2=
-\frac{1}{2}+\frac{8 \tilde{N}_{0} \tilde{N}_{2} \tilde{N}_{4}-12 \tilde{N}_{0} \tilde{N}_{2}^{2} \tilde{h}_{1}^{2}-12 \tilde{N}_{0} \tilde{N}_{2} \tilde{N}_{3} \tilde{h}_{1}-9 \tilde{N}_{0} \tilde{N}_{3}^{2}-8 \tilde{N}_{2}^{3}}{2 \tilde{N}_{0} \tilde{N}_{2} \,(4 \tilde{N}_{2} \tilde{h}_{1}+3 \tilde{N}_{3})}\,\frac1r+{\cal O}(r^{-2})\,,
\label{cOdA}
\ee
\end{widetext}
whose leading-order term is negative. 
Hence the nonsingular SSS object 
without the horizon is excluded by 
angular Laplacian 
instability at large distances. 
We note that, even without imposing 
the condition $\tilde{N}_0=1$ for asymptotic 
flatness, the leading-order contribution to $c_{\Omega,\delta A}^2$ 
is $-1/2$. The above results show that, 
as $r$ increases from 0, 
$c_{\Omega,\delta A}^2$ enters 
the regime $c_{\Omega,\delta A}^2<0$ to 
approach the asymptotic value $-1/2$. 
Indeed, we have numerically confirmed 
this property\footnote{This behavior of 
$c_{\Omega,\delta A}^2$ is similar to 
what happens for hairy BHs present 
in cubic vector Galileon theories \cite{DeFelice:2024bdq}.} 
by choosing $N(r)$ of 
the form (\ref{Nexample}).

\section{Conclusions} 
\label{consec}

In this paper, we extended our previous 
analysis of the nonsingular BHs 
in NED \cite{DeFelice:2024seu} 
to more general theories characterized by 
the matter Lagrangian ${\cal L}(F,\phi,X)$ 
with an Einstein-Hilbert term. 
In NED with the Lagrangian ${\cal L}(F)$, 
the charged SSS objects that are nonsingular at their centers exhibit angular Laplacian instabilities arising from vector-field perturbations. 
The primary objective of our study was to investigate whether this property persists when a scalar field $\phi$ is introduced into the theory.

In Sec.~\ref{backsec}, we first showed that 
the background solutions with mixed electric and magnetic charges do not exist. 
Then, as in the case of NED, we could 
separate the analysis into either 
electrically or magnetically charged objects. 
For the electric configuration, both $A_0'$ 
and $\phi'$ are generally nonvanishing, 
while $A_0'=0$ for the magnetic 
configuration.

In Sec.~\ref{persec}, we derived the 
second-order action of perturbations by 
taking into account both electric and 
magnetic charges. For the electric configuration, 
the action consists of 
dynamical perturbations in the odd-parity 
sector A and the even-parity sector B. 
In this case, we showed that there are 
neither ghosts nor Laplacian instabilities 
under the conditions 
${\cal L}_{,X}>0$, ${\cal L}_{,F}>0$, 
$c_{r, \delta \phi}^2>0$, and 
$c_{\Omega,\delta A}^2>0$, 
where $c_{r, \delta \phi}^2$ and 
$c_{\Omega,V}^2$ are given, 
respectively by Eqs.~(\ref{crdelta}) 
and (\ref{cO2}).
For the magnetic case, the action can be 
decomposed into two sectors C and D, 
both of which contain the contributions of 
odd- and even-parity dynamical perturbations. 
In this case, we obtained the linear stability conditions ${\cal L}_{,X}>0$, ${\cal L}_{,F}>0$, $c_{r, \delta \phi}^2>0$, and 
$c_{\Omega,\delta A}^2>0$, where 
$c_{r, \delta \phi}^2$ and 
$c_{\Omega,\delta A}^2$ are given, 
respectively by Eqs.~(\ref{crdelta2}) and 
(\ref{cOdelta1}).

In Sec.~\ref{kessencesec}, we applied the linear stability conditions to NED with 
a k-essence scalar field. For nonsingular 
electric SSS objects, we showed that 
the even-parity vector-field perturbation 
$V$ is subject to angular Laplacian 
instability around $r=0$ 
due to the negative 
leading term of $c_{\Omega,V}^2$. 
For the nonsingular magnetic configuration, 
the same angular Laplacian instability arises 
for the odd-parity vector-field perturbation 
$\delta A$. Since the fields $V$ and 
$\delta A$ are coupled to the even-parity 
gravitational perturbation $\psi$, 
the background metric regular at $r=0$ 
cannot remain in a steady state. 
This means that, in theories 
given by the Lagrangian 
$\tilde{\cal L}(F)+K(\phi,X)$, 
the nonsingular SSS objects 
(including regular BHs) 
do not exist as stable configurations.

In Sec.~\ref{muphisec}, we analyzed regular solutions for the Lagrangian of the form ${\cal L}=X +\mu(\phi) F^n$. For both electric and magnetic 
SSS objects, we showed that the absence of angular Laplacian instabilities imposes the condition 
$n>1/2$. Under this inequality, the 
metric components are positive at 
any distance $r$, so that there 
are no regular BHs with 
apparent  horizons.

In Sec.~\ref{muphisec}, we also studied the linear stability of nonsingular horizonless SSS objects further. For a given function of $N(r)$, the profile of $\phi(r)$ is identical for both the electric and magnetic configurations. 
To avoid the formation of cusp-like structures, we impose the regular behavior $\phi'(r) \propto r$ 
around $r=0$. 
One of the examples for $N(r)$ consistent with this boundary condition and asymptotic flatness 
is given by Eq.~(\ref{Nexample}).

For the electric horizonless SSS 
configuration studied in Sec.~\ref{muphisec}, 
the solutions can be described by regular compact objects if $n<3/2$. 
In this case, 
the absence of ghosts requires that 
$n_{\rm min} \leq n$, where 
$n_{\rm min}$ is the minimum value 
of $n$ larger than $1/2$.
As shown in Fig.~\ref{fig5}, 
$n_{\rm min}$ mildly depends on 
$N_0$, where $N_0$ is a constant 
appearing in Eq.~(\ref{Nexample}).
Therefore, the allowed parameter 
space of electric regular 
compact objects is given by 
$n_{\rm min}\leq n < 3/2$. 
Since there is a duality relation 
$n_{\rm e}=n_{\rm m}/(2n_{\rm m}-1)$ 
between electric and magnetic cases, 
the existence of linearly stable 
magnetic compact objects 
requires the condition $3/4<n \le 
n_{\rm max}=n_{\rm e, min}
/(2n_{\rm e, min}-1)$.

Finally, we investigated 
Lagrangians of the form 
${\cal L}=X\kappa(F)$ 
in Sec.~\ref{kappasec}. 
Similar to the previous case, 
these theories do not support the SSS configurations with apparent  horizons, precluding the existence of nonsingular BHs. 
Unlike theories discussed in Sec.~\ref{muphisec}, 
all regular SSS solutions obtained in this 
framework (including those without 
horizons) suffer from angular Laplacian instabilities. Specifically, for electric  solutions, such instabilities arise 
around the regular center. 
In contrast, for magnetic solutions, 
they manifest at large distances away 
from the origin.

To summarize, we have not found any linearly stable nonsingular BHs for 
theories studied in this paper.
To reach this conclusion, we have 
made three assumptions: 
(i) the choice of three subsets of 
${\cal L}(F, \phi, X)$ theories, 
(ii) the spherically symmetric and static regular background with asymptotic 
flatness, and 
(iii) the existence of the BH apparent  horizon. 
To circumvent the no-go result, one could either modify the theories---by extending 
the analysis to other subclasses 
of ${\cal L}(F,\phi, X)$ theories and more 
general theories---or alter the spacetime background. For instance, rotating regular 
BH solutions, if they exist, might offer a stable alternative.

Our results show that the construction of 
regular BHs in the framework of 
classical field theories is highly complex 
and challenging. 
In other words, Penrose's singularity theorem holds 
even in scenarios where some of its original assumptions are violated. 
If this property generally persists for 
local classical four-dimensional actions, we may 
need to resort to non-local theories of gravity
or higher-dimensional theories (see, e.g., 
\cite{Modesto:2011kw,Biswas:2011ar,Modesto:2014lga,Tomboulis:2015gfa,Bueno:2024dgm,Bueno:2024zsx}).
On the other hand, our analysis 
in Sec.~\ref{muphisec} 
has uncovered linearly stable regular 
SSS solutions without horizons that, in principle, could correspond to physically realizable configurations in nature. 
The exploration of physical and geometric properties of such stable compact objects 
without horizons, including the comparisons with boson and 
Proca stars \cite{Kaup:1968zz,Brito:2015pxa}, 
remains an intriguing avenue for future research.

\section*{Acknowledgements}

We thank Pablo Cano, Vitor Cardoso, Hiroki Takeda, and Kent Yagi for useful discussions. 
The Japan Society for the Promotion of Science supported ADF's work through Grants-in-Aid for Scientific Research No.~20K03969.
ST was supported by the Grant-in-Aid for Scientific Research Fund of the JSPS No.~22K03642 and Waseda University Special Research Project No.~2024C-474. 

\section*{Appendix~A:~Coefficients of 
the second-order action}
\renewcommand{\theequation}{A.\arabic{equation}} 
\setcounter{equation}{0}
\label{Appencoeff}

The explicit forms of coefficients appearing 
in the Lagrangians (\ref{Lag1}) and (\ref{Lag2}) are given by 
\begin{widetext}
\ba
& &
a_0=\frac{r^2 A_0'^2}{8N^{3/2}}
(N{\cal L}_{,F}+A_0'^2 {\cal L}_{,FF}),\qquad
a_1=-\frac{\Mpl^2 r h\sqrt{N}}{2},\qquad
a_2=\frac{\Mpl^2 h\sqrt{N}}{2},\nonumber \\
& &
a_3=\frac{
Nr^2 A_0'^2 ({\cal L}_{,F}
+h\phi'^2 {\cal L}_{,XF})
-r^2 A_0'^4 {\cal L}_{,FF}
-N^2[2\Mpl^2+r^2(2{\cal L}
+h\phi'^2 {\cal L}_{,X})]}{4N^{3/2}} 
\qquad
a_4=-\frac{\Mpl^2 \sqrt{N}}{4},\nonumber \\
& &
a_5=\frac{\Mpl^2 N(h+1)
+r^2(N{\cal L}-A_0'^2 {\cal L}_{,F})}
{4\sqrt{N}r},
\qquad
a_6=\frac{q_M(N {\cal L}_{,F}-A_0'^2 {\cal L}_{,FF})}{2\sqrt{N}\,r^2},
\qquad 
a_7=\frac{h r^2 \phi'(N {\cal L}_{,X}
-A_0'^2 {\cal L}_{,XF})}{2\sqrt{N}},\nonumber \\
& &
a_8=\frac{r^2 (A_0'^2{\cal L}_{,\phi F}
-N {\cal L}_{,\phi})}{2\sqrt{N}},\qquad
b_1=\frac{\Mpl^2}{4\sqrt{N}},\qquad
b_2=\frac{\Mpl^2 r}{\sqrt{N}},\qquad b_3=-2b_1\,,\qquad 
b_4=-\frac{r^2 \phi'{\cal L}_{,X}}{\sqrt{N}}\,,\nonumber \\
& &
c_0=-\frac{a_3}{2}
+\frac{h r^2 \phi'^2  
(h \phi'^2 N {\cal L}_{,XX} 
-A_0'^2 {\cal L}_{,XF})}
{8\sqrt{N}},\qquad
c_1=-a_5-\frac{h r\phi'^2 \sqrt{N} 
{\cal L}_{,X}}{4},\qquad
c_2=-a_6-\frac{q_M h \phi'^2 
\sqrt{N}{\cal L}_{,XF}}{2r^2},\nonumber \\
& &
c_3=\frac{h r^2 \phi' 
[N({\cal L}_{,X}-h \phi'^2 {\cal L}_{,XX})+
A_0'^2 {\cal L}_{,XF}]}{2\sqrt{N}},\qquad
c_4=\frac{r^2[N({\cal L}_{,\phi}
+h \phi'^2 {\cal L}_{,\phi X})
-A_0'^2 {\cal L}_{,\phi F}]}{2\sqrt{N}},
\qquad \nonumber \\
& &
d_0=b_1,\qquad 
d_1=\frac{h \sqrt{N}(\Mpl^2 r^2-q_M^2{\cal L}_{,F})}{2r^4}, \qquad
d_2=-\frac{A_0'{\cal L}_{,F}}{\sqrt{N}},\qquad 
d_3=-\frac{q_M h \sqrt{N}
{\cal L}_{,F}}{r^2},\qquad 
d_4=h \phi'\sqrt{N}{\cal L}_{,X}\,,
\nonumber \\
& &
s_1=\frac{r^2( N{\cal L}_{,F}+A_0'^2 {\cal L}_{,FF})}{2N^{3/2}},\qquad 
s_2=A_0' s_1,\qquad 
s_3=-s_2+\frac{h r^2 \phi'^2 A_0'
{\cal L}_{,XF}}{2\sqrt{N}},\qquad
s_4=-\frac{q_M A_0'{\cal L}_{,FF}}{\sqrt{N}\,r^2},\nonumber \\
& &
s_5=-\frac{h r^2 \phi' A_0' 
{\cal L}_{,XF}}{\sqrt{N}},\qquad
s_6=\frac{r^2 A_0'{\cal L}_{,\phi F}}{\sqrt{N}},\qquad 
s_7=\frac{{\cal L}_{,F}}{2h \sqrt{N}},\qquad 
s_8=-\frac{h \sqrt{N} {\cal L}_{,F}}{2}\,,\nonumber \\
& &
u_1=\frac{r^2 {\cal L}_{,X}}
{2h \sqrt{N}},\qquad
u_2=\frac{r^2 h \sqrt{N}(h \phi'^2 
{\cal L}_{,XX}-{\cal L}_{,X})}{2},\qquad
u_3=-\frac{\sqrt{N} {\cal L}_{,X}}{2},\qquad
\tilde{u}_3=\frac{1}{2} 
\frac{\partial {\cal E}_{\phi}}{\partial \phi},\nonumber \\
& &
u_4=\frac{Lq_M h\phi' \sqrt{N}{\cal L}_{,XF}}{r^2},\qquad 
u_5=-\frac{L q_M \sqrt{N}{\cal L}_{,\phi F}}{r^2}\,,
\label{coeff1}\\
& & p_1=\frac{\Mpl^2}{4\sqrt{N}\,r^2},\qquad p_2=\frac{d_2}{r},\qquad p_3=s_7,\qquad 
p_4=s_8,\qquad p_5=\frac{\sqrt{N}
(q_M^2{\cal L}_{,FF}-r^4{\cal L}_{,F})}{2r^6},
\qquad p_6=-\frac{\Mpl^2 h \sqrt{N}}{4r^2},\nonumber \\
& & 
p_7=d_1,\qquad p_8=\frac{\Mpl^2}{4r^2h \sqrt{N}},\qquad 
p_9=-\frac{d_1}{h^2 N},\qquad p_{10}=-\frac{d_3}{h^2 N},\qquad 
p_{11}=-A_0'h p_{10},\qquad p_{12}=-h^2 N p_{10}\,,\label{coeff2}
\ea
\end{widetext}
where ${\cal E}_{\phi}$ is defined 
in Eq.~(\ref{back4}).

\bibliographystyle{mybibstyle}
\bibliography{bib}

\begin{thebibliography}{72}%
\makeatletter
\providecommand \@ifxundefined [1]{%
 \@ifx{#1\undefined}
}%
\providecommand \@ifnum [1]{%
 \ifnum #1\expandafter \@firstoftwo
 \else \expandafter \@secondoftwo
 \fi
}%
\providecommand \@ifx [1]{%
 \ifx #1\expandafter \@firstoftwo
 \else \expandafter \@secondoftwo
 \fi
}%
\providecommand \natexlab [1]{#1}%
\providecommand \enquote  [1]{``#1''}%
\providecommand \bibnamefont  [1]{#1}%
\providecommand \bibfnamefont [1]{#1}%
\providecommand \citenamefont [1]{#1}%
\providecommand \href@noop [0]{\@secondoftwo}%
\providecommand \href [0]{\begingroup \@sanitize@url \@href}%
\providecommand \@href[1]{\@@startlink{#1}\@@href}%
\providecommand \@@href[1]{\endgroup#1\@@endlink}%
\providecommand \@sanitize@url [0]{\catcode `\\12\catcode `\$12\catcode
  `\&12\catcode `\#12\catcode `\^12\catcode `\_12\catcode `\%12\relax}%
\providecommand \@@startlink[1]{}%
\providecommand \@@endlink[0]{}%
\providecommand \url  [0]{\begingroup\@sanitize@url \@url }%
\providecommand \@url [1]{\endgroup\@href {#1}{\urlprefix }}%
\providecommand \urlprefix  [0]{URL }%
\providecommand \Eprint [0]{\href }%
\providecommand \doibase [0]{http://dx.doi.org/}%
\providecommand \selectlanguage [0]{\@gobble}%
\providecommand \bibinfo  [0]{\@secondoftwo}%
\providecommand \bibfield  [0]{\@secondoftwo}%
\providecommand \translation [1]{[#1]}%
\providecommand \BibitemOpen [0]{}%
\providecommand \bibitemStop [0]{}%
\providecommand \bibitemNoStop [0]{.\EOS\space}%
\providecommand \EOS [0]{\spacefactor3000\relax}%
\providecommand \BibitemShut  [1]{\csname bibitem#1\endcsname}%
\let\auto@bib@innerbib\@empty
\bibitem [{\citenamefont {Penrose}(1965)}]{Penrose:1964wq}%
  \BibitemOpen
  \bibfield  {author} {\bibinfo {author} {\bibfnamefont {R.}~\bibnamefont
  {Penrose}},\ }\href {\doibase 10.1103/PhysRevLett.14.57} {\bibfield
  {journal} {\bibinfo  {journal} {\emph {Phys. Rev. Lett.}}\ }\textbf {\bibinfo
  {volume} {14}},\ \bibinfo {pages} {57} (\bibinfo {year} {1965})}\BibitemShut
  {NoStop}%
\bibitem [{\citenamefont {Bardeen}()}]{Bardeen:1968}%
  \BibitemOpen
  \bibfield  {author} {\bibinfo {author} {\bibfnamefont {J.}~\bibnamefont
  {Bardeen}},\ }\href@noop {} {\bibinfo  {journal} {\emph {Proceedings of
  International Conference GR5 (Tbilisi, USSR, 1968)}}\ }\BibitemShut {NoStop}%
\bibitem [{\citenamefont {Dymnikova}(1992)}]{Dymnikova:1992ux}%
  \BibitemOpen
\bibfield  {journal} {  }\bibfield  {author} {\bibinfo {author} {\bibfnamefont
  {I.}~\bibnamefont {Dymnikova}},\ }\href {\doibase 10.1007/BF00760226}
  {\bibfield  {journal} {\bibinfo  {journal} {\emph {Gen. Rel. Grav.}}\
  }\textbf {\bibinfo {volume} {24}},\ \bibinfo {pages} {235} (\bibinfo {year}
  {1992})}\BibitemShut {NoStop}%
\bibitem [{\citenamefont {Dymnikova}(2004)}]{Dymnikova:2004zc}%
  \BibitemOpen
  \bibfield  {author} {\bibinfo {author} {\bibfnamefont {I.}~\bibnamefont
  {Dymnikova}},\ }\href {\doibase 10.1088/0264-9381/21/18/009} {\bibfield
  {journal} {\bibinfo  {journal} {\emph {Class. Quant. Grav.}}\ }\textbf
  {\bibinfo {volume} {21}},\ \bibinfo {pages} {4417} (\bibinfo {year}
  {2004})},\ \Eprint {http://arxiv.org/abs/gr-qc/0407072} {arXiv:gr-qc/0407072}
  \BibitemShut {NoStop}%
\bibitem [{\citenamefont {Hayward}(2006)}]{Hayward:2005gi}%
  \BibitemOpen
  \bibfield  {author} {\bibinfo {author} {\bibfnamefont {S.~A.}\ \bibnamefont
  {Hayward}},\ }\href {\doibase 10.1103/PhysRevLett.96.031103} {\bibfield
  {journal} {\bibinfo  {journal} {\emph {Phys. Rev. Lett.}}\ }\textbf {\bibinfo
  {volume} {96}},\ \bibinfo {pages} {031103} (\bibinfo {year} {2006})},\
  \Eprint {http://arxiv.org/abs/gr-qc/0506126} {arXiv:gr-qc/0506126}
  \BibitemShut {NoStop}%
\bibitem [{\citenamefont {Ansoldi}(2008)}]{Ansoldi:2008jw}%
  \BibitemOpen
  \bibfield  {author} {\bibinfo {author} {\bibfnamefont {S.}~\bibnamefont
  {Ansoldi}},\ }in\ \href@noop {} {\emph {\bibinfo {booktitle} {{Conference on
  Black Holes and Naked Singularities}}}}\ (\bibinfo {year} {2008})\ \Eprint
  {http://arxiv.org/abs/0802.0330} {arXiv:0802.0330 [gr-qc]} \BibitemShut
  {NoStop}%
\bibitem [{\citenamefont {Balart}\ and\ \citenamefont
  {Vagenas}(2014)}]{Balart:2014cga}%
  \BibitemOpen
  \bibfield  {author} {\bibinfo {author} {\bibfnamefont {L.}~\bibnamefont
  {Balart}} and \bibinfo {author} {\bibfnamefont {E.~C.}\ \bibnamefont
  {Vagenas}},\ }\href {\doibase 10.1103/PhysRevD.90.124045} {\bibfield
  {journal} {\bibinfo  {journal} {\emph {Phys. Rev. D}}\ }\textbf {\bibinfo
  {volume} {90}},\ \bibinfo {pages} {124045} (\bibinfo {year} {2014})},\
  \Eprint {http://arxiv.org/abs/1408.0306} {arXiv:1408.0306 [gr-qc]}
  \BibitemShut {NoStop}%
\bibitem [{\citenamefont {Fan}\ and\ \citenamefont {Wang}(2016)}]{Fan:2016hvf}%
  \BibitemOpen
  \bibfield  {author} {\bibinfo {author} {\bibfnamefont {Z.-Y.}\ \bibnamefont
  {Fan}} and \bibinfo {author} {\bibfnamefont {X.}~\bibnamefont {Wang}},\
  }\href {\doibase 10.1103/PhysRevD.94.124027} {\bibfield  {journal} {\bibinfo
  {journal} {\emph {Phys. Rev. D}}\ }\textbf {\bibinfo {volume} {94}},\
  \bibinfo {pages} {124027} (\bibinfo {year} {2016})},\ \Eprint
  {http://arxiv.org/abs/1610.02636} {arXiv:1610.02636 [gr-qc]} \BibitemShut
  {NoStop}%
\bibitem [{\citenamefont {Simpson}\ and\ \citenamefont
  {Visser}(2019)}]{Simpson:2018tsi}%
  \BibitemOpen
  \bibfield  {author} {\bibinfo {author} {\bibfnamefont {A.}~\bibnamefont
  {Simpson}} and \bibinfo {author} {\bibfnamefont {M.}~\bibnamefont {Visser}},\
  }\href {\doibase 10.1088/1475-7516/2019/02/042} {\bibfield  {journal}
  {\bibinfo  {journal} {\emph {JCAP}}\ }\textbf {\bibinfo {volume} {02}},\
  \bibinfo {pages} {042} (\bibinfo {year} {2019})},\ \Eprint
  {http://arxiv.org/abs/1812.07114} {arXiv:1812.07114 [gr-qc]} \BibitemShut
  {NoStop}%
\bibitem [{\citenamefont {Fujii}\ and\ \citenamefont
  {Maeda}(2007)}]{Fujii:2003pa}%
  \BibitemOpen
  \bibfield  {author} {\bibinfo {author} {\bibfnamefont {Y.}~\bibnamefont
  {Fujii}} and \bibinfo {author} {\bibfnamefont {K.}~\bibnamefont {Maeda}},\
  }\href {\doibase 10.1017/CBO9780511535093} {\emph {\bibinfo {title} {{The
  scalar-tensor theory of gravitation}}}},\ Cambridge Monographs on
  Mathematical Physics\ (\bibinfo  {publisher} {Cambridge University Press},\
  \bibinfo {year} {2007})\BibitemShut {NoStop}%
\bibitem [{\citenamefont {Horndeski}(1974)}]{Horndeski:1974wa}%
  \BibitemOpen
  \bibfield  {author} {\bibinfo {author} {\bibfnamefont {G.~W.}\ \bibnamefont
  {Horndeski}},\ }\href {\doibase 10.1007/BF01807638} {\bibfield  {journal}
  {\bibinfo  {journal} {\emph {Int. J. Theor. Phys.}}\ }\textbf {\bibinfo
  {volume} {10}},\ \bibinfo {pages} {363} (\bibinfo {year} {1974})}\BibitemShut
  {NoStop}%
\bibitem [{\citenamefont {Deffayet}\ \emph {et~al.}(2011)\citenamefont
  {Deffayet}, \citenamefont {Gao}, \citenamefont {Steer},\ and\ \citenamefont
  {Zahariade}}]{Deffayet:2011gz}%
  \BibitemOpen
  \bibfield  {author} {\bibinfo {author} {\bibfnamefont {C.}~\bibnamefont
  {Deffayet}}, \bibinfo {author} {\bibfnamefont {X.}~\bibnamefont {Gao}},
  \bibinfo {author} {\bibfnamefont {D.~A.}\ \bibnamefont {Steer}},  and
  \bibinfo {author} {\bibfnamefont {G.}~\bibnamefont {Zahariade}},\ }\href
  {\doibase 10.1103/PhysRevD.84.064039} {\bibfield  {journal} {\bibinfo
  {journal} {\emph {Phys. Rev. D}}\ }\textbf {\bibinfo {volume} {84}},\
  \bibinfo {pages} {064039} (\bibinfo {year} {2011})},\ \Eprint
  {http://arxiv.org/abs/1103.3260} {arXiv:1103.3260 [hep-th]} \BibitemShut
  {NoStop}%
\bibitem [{\citenamefont {Kobayashi}\ \emph {et~al.}(2011)\citenamefont
  {Kobayashi}, \citenamefont {Yamaguchi},\ and\ \citenamefont
  {Yokoyama}}]{Kobayashi:2011nu}%
  \BibitemOpen
  \bibfield  {author} {\bibinfo {author} {\bibfnamefont {T.}~\bibnamefont
  {Kobayashi}}, \bibinfo {author} {\bibfnamefont {M.}~\bibnamefont
  {Yamaguchi}},  and \bibinfo {author} {\bibfnamefont {J.}~\bibnamefont
  {Yokoyama}},\ }\href {\doibase 10.1143/PTP.126.511} {\bibfield  {journal}
  {\bibinfo  {journal} {\emph {Prog. Theor. Phys.}}\ }\textbf {\bibinfo
  {volume} {126}},\ \bibinfo {pages} {511} (\bibinfo {year} {2011})},\ \Eprint
  {http://arxiv.org/abs/1105.5723} {arXiv:1105.5723 [hep-th]} \BibitemShut
  {NoStop}%
\bibitem [{\citenamefont {Charmousis}\ \emph {et~al.}(2012)\citenamefont
  {Charmousis}, \citenamefont {Copeland}, \citenamefont {Padilla},\ and\
  \citenamefont {Saffin}}]{Charmousis:2011bf}%
  \BibitemOpen
  \bibfield  {author} {\bibinfo {author} {\bibfnamefont {C.}~\bibnamefont
  {Charmousis}}, \bibinfo {author} {\bibfnamefont {E.~J.}\ \bibnamefont
  {Copeland}}, \bibinfo {author} {\bibfnamefont {A.}~\bibnamefont {Padilla}},
  and \bibinfo {author} {\bibfnamefont {P.~M.}\ \bibnamefont {Saffin}},\ }\href
  {\doibase 10.1103/PhysRevLett.108.051101} {\bibfield  {journal} {\bibinfo
  {journal} {\emph {Phys. Rev. Lett.}}\ }\textbf {\bibinfo {volume} {108}},\
  \bibinfo {pages} {051101} (\bibinfo {year} {2012})},\ \Eprint
  {http://arxiv.org/abs/1106.2000} {arXiv:1106.2000 [hep-th]} \BibitemShut
  {NoStop}%
\bibitem [{\citenamefont {Hui}\ and\ \citenamefont
  {Nicolis}(2013)}]{Hui:2012qt}%
  \BibitemOpen
  \bibfield  {author} {\bibinfo {author} {\bibfnamefont {L.}~\bibnamefont
  {Hui}} and \bibinfo {author} {\bibfnamefont {A.}~\bibnamefont {Nicolis}},\
  }\href {\doibase 10.1103/PhysRevLett.110.241104} {\bibfield  {journal}
  {\bibinfo  {journal} {\emph {Phys. Rev. Lett.}}\ }\textbf {\bibinfo {volume}
  {110}},\ \bibinfo {pages} {241104} (\bibinfo {year} {2013})},\ \Eprint
  {http://arxiv.org/abs/1202.1296} {arXiv:1202.1296 [hep-th]} \BibitemShut
  {NoStop}%
\bibitem [{\citenamefont {Creminelli}\ \emph {et~al.}(2020)\citenamefont
  {Creminelli}, \citenamefont {Loayza}, \citenamefont {Serra}, \citenamefont
  {Trincherini},\ and\ \citenamefont {Trombetta}}]{Creminelli:2020lxn}%
  \BibitemOpen
  \bibfield  {author} {\bibinfo {author} {\bibfnamefont {P.}~\bibnamefont
  {Creminelli}}, \bibinfo {author} {\bibfnamefont {N.}~\bibnamefont {Loayza}},
  \bibinfo {author} {\bibfnamefont {F.}~\bibnamefont {Serra}}, \bibinfo
  {author} {\bibfnamefont {E.}~\bibnamefont {Trincherini}},  and \bibinfo
  {author} {\bibfnamefont {L.~G.}\ \bibnamefont {Trombetta}},\ }\href {\doibase
  10.1007/JHEP08(2020)045} {\bibfield  {journal} {\bibinfo  {journal} {\emph
  {JHEP}}\ }\textbf {\bibinfo {volume} {08}},\ \bibinfo {pages} {045} (\bibinfo
  {year} {2020})},\ \Eprint {http://arxiv.org/abs/2004.02893} {arXiv:2004.02893
  [hep-th]} \BibitemShut {NoStop}%
\bibitem [{\citenamefont {Minamitsuji}\ \emph
  {et~al.}(2022{\natexlab{a}})\citenamefont {Minamitsuji}, \citenamefont
  {Takahashi},\ and\ \citenamefont {Tsujikawa}}]{Minamitsuji:2022mlv}%
  \BibitemOpen
  \bibfield  {author} {\bibinfo {author} {\bibfnamefont {M.}~\bibnamefont
  {Minamitsuji}}, \bibinfo {author} {\bibfnamefont {K.}~\bibnamefont
  {Takahashi}},  and \bibinfo {author} {\bibfnamefont {S.}~\bibnamefont
  {Tsujikawa}},\ }\href {\doibase 10.1103/PhysRevD.105.104001} {\bibfield
  {journal} {\bibinfo  {journal} {\emph {Phys. Rev. D}}\ }\textbf {\bibinfo
  {volume} {105}},\ \bibinfo {pages} {104001} (\bibinfo {year}
  {2022}{\natexlab{a}})},\ \Eprint {http://arxiv.org/abs/2201.09687}
  {arXiv:2201.09687 [gr-qc]} \BibitemShut {NoStop}%
\bibitem [{\citenamefont {Minamitsuji}\ \emph
  {et~al.}(2022{\natexlab{b}})\citenamefont {Minamitsuji}, \citenamefont
  {Takahashi},\ and\ \citenamefont {Tsujikawa}}]{Minamitsuji:2022vbi}%
  \BibitemOpen
  \bibfield  {author} {\bibinfo {author} {\bibfnamefont {M.}~\bibnamefont
  {Minamitsuji}}, \bibinfo {author} {\bibfnamefont {K.}~\bibnamefont
  {Takahashi}},  and \bibinfo {author} {\bibfnamefont {S.}~\bibnamefont
  {Tsujikawa}},\ }\href {\doibase 10.1103/PhysRevD.106.044003} {\bibfield
  {journal} {\bibinfo  {journal} {\emph {Phys. Rev. D}}\ }\textbf {\bibinfo
  {volume} {106}},\ \bibinfo {pages} {044003} (\bibinfo {year}
  {2022}{\natexlab{b}})},\ \Eprint {http://arxiv.org/abs/2204.13837}
  {arXiv:2204.13837 [gr-qc]} \BibitemShut {NoStop}%
\bibitem [{\citenamefont {Kanti}\ \emph {et~al.}(1996)\citenamefont {Kanti},
  \citenamefont {Mavromatos}, \citenamefont {Rizos}, \citenamefont {Tamvakis},\
  and\ \citenamefont {Winstanley}}]{Kanti:1995vq}%
  \BibitemOpen
  \bibfield  {author} {\bibinfo {author} {\bibfnamefont {P.}~\bibnamefont
  {Kanti}}, \bibinfo {author} {\bibfnamefont {N.~E.}\ \bibnamefont
  {Mavromatos}}, \bibinfo {author} {\bibfnamefont {J.}~\bibnamefont {Rizos}},
  \bibinfo {author} {\bibfnamefont {K.}~\bibnamefont {Tamvakis}},  and \bibinfo
  {author} {\bibfnamefont {E.}~\bibnamefont {Winstanley}},\ }\href {\doibase
  10.1103/PhysRevD.54.5049} {\bibfield  {journal} {\bibinfo  {journal} {\emph
  {Phys. Rev. D}}\ }\textbf {\bibinfo {volume} {54}},\ \bibinfo {pages} {5049}
  (\bibinfo {year} {1996})},\ \Eprint {http://arxiv.org/abs/hep-th/9511071}
  {arXiv:hep-th/9511071} \BibitemShut {NoStop}%
\bibitem [{\citenamefont {Torii}\ \emph {et~al.}(1997)\citenamefont {Torii},
  \citenamefont {Yajima},\ and\ \citenamefont {Maeda}}]{Torii:1996yi}%
  \BibitemOpen
  \bibfield  {author} {\bibinfo {author} {\bibfnamefont {T.}~\bibnamefont
  {Torii}}, \bibinfo {author} {\bibfnamefont {H.}~\bibnamefont {Yajima}},  and
  \bibinfo {author} {\bibfnamefont {K.-i.}\ \bibnamefont {Maeda}},\ }\href
  {\doibase 10.1103/PhysRevD.55.739} {\bibfield  {journal} {\bibinfo  {journal}
  {\emph {Phys. Rev. D}}\ }\textbf {\bibinfo {volume} {55}},\ \bibinfo {pages}
  {739} (\bibinfo {year} {1997})},\ \Eprint
  {http://arxiv.org/abs/gr-qc/9606034} {arXiv:gr-qc/9606034} \BibitemShut
  {NoStop}%
\bibitem [{\citenamefont {Sotiriou}\ and\ \citenamefont
  {Zhou}(2014)}]{Sotiriou:2013qea}%
  \BibitemOpen
  \bibfield  {author} {\bibinfo {author} {\bibfnamefont {T.~P.}\ \bibnamefont
  {Sotiriou}} and \bibinfo {author} {\bibfnamefont {S.-Y.}\ \bibnamefont
  {Zhou}},\ }\href {\doibase 10.1103/PhysRevLett.112.251102} {\bibfield
  {journal} {\bibinfo  {journal} {\emph {Phys. Rev. Lett.}}\ }\textbf {\bibinfo
  {volume} {112}},\ \bibinfo {pages} {251102} (\bibinfo {year} {2014})},\
  \Eprint {http://arxiv.org/abs/1312.3622} {arXiv:1312.3622 [gr-qc]}
  \BibitemShut {NoStop}%
\bibitem [{\citenamefont {Doneva}\ and\ \citenamefont
  {Yazadjiev}(2018)}]{Doneva:2017bvd}%
  \BibitemOpen
  \bibfield  {author} {\bibinfo {author} {\bibfnamefont {D.~D.}\ \bibnamefont
  {Doneva}} and \bibinfo {author} {\bibfnamefont {S.~S.}\ \bibnamefont
  {Yazadjiev}},\ }\href {\doibase 10.1103/PhysRevLett.120.131103} {\bibfield
  {journal} {\bibinfo  {journal} {\emph {Phys. Rev. Lett.}}\ }\textbf {\bibinfo
  {volume} {120}},\ \bibinfo {pages} {131103} (\bibinfo {year} {2018})},\
  \Eprint {http://arxiv.org/abs/1711.01187} {arXiv:1711.01187 [gr-qc]}
  \BibitemShut {NoStop}%
\bibitem [{\citenamefont {Silva}\ \emph {et~al.}(2018)\citenamefont {Silva},
  \citenamefont {Sakstein}, \citenamefont {Gualtieri}, \citenamefont
  {Sotiriou},\ and\ \citenamefont {Berti}}]{Silva:2017uqg}%
  \BibitemOpen
  \bibfield  {author} {\bibinfo {author} {\bibfnamefont {H.~O.}\ \bibnamefont
  {Silva}}, \bibinfo {author} {\bibfnamefont {J.}~\bibnamefont {Sakstein}},
  \bibinfo {author} {\bibfnamefont {L.}~\bibnamefont {Gualtieri}}, \bibinfo
  {author} {\bibfnamefont {T.~P.}\ \bibnamefont {Sotiriou}},  and \bibinfo
  {author} {\bibfnamefont {E.}~\bibnamefont {Berti}},\ }\href {\doibase
  10.1103/PhysRevLett.120.131104} {\bibfield  {journal} {\bibinfo  {journal}
  {\emph {Phys. Rev. Lett.}}\ }\textbf {\bibinfo {volume} {120}},\ \bibinfo
  {pages} {131104} (\bibinfo {year} {2018})},\ \Eprint
  {http://arxiv.org/abs/1711.02080} {arXiv:1711.02080 [gr-qc]} \BibitemShut
  {NoStop}%
\bibitem [{\citenamefont {Antoniou}\ \emph {et~al.}(2018)\citenamefont
  {Antoniou}, \citenamefont {Bakopoulos},\ and\ \citenamefont
  {Kanti}}]{Antoniou:2017acq}%
  \BibitemOpen
  \bibfield  {author} {\bibinfo {author} {\bibfnamefont {G.}~\bibnamefont
  {Antoniou}}, \bibinfo {author} {\bibfnamefont {A.}~\bibnamefont
  {Bakopoulos}},  and \bibinfo {author} {\bibfnamefont {P.}~\bibnamefont
  {Kanti}},\ }\href {\doibase 10.1103/PhysRevLett.120.131102} {\bibfield
  {journal} {\bibinfo  {journal} {\emph {Phys. Rev. Lett.}}\ }\textbf {\bibinfo
  {volume} {120}},\ \bibinfo {pages} {131102} (\bibinfo {year} {2018})},\
  \Eprint {http://arxiv.org/abs/1711.03390} {arXiv:1711.03390 [hep-th]}
  \BibitemShut {NoStop}%
\bibitem [{\citenamefont {Ayon-Beato}\ and\ \citenamefont
  {Garcia}(1998)}]{Ayon-Beato:1998hmi}%
  \BibitemOpen
  \bibfield  {author} {\bibinfo {author} {\bibfnamefont {E.}~\bibnamefont
  {Ayon-Beato}} and \bibinfo {author} {\bibfnamefont {A.}~\bibnamefont
  {Garcia}},\ }\href {\doibase 10.1103/PhysRevLett.80.5056} {\bibfield
  {journal} {\bibinfo  {journal} {\emph {Phys. Rev. Lett.}}\ }\textbf {\bibinfo
  {volume} {80}},\ \bibinfo {pages} {5056} (\bibinfo {year} {1998})},\ \Eprint
  {http://arxiv.org/abs/gr-qc/9911046} {arXiv:gr-qc/9911046} \BibitemShut
  {NoStop}%
\bibitem [{\citenamefont {Ayon-Beato}\ and\ \citenamefont
  {Garcia}(1999)}]{Ayon-Beato:1999kuh}%
  \BibitemOpen
  \bibfield  {author} {\bibinfo {author} {\bibfnamefont {E.}~\bibnamefont
  {Ayon-Beato}} and \bibinfo {author} {\bibfnamefont {A.}~\bibnamefont
  {Garcia}},\ }\href {\doibase 10.1016/S0370-2693(99)01038-2} {\bibfield
  {journal} {\bibinfo  {journal} {\emph {Phys. Lett. B}}\ }\textbf {\bibinfo
  {volume} {464}},\ \bibinfo {pages} {25} (\bibinfo {year} {1999})},\ \Eprint
  {http://arxiv.org/abs/hep-th/9911174} {arXiv:hep-th/9911174} \BibitemShut
  {NoStop}%
\bibitem [{\citenamefont {Ayon-Beato}\ and\ \citenamefont
  {Garcia}(2000)}]{Ayon-Beato:2000mjt}%
  \BibitemOpen
  \bibfield  {author} {\bibinfo {author} {\bibfnamefont {E.}~\bibnamefont
  {Ayon-Beato}} and \bibinfo {author} {\bibfnamefont {A.}~\bibnamefont
  {Garcia}},\ }\href {\doibase 10.1016/S0370-2693(00)01125-4} {\bibfield
  {journal} {\bibinfo  {journal} {\emph {Phys. Lett. B}}\ }\textbf {\bibinfo
  {volume} {493}},\ \bibinfo {pages} {149} (\bibinfo {year} {2000})},\ \Eprint
  {http://arxiv.org/abs/gr-qc/0009077} {arXiv:gr-qc/0009077} \BibitemShut
  {NoStop}%
\bibitem [{\citenamefont {Bronnikov}(2001)}]{Bronnikov:2000vy}%
  \BibitemOpen
  \bibfield  {author} {\bibinfo {author} {\bibfnamefont {K.~A.}\ \bibnamefont
  {Bronnikov}},\ }\href {\doibase 10.1103/PhysRevD.63.044005} {\bibfield
  {journal} {\bibinfo  {journal} {\emph {Phys. Rev. D}}\ }\textbf {\bibinfo
  {volume} {63}},\ \bibinfo {pages} {044005} (\bibinfo {year} {2001})},\
  \Eprint {http://arxiv.org/abs/gr-qc/0006014} {arXiv:gr-qc/0006014}
  \BibitemShut {NoStop}%
\bibitem [{\citenamefont {Rodrigues}\ and\ \citenamefont
  {de~Sousa~Silva}(2018)}]{Rodrigues:2018bdc}%
  \BibitemOpen
  \bibfield  {author} {\bibinfo {author} {\bibfnamefont {M.~E.}\ \bibnamefont
  {Rodrigues}} and \bibinfo {author} {\bibfnamefont {M.~V.}\ \bibnamefont
  {de~Sousa~Silva}},\ }\href {\doibase 10.1088/1475-7516/2018/06/025}
  {\bibfield  {journal} {\bibinfo  {journal} {\emph {JCAP}}\ }\textbf {\bibinfo
  {volume} {06}},\ \bibinfo {pages} {025} (\bibinfo {year} {2018})},\ \Eprint
  {http://arxiv.org/abs/1802.05095} {arXiv:1802.05095 [gr-qc]} \BibitemShut
  {NoStop}%
\bibitem [{\citenamefont {Maeda}(2022)}]{Maeda:2021jdc}%
  \BibitemOpen
  \bibfield  {author} {\bibinfo {author} {\bibfnamefont {H.}~\bibnamefont
  {Maeda}},\ }\href {\doibase 10.1007/JHEP11(2022)108} {\bibfield  {journal}
  {\bibinfo  {journal} {\emph {JHEP}}\ }\textbf {\bibinfo {volume} {11}},\
  \bibinfo {pages} {108} (\bibinfo {year} {2022})},\ \Eprint
  {http://arxiv.org/abs/2107.04791} {arXiv:2107.04791 [gr-qc]} \BibitemShut
  {NoStop}%
\bibitem [{\citenamefont {Heisenberg}\ and\ \citenamefont
  {Euler}(1936)}]{Heisenberg:1936nmg}%
  \BibitemOpen
  \bibfield  {author} {\bibinfo {author} {\bibfnamefont {W.}~\bibnamefont
  {Heisenberg}} and \bibinfo {author} {\bibfnamefont {H.}~\bibnamefont
  {Euler}},\ }\href {\doibase 10.1007/BF01343663} {\bibfield  {journal}
  {\bibinfo  {journal} {\emph {Z. Phys.}}\ }\textbf {\bibinfo {volume} {98}},\
  \bibinfo {pages} {714} (\bibinfo {year} {1936})},\ \Eprint
  {http://arxiv.org/abs/physics/0605038} {arXiv:physics/0605038} \BibitemShut
  {NoStop}%
\bibitem [{\citenamefont {Born}\ and\ \citenamefont
  {Infeld}(1934)}]{Born:1934gh}%
  \BibitemOpen
  \bibfield  {author} {\bibinfo {author} {\bibfnamefont {M.}~\bibnamefont
  {Born}} and \bibinfo {author} {\bibfnamefont {L.}~\bibnamefont {Infeld}},\
  }\href {\doibase 10.1098/rspa.1934.0059} {\bibfield  {journal} {\bibinfo
  {journal} {\emph {Proc. Roy. Soc. Lond. A}}\ }\textbf {\bibinfo {volume}
  {144}},\ \bibinfo {pages} {425} (\bibinfo {year} {1934})}\BibitemShut
  {NoStop}%
\bibitem [{\citenamefont {Yajima}\ and\ \citenamefont
  {Tamaki}(2001)}]{Yajima:2000kw}%
  \BibitemOpen
  \bibfield  {author} {\bibinfo {author} {\bibfnamefont {H.}~\bibnamefont
  {Yajima}} and \bibinfo {author} {\bibfnamefont {T.}~\bibnamefont {Tamaki}},\
  }\href {\doibase 10.1103/PhysRevD.63.064007} {\bibfield  {journal} {\bibinfo
  {journal} {\emph {Phys. Rev. D}}\ }\textbf {\bibinfo {volume} {63}},\
  \bibinfo {pages} {064007} (\bibinfo {year} {2001})},\ \Eprint
  {http://arxiv.org/abs/gr-qc/0005016} {arXiv:gr-qc/0005016} \BibitemShut
  {NoStop}%
\bibitem [{\citenamefont {Fernando}\ and\ \citenamefont
  {Krug}(2003)}]{Fernando:2003tz}%
  \BibitemOpen
  \bibfield  {author} {\bibinfo {author} {\bibfnamefont {S.}~\bibnamefont
  {Fernando}} and \bibinfo {author} {\bibfnamefont {D.}~\bibnamefont {Krug}},\
  }\href {\doibase 10.1023/A:1021315214180} {\bibfield  {journal} {\bibinfo
  {journal} {\emph {Gen. Rel. Grav.}}\ }\textbf {\bibinfo {volume} {35}},\
  \bibinfo {pages} {129} (\bibinfo {year} {2003})},\ \Eprint
  {http://arxiv.org/abs/hep-th/0306120} {arXiv:hep-th/0306120} \BibitemShut
  {NoStop}%
\bibitem [{\citenamefont {Cai}\ \emph {et~al.}(2004)\citenamefont {Cai},
  \citenamefont {Pang},\ and\ \citenamefont {Wang}}]{Cai:2004eh}%
  \BibitemOpen
  \bibfield  {author} {\bibinfo {author} {\bibfnamefont {R.-G.}\ \bibnamefont
  {Cai}}, \bibinfo {author} {\bibfnamefont {D.-W.}\ \bibnamefont {Pang}},  and
  \bibinfo {author} {\bibfnamefont {A.}~\bibnamefont {Wang}},\ }\href {\doibase
  10.1103/PhysRevD.70.124034} {\bibfield  {journal} {\bibinfo  {journal} {\emph
  {Phys. Rev. D}}\ }\textbf {\bibinfo {volume} {70}},\ \bibinfo {pages}
  {124034} (\bibinfo {year} {2004})},\ \Eprint
  {http://arxiv.org/abs/hep-th/0410158} {arXiv:hep-th/0410158} \BibitemShut
  {NoStop}%
\bibitem [{\citenamefont {Dey}(2004)}]{Dey:2004yt}%
  \BibitemOpen
  \bibfield  {author} {\bibinfo {author} {\bibfnamefont {T.~K.}\ \bibnamefont
  {Dey}},\ }\href {\doibase 10.1016/j.physletb.2004.06.047} {\bibfield
  {journal} {\bibinfo  {journal} {\emph {Phys. Lett. B}}\ }\textbf {\bibinfo
  {volume} {595}},\ \bibinfo {pages} {484} (\bibinfo {year} {2004})},\ \Eprint
  {http://arxiv.org/abs/hep-th/0406169} {arXiv:hep-th/0406169} \BibitemShut
  {NoStop}%
\bibitem [{\citenamefont {Moreno}\ and\ \citenamefont
  {Sarbach}(2003)}]{Moreno:2002gg}%
  \BibitemOpen
  \bibfield  {author} {\bibinfo {author} {\bibfnamefont {C.}~\bibnamefont
  {Moreno}} and \bibinfo {author} {\bibfnamefont {O.}~\bibnamefont {Sarbach}},\
  }\href {\doibase 10.1103/PhysRevD.67.024028} {\bibfield  {journal} {\bibinfo
  {journal} {\emph {Phys. Rev. D}}\ }\textbf {\bibinfo {volume} {67}},\
  \bibinfo {pages} {024028} (\bibinfo {year} {2003})},\ \Eprint
  {http://arxiv.org/abs/gr-qc/0208090} {arXiv:gr-qc/0208090} \BibitemShut
  {NoStop}%
\bibitem [{\citenamefont {Toshmatov}\ \emph
  {et~al.}(2018{\natexlab{a}})\citenamefont {Toshmatov}, \citenamefont
  {Stuchl\'\i{}k}, \citenamefont {Schee},\ and\ \citenamefont
  {Ahmedov}}]{Toshmatov:2018tyo}%
  \BibitemOpen
  \bibfield  {author} {\bibinfo {author} {\bibfnamefont {B.}~\bibnamefont
  {Toshmatov}}, \bibinfo {author} {\bibfnamefont {Z.}~\bibnamefont
  {Stuchl\'\i{}k}}, \bibinfo {author} {\bibfnamefont {J.}~\bibnamefont
  {Schee}},  and \bibinfo {author} {\bibfnamefont {B.}~\bibnamefont
  {Ahmedov}},\ }\href {\doibase 10.1103/PhysRevD.97.084058} {\bibfield
  {journal} {\bibinfo  {journal} {\emph {Phys. Rev. D}}\ }\textbf {\bibinfo
  {volume} {97}},\ \bibinfo {pages} {084058} (\bibinfo {year}
  {2018}{\natexlab{a}})},\ \Eprint {http://arxiv.org/abs/1805.00240}
  {arXiv:1805.00240 [gr-qc]} \BibitemShut {NoStop}%
\bibitem [{\citenamefont {Toshmatov}\ \emph
  {et~al.}(2018{\natexlab{b}})\citenamefont {Toshmatov}, \citenamefont
  {Stuchl\'\i{}k},\ and\ \citenamefont {Ahmedov}}]{Toshmatov:2018ell}%
  \BibitemOpen
  \bibfield  {author} {\bibinfo {author} {\bibfnamefont {B.}~\bibnamefont
  {Toshmatov}}, \bibinfo {author} {\bibfnamefont {Z.}~\bibnamefont
  {Stuchl\'\i{}k}},  and \bibinfo {author} {\bibfnamefont {B.}~\bibnamefont
  {Ahmedov}},\ }\href {\doibase 10.1103/PhysRevD.98.085021} {\bibfield
  {journal} {\bibinfo  {journal} {\emph {Phys. Rev. D}}\ }\textbf {\bibinfo
  {volume} {98}},\ \bibinfo {pages} {085021} (\bibinfo {year}
  {2018}{\natexlab{b}})},\ \Eprint {http://arxiv.org/abs/1810.06383}
  {arXiv:1810.06383 [gr-qc]} \BibitemShut {NoStop}%
\bibitem [{\citenamefont {Toshmatov}\ \emph {et~al.}(2019)\citenamefont
  {Toshmatov}, \citenamefont {Stuchl\'\i{}k}, \citenamefont {Ahmedov},\ and\
  \citenamefont {Malafarina}}]{Toshmatov:2019gxg}%
  \BibitemOpen
  \bibfield  {author} {\bibinfo {author} {\bibfnamefont {B.}~\bibnamefont
  {Toshmatov}}, \bibinfo {author} {\bibfnamefont {Z.}~\bibnamefont
  {Stuchl\'\i{}k}}, \bibinfo {author} {\bibfnamefont {B.}~\bibnamefont
  {Ahmedov}},  and \bibinfo {author} {\bibfnamefont {D.}~\bibnamefont
  {Malafarina}},\ }\href {\doibase 10.1103/PhysRevD.99.064043} {\bibfield
  {journal} {\bibinfo  {journal} {\emph {Phys. Rev. D}}\ }\textbf {\bibinfo
  {volume} {99}},\ \bibinfo {pages} {064043} (\bibinfo {year} {2019})},\
  \Eprint {http://arxiv.org/abs/1903.03778} {arXiv:1903.03778 [gr-qc]}
  \BibitemShut {NoStop}%
\bibitem [{\citenamefont {Nomura}\ \emph {et~al.}(2020)\citenamefont {Nomura},
  \citenamefont {Yoshida},\ and\ \citenamefont {Soda}}]{Nomura:2020tpc}%
  \BibitemOpen
  \bibfield  {author} {\bibinfo {author} {\bibfnamefont {K.}~\bibnamefont
  {Nomura}}, \bibinfo {author} {\bibfnamefont {D.}~\bibnamefont {Yoshida}},
  and \bibinfo {author} {\bibfnamefont {J.}~\bibnamefont {Soda}},\ }\href
  {\doibase 10.1103/PhysRevD.101.124026} {\bibfield  {journal} {\bibinfo
  {journal} {\emph {Phys. Rev. D}}\ }\textbf {\bibinfo {volume} {101}},\
  \bibinfo {pages} {124026} (\bibinfo {year} {2020})},\ \Eprint
  {http://arxiv.org/abs/2004.07560} {arXiv:2004.07560 [gr-qc]} \BibitemShut
  {NoStop}%
\bibitem [{\citenamefont {De~Felice}\ and\ \citenamefont
  {Tsujikawa}(2025)}]{DeFelice:2024seu}%
  \BibitemOpen
  \bibfield  {author} {\bibinfo {author} {\bibfnamefont {A.}~\bibnamefont
  {De~Felice}} and \bibinfo {author} {\bibfnamefont {S.}~\bibnamefont
  {Tsujikawa}},\ }\href {\doibase 10.1103/PhysRevLett.134.081401} {\bibfield
  {journal} {\bibinfo  {journal} {\emph {Phys. Rev. Lett.}}\ }\textbf {\bibinfo
  {volume} {134}},\ \bibinfo {pages} {081401} (\bibinfo {year} {2025})},\
  \Eprint {http://arxiv.org/abs/2410.00314} {arXiv:2410.00314 [gr-qc]}
  \BibitemShut {NoStop}%
\bibitem [{\citenamefont {Gannouji}\ and\ \citenamefont
  {Baez}(2022)}]{Gannouji:2021oqz}%
  \BibitemOpen
  \bibfield  {author} {\bibinfo {author} {\bibfnamefont {R.}~\bibnamefont
  {Gannouji}} and \bibinfo {author} {\bibfnamefont {Y.~R.}\ \bibnamefont
  {Baez}},\ }\href {\doibase 10.1007/JHEP02(2022)020} {\bibfield  {journal}
  {\bibinfo  {journal} {\emph {JHEP}}\ }\textbf {\bibinfo {volume} {02}},\
  \bibinfo {pages} {020} (\bibinfo {year} {2022})},\ \Eprint
  {http://arxiv.org/abs/2112.00109} {arXiv:2112.00109 [gr-qc]} \BibitemShut
  {NoStop}%
\bibitem [{\citenamefont {Kase}\ and\ \citenamefont
  {Tsujikawa}(2023)}]{Kase:2023kvq}%
  \BibitemOpen
  \bibfield  {author} {\bibinfo {author} {\bibfnamefont {R.}~\bibnamefont
  {Kase}} and \bibinfo {author} {\bibfnamefont {S.}~\bibnamefont {Tsujikawa}},\
  }\href {\doibase 10.1103/PhysRevD.107.104045} {\bibfield  {journal} {\bibinfo
   {journal} {\emph {Phys. Rev. D}}\ }\textbf {\bibinfo {volume} {107}},\
  \bibinfo {pages} {104045} (\bibinfo {year} {2023})},\ \Eprint
  {http://arxiv.org/abs/2301.10362} {arXiv:2301.10362 [gr-qc]} \BibitemShut
  {NoStop}%
\bibitem [{\citenamefont {Frolov}(2016)}]{Frolov:2016pav}%
  \BibitemOpen
  \bibfield  {author} {\bibinfo {author} {\bibfnamefont {V.~P.}\ \bibnamefont
  {Frolov}},\ }\href {\doibase 10.1103/PhysRevD.94.104056} {\bibfield
  {journal} {\bibinfo  {journal} {\emph {Phys. Rev. D}}\ }\textbf {\bibinfo
  {volume} {94}},\ \bibinfo {pages} {104056} (\bibinfo {year} {2016})},\
  \Eprint {http://arxiv.org/abs/1609.01758} {arXiv:1609.01758 [gr-qc]}
  \BibitemShut {NoStop}%
\bibitem [{\citenamefont {Regge}\ and\ \citenamefont
  {Wheeler}(1957)}]{Regge:1957td}%
  \BibitemOpen
  \bibfield  {author} {\bibinfo {author} {\bibfnamefont {T.}~\bibnamefont
  {Regge}} and \bibinfo {author} {\bibfnamefont {J.~A.}\ \bibnamefont
  {Wheeler}},\ }\href {\doibase 10.1103/PhysRev.108.1063} {\bibfield  {journal}
  {\bibinfo  {journal} {\emph {Phys. Rev.}}\ }\textbf {\bibinfo {volume}
  {108}},\ \bibinfo {pages} {1063} (\bibinfo {year} {1957})}\BibitemShut
  {NoStop}%
\bibitem [{\citenamefont {Zerilli}(1970)}]{Zerilli:1970se}%
  \BibitemOpen
  \bibfield  {author} {\bibinfo {author} {\bibfnamefont {F.~J.}\ \bibnamefont
  {Zerilli}},\ }\href {\doibase 10.1103/PhysRevLett.24.737} {\bibfield
  {journal} {\bibinfo  {journal} {\emph {Phys. Rev. Lett.}}\ }\textbf {\bibinfo
  {volume} {24}},\ \bibinfo {pages} {737} (\bibinfo {year} {1970})}\BibitemShut
  {NoStop}%
\bibitem [{\citenamefont {Moncrief}(1974)}]{Moncrief:1974ng}%
  \BibitemOpen
  \bibfield  {author} {\bibinfo {author} {\bibfnamefont {V.}~\bibnamefont
  {Moncrief}},\ }\href {\doibase 10.1103/PhysRevD.10.1057} {\bibfield
  {journal} {\bibinfo  {journal} {\emph {Phys. Rev. D}}\ }\textbf {\bibinfo
  {volume} {10}},\ \bibinfo {pages} {1057} (\bibinfo {year}
  {1974})}\BibitemShut {NoStop}%
\bibitem [{\citenamefont {Zerilli}(1974)}]{Zerilli:1974ai}%
  \BibitemOpen
  \bibfield  {author} {\bibinfo {author} {\bibfnamefont {F.~J.}\ \bibnamefont
  {Zerilli}},\ }\href {\doibase 10.1103/PhysRevD.9.860} {\bibfield  {journal}
  {\bibinfo  {journal} {\emph {Phys. Rev. D}}\ }\textbf {\bibinfo {volume}
  {9}},\ \bibinfo {pages} {860} (\bibinfo {year} {1974})}\BibitemShut {NoStop}%
\bibitem [{\citenamefont {De~Felice}\ \emph {et~al.}(2011)\citenamefont
  {De~Felice}, \citenamefont {Suyama},\ and\ \citenamefont
  {Tanaka}}]{DeFelice:2011ka}%
  \BibitemOpen
  \bibfield  {author} {\bibinfo {author} {\bibfnamefont {A.}~\bibnamefont
  {De~Felice}}, \bibinfo {author} {\bibfnamefont {T.}~\bibnamefont {Suyama}},
  and \bibinfo {author} {\bibfnamefont {T.}~\bibnamefont {Tanaka}},\ }\href
  {\doibase 10.1103/PhysRevD.83.104035} {\bibfield  {journal} {\bibinfo
  {journal} {\emph {Phys. Rev. D}}\ }\textbf {\bibinfo {volume} {83}},\
  \bibinfo {pages} {104035} (\bibinfo {year} {2011})},\ \Eprint
  {http://arxiv.org/abs/1102.1521} {arXiv:1102.1521 [gr-qc]} \BibitemShut
  {NoStop}%
\bibitem [{\citenamefont {Kobayashi}\ \emph {et~al.}(2012)\citenamefont
  {Kobayashi}, \citenamefont {Motohashi},\ and\ \citenamefont
  {Suyama}}]{Kobayashi:2012kh}%
  \BibitemOpen
  \bibfield  {author} {\bibinfo {author} {\bibfnamefont {T.}~\bibnamefont
  {Kobayashi}}, \bibinfo {author} {\bibfnamefont {H.}~\bibnamefont
  {Motohashi}},  and \bibinfo {author} {\bibfnamefont {T.}~\bibnamefont
  {Suyama}},\ }\href {\doibase 10.1103/PhysRevD.85.084025} {\bibfield
  {journal} {\bibinfo  {journal} {\emph {Phys. Rev. D}}\ }\textbf {\bibinfo
  {volume} {85}},\ \bibinfo {pages} {084025} (\bibinfo {year} {2012})},\
  \bibinfo {note} {[Erratum: Phys.Rev.D 96, 109903 (2017)]},\ \Eprint
  {http://arxiv.org/abs/1202.4893} {arXiv:1202.4893 [gr-qc]} \BibitemShut
  {NoStop}%
\bibitem [{\citenamefont {Kobayashi}\ \emph {et~al.}(2014)\citenamefont
  {Kobayashi}, \citenamefont {Motohashi},\ and\ \citenamefont
  {Suyama}}]{Kobayashi:2014wsa}%
  \BibitemOpen
  \bibfield  {author} {\bibinfo {author} {\bibfnamefont {T.}~\bibnamefont
  {Kobayashi}}, \bibinfo {author} {\bibfnamefont {H.}~\bibnamefont
  {Motohashi}},  and \bibinfo {author} {\bibfnamefont {T.}~\bibnamefont
  {Suyama}},\ }\href {\doibase 10.1103/PhysRevD.89.084042} {\bibfield
  {journal} {\bibinfo  {journal} {\emph {Phys. Rev. D}}\ }\textbf {\bibinfo
  {volume} {89}},\ \bibinfo {pages} {084042} (\bibinfo {year} {2014})},\
  \Eprint {http://arxiv.org/abs/1402.6740} {arXiv:1402.6740 [gr-qc]}
  \BibitemShut {NoStop}%
\bibitem [{\citenamefont {Armendariz-Picon}\ \emph {et~al.}(1999)\citenamefont
  {Armendariz-Picon}, \citenamefont {Damour},\ and\ \citenamefont
  {Mukhanov}}]{Armendariz-Picon:1999hyi}%
  \BibitemOpen
  \bibfield  {author} {\bibinfo {author} {\bibfnamefont {C.}~\bibnamefont
  {Armendariz-Picon}}, \bibinfo {author} {\bibfnamefont {T.}~\bibnamefont
  {Damour}},  and \bibinfo {author} {\bibfnamefont {V.~F.}\ \bibnamefont
  {Mukhanov}},\ }\href {\doibase 10.1016/S0370-2693(99)00603-6} {\bibfield
  {journal} {\bibinfo  {journal} {\emph {Phys. Lett. B}}\ }\textbf {\bibinfo
  {volume} {458}},\ \bibinfo {pages} {209} (\bibinfo {year} {1999})},\ \Eprint
  {http://arxiv.org/abs/hep-th/9904075} {arXiv:hep-th/9904075} \BibitemShut
  {NoStop}%
\bibitem [{\citenamefont {Chiba}\ \emph {et~al.}(2000)\citenamefont {Chiba},
  \citenamefont {Okabe},\ and\ \citenamefont {Yamaguchi}}]{Chiba:1999ka}%
  \BibitemOpen
  \bibfield  {author} {\bibinfo {author} {\bibfnamefont {T.}~\bibnamefont
  {Chiba}}, \bibinfo {author} {\bibfnamefont {T.}~\bibnamefont {Okabe}},  and
  \bibinfo {author} {\bibfnamefont {M.}~\bibnamefont {Yamaguchi}},\ }\href
  {\doibase 10.1103/PhysRevD.62.023511} {\bibfield  {journal} {\bibinfo
  {journal} {\emph {Phys. Rev. D}}\ }\textbf {\bibinfo {volume} {62}},\
  \bibinfo {pages} {023511} (\bibinfo {year} {2000})},\ \Eprint
  {http://arxiv.org/abs/astro-ph/9912463} {arXiv:astro-ph/9912463} \BibitemShut
  {NoStop}%
\bibitem [{\citenamefont {Armendariz-Picon}\ \emph {et~al.}(2000)\citenamefont
  {Armendariz-Picon}, \citenamefont {Mukhanov},\ and\ \citenamefont
  {Steinhardt}}]{Armendariz-Picon:2000nqq}%
  \BibitemOpen
  \bibfield  {author} {\bibinfo {author} {\bibfnamefont {C.}~\bibnamefont
  {Armendariz-Picon}}, \bibinfo {author} {\bibfnamefont {V.~F.}\ \bibnamefont
  {Mukhanov}},  and \bibinfo {author} {\bibfnamefont {P.~J.}\ \bibnamefont
  {Steinhardt}},\ }\href {\doibase 10.1103/PhysRevLett.85.4438} {\bibfield
  {journal} {\bibinfo  {journal} {\emph {Phys. Rev. Lett.}}\ }\textbf {\bibinfo
  {volume} {85}},\ \bibinfo {pages} {4438} (\bibinfo {year} {2000})},\ \Eprint
  {http://arxiv.org/abs/astro-ph/0004134} {arXiv:astro-ph/0004134} \BibitemShut
  {NoStop}%
\bibitem [{\citenamefont {Pereira}\ \emph {et~al.}(2024)\citenamefont
  {Pereira}, \citenamefont {C.~Rodrigues}, \citenamefont {Silva}, \citenamefont
  {Fabris}, \citenamefont {Rodrigues},\ and\ \citenamefont
  {Belich}}]{Pereira:2024rtv}%
  \BibitemOpen
  \bibfield  {author} {\bibinfo {author} {\bibfnamefont {C.~F.~S.}\
  \bibnamefont {Pereira}}, \bibinfo {author} {\bibfnamefont {D.}~\bibnamefont
  {C.~Rodrigues}}, \bibinfo {author} {\bibfnamefont {M.~V. d.~S.}\ \bibnamefont
  {Silva}}, \bibinfo {author} {\bibfnamefont {J.~C.}\ \bibnamefont {Fabris}},
  \bibinfo {author} {\bibfnamefont {M.~E.}\ \bibnamefont {Rodrigues}},  and
  \bibinfo {author} {\bibfnamefont {H.}~\bibnamefont {Belich}},\ }\Eprint
  {http://arxiv.org/abs/2409.09182} {arXiv:2409.09182 [gr-qc]} \BibitemShut
  {NoStop}%
\bibitem [{\citenamefont {Gasperini}\ and\ \citenamefont
  {Veneziano}(2003)}]{Gasperini:2002bn}%
  \BibitemOpen
  \bibfield  {author} {\bibinfo {author} {\bibfnamefont {M.}~\bibnamefont
  {Gasperini}} and \bibinfo {author} {\bibfnamefont {G.}~\bibnamefont
  {Veneziano}},\ }\href {\doibase 10.1016/S0370-1573(02)00389-7} {\bibfield
  {journal} {\bibinfo  {journal} {\emph {Phys. Rept.}}\ }\textbf {\bibinfo
  {volume} {373}},\ \bibinfo {pages} {1} (\bibinfo {year} {2003})},\ \Eprint
  {http://arxiv.org/abs/hep-th/0207130} {arXiv:hep-th/0207130} \BibitemShut
  {NoStop}%
\bibitem [{\citenamefont {Gibbons}\ and\ \citenamefont
  {Maeda}(1988)}]{Gibbons:1987ps}%
  \BibitemOpen
  \bibfield  {author} {\bibinfo {author} {\bibfnamefont {G.~W.}\ \bibnamefont
  {Gibbons}} and \bibinfo {author} {\bibfnamefont {K.-i.}\ \bibnamefont
  {Maeda}},\ }\href {\doibase 10.1016/0550-3213(88)90006-5} {\bibfield
  {journal} {\bibinfo  {journal} {\emph {Nucl. Phys. B}}\ }\textbf {\bibinfo
  {volume} {298}},\ \bibinfo {pages} {741} (\bibinfo {year}
  {1988})}\BibitemShut {NoStop}%
\bibitem [{\citenamefont {Garfinkle}\ \emph {et~al.}(1991)\citenamefont
  {Garfinkle}, \citenamefont {Horowitz},\ and\ \citenamefont
  {Strominger}}]{Garfinkle:1990qj}%
  \BibitemOpen
  \bibfield  {author} {\bibinfo {author} {\bibfnamefont {D.}~\bibnamefont
  {Garfinkle}}, \bibinfo {author} {\bibfnamefont {G.~T.}\ \bibnamefont
  {Horowitz}},  and \bibinfo {author} {\bibfnamefont {A.}~\bibnamefont
  {Strominger}},\ }\href {\doibase 10.1103/PhysRevD.43.3140} {\bibfield
  {journal} {\bibinfo  {journal} {\emph {Phys. Rev. D}}\ }\textbf {\bibinfo
  {volume} {43}},\ \bibinfo {pages} {3140} (\bibinfo {year} {1991})},\ \bibinfo
  {note} {[Erratum: Phys.Rev.D 45, 3888 (1992)]}\BibitemShut {NoStop}%
\bibitem [{\citenamefont {Herdeiro}\ \emph {et~al.}(2018)\citenamefont
  {Herdeiro}, \citenamefont {Radu}, \citenamefont {Sanchis-Gual},\ and\
  \citenamefont {Font}}]{Herdeiro:2018wub}%
  \BibitemOpen
  \bibfield  {author} {\bibinfo {author} {\bibfnamefont {C.~A.~R.}\
  \bibnamefont {Herdeiro}}, \bibinfo {author} {\bibfnamefont {E.}~\bibnamefont
  {Radu}}, \bibinfo {author} {\bibfnamefont {N.}~\bibnamefont {Sanchis-Gual}},
  and \bibinfo {author} {\bibfnamefont {J.~A.}\ \bibnamefont {Font}},\ }\href
  {\doibase 10.1103/PhysRevLett.121.101102} {\bibfield  {journal} {\bibinfo
  {journal} {\emph {Phys. Rev. Lett.}}\ }\textbf {\bibinfo {volume} {121}},\
  \bibinfo {pages} {101102} (\bibinfo {year} {2018})},\ \Eprint
  {http://arxiv.org/abs/1806.05190} {arXiv:1806.05190 [gr-qc]} \BibitemShut
  {NoStop}%
\bibitem [{\citenamefont {Fernandes}\ \emph {et~al.}(2019)\citenamefont
  {Fernandes}, \citenamefont {Herdeiro}, \citenamefont {Pombo}, \citenamefont
  {Radu},\ and\ \citenamefont {Sanchis-Gual}}]{Fernandes:2019rez}%
  \BibitemOpen
  \bibfield  {author} {\bibinfo {author} {\bibfnamefont {P.~G.~S.}\
  \bibnamefont {Fernandes}}, \bibinfo {author} {\bibfnamefont {C.~A.~R.}\
  \bibnamefont {Herdeiro}}, \bibinfo {author} {\bibfnamefont {A.~M.}\
  \bibnamefont {Pombo}}, \bibinfo {author} {\bibfnamefont {E.}~\bibnamefont
  {Radu}},  and \bibinfo {author} {\bibfnamefont {N.}~\bibnamefont
  {Sanchis-Gual}},\ }\href {\doibase 10.1088/1361-6382/ab23a1} {\bibfield
  {journal} {\bibinfo  {journal} {\emph {Class. Quant. Grav.}}\ }\textbf
  {\bibinfo {volume} {36}},\ \bibinfo {pages} {134002} (\bibinfo {year}
  {2019})},\ \bibinfo {note} {[Erratum: Class.Quant.Grav. 37, 049501 (2020)]},\
  \Eprint {http://arxiv.org/abs/1902.05079} {arXiv:1902.05079 [gr-qc]}
  \BibitemShut {NoStop}%
\bibitem [{\citenamefont {Myung}\ and\ \citenamefont
  {Zou}(2019)}]{Myung:2018jvi}%
  \BibitemOpen
  \bibfield  {author} {\bibinfo {author} {\bibfnamefont {Y.~S.}\ \bibnamefont
  {Myung}} and \bibinfo {author} {\bibfnamefont {D.-C.}\ \bibnamefont {Zou}},\
  }\href {\doibase 10.1016/j.physletb.2019.01.046} {\bibfield  {journal}
  {\bibinfo  {journal} {\emph {Phys. Lett. B}}\ }\textbf {\bibinfo {volume}
  {790}},\ \bibinfo {pages} {400} (\bibinfo {year} {2019})},\ \Eprint
  {http://arxiv.org/abs/1812.03604} {arXiv:1812.03604 [gr-qc]} \BibitemShut
  {NoStop}%
\bibitem [{\citenamefont {Bl\'azquez-Salcedo}\ \emph
  {et~al.}(2020)\citenamefont {Bl\'azquez-Salcedo}, \citenamefont {Herdeiro},
  \citenamefont {Kunz}, \citenamefont {Pombo},\ and\ \citenamefont
  {Radu}}]{Blazquez-Salcedo:2020nhs}%
  \BibitemOpen
  \bibfield  {author} {\bibinfo {author} {\bibfnamefont {J.~L.}\ \bibnamefont
  {Bl\'azquez-Salcedo}}, \bibinfo {author} {\bibfnamefont {C.~A.~R.}\
  \bibnamefont {Herdeiro}}, \bibinfo {author} {\bibfnamefont {J.}~\bibnamefont
  {Kunz}}, \bibinfo {author} {\bibfnamefont {A.~M.}\ \bibnamefont {Pombo}},
  and \bibinfo {author} {\bibfnamefont {E.}~\bibnamefont {Radu}},\ }\href
  {\doibase 10.1016/j.physletb.2020.135493} {\bibfield  {journal} {\bibinfo
  {journal} {\emph {Phys. Lett. B}}\ }\textbf {\bibinfo {volume} {806}},\
  \bibinfo {pages} {135493} (\bibinfo {year} {2020})},\ \Eprint
  {http://arxiv.org/abs/2002.00963} {arXiv:2002.00963 [gr-qc]} \BibitemShut
  {NoStop}%
\bibitem [{\citenamefont {De~Felice}\ \emph {et~al.}(2024)\citenamefont
  {De~Felice}, \citenamefont {Kase},\ and\ \citenamefont
  {Tsujikawa}}]{DeFelice:2024bdq}%
  \BibitemOpen
  \bibfield  {author} {\bibinfo {author} {\bibfnamefont {A.}~\bibnamefont
  {De~Felice}}, \bibinfo {author} {\bibfnamefont {R.}~\bibnamefont {Kase}},
  and \bibinfo {author} {\bibfnamefont {S.}~\bibnamefont {Tsujikawa}},\ }\href
  {\doibase 10.1088/1475-7516/2024/10/072} {\bibfield  {journal} {\bibinfo
  {journal} {\emph {JCAP}}\ }\textbf {\bibinfo {volume} {10}},\ \bibinfo
  {pages} {072} (\bibinfo {year} {2024})},\ \Eprint
  {http://arxiv.org/abs/2409.15606} {arXiv:2409.15606 [gr-qc]} \BibitemShut
  {NoStop}%
\bibitem [{\citenamefont {Modesto}(2012)}]{Modesto:2011kw}%
  \BibitemOpen
  \bibfield  {author} {\bibinfo {author} {\bibfnamefont {L.}~\bibnamefont
  {Modesto}},\ }\href {\doibase 10.1103/PhysRevD.86.044005} {\bibfield
  {journal} {\bibinfo  {journal} {\emph {Phys. Rev. D}}\ }\textbf {\bibinfo
  {volume} {86}},\ \bibinfo {pages} {044005} (\bibinfo {year} {2012})},\
  \Eprint {http://arxiv.org/abs/1107.2403} {arXiv:1107.2403 [hep-th]}
  \BibitemShut {NoStop}%
\bibitem [{\citenamefont {Biswas}\ \emph {et~al.}(2012)\citenamefont {Biswas},
  \citenamefont {Gerwick}, \citenamefont {Koivisto},\ and\ \citenamefont
  {Mazumdar}}]{Biswas:2011ar}%
  \BibitemOpen
  \bibfield  {author} {\bibinfo {author} {\bibfnamefont {T.}~\bibnamefont
  {Biswas}}, \bibinfo {author} {\bibfnamefont {E.}~\bibnamefont {Gerwick}},
  \bibinfo {author} {\bibfnamefont {T.}~\bibnamefont {Koivisto}},  and \bibinfo
  {author} {\bibfnamefont {A.}~\bibnamefont {Mazumdar}},\ }\href {\doibase
  10.1103/PhysRevLett.108.031101} {\bibfield  {journal} {\bibinfo  {journal}
  {\emph {Phys. Rev. Lett.}}\ }\textbf {\bibinfo {volume} {108}},\ \bibinfo
  {pages} {031101} (\bibinfo {year} {2012})},\ \Eprint
  {http://arxiv.org/abs/1110.5249} {arXiv:1110.5249 [gr-qc]} \BibitemShut
  {NoStop}%
\bibitem [{\citenamefont {Modesto}\ and\ \citenamefont
  {Rachwal}(2014)}]{Modesto:2014lga}%
  \BibitemOpen
  \bibfield  {author} {\bibinfo {author} {\bibfnamefont {L.}~\bibnamefont
  {Modesto}} and \bibinfo {author} {\bibfnamefont {L.}~\bibnamefont
  {Rachwal}},\ }\href {\doibase 10.1016/j.nuclphysb.2014.10.015} {\bibfield
  {journal} {\bibinfo  {journal} {\emph {Nucl. Phys. B}}\ }\textbf {\bibinfo
  {volume} {889}},\ \bibinfo {pages} {228} (\bibinfo {year} {2014})},\ \Eprint
  {http://arxiv.org/abs/1407.8036} {arXiv:1407.8036 [hep-th]} \BibitemShut
  {NoStop}%
\bibitem [{\citenamefont {Tomboulis}(2015)}]{Tomboulis:2015gfa}%
  \BibitemOpen
  \bibfield  {author} {\bibinfo {author} {\bibfnamefont {E.~T.}\ \bibnamefont
  {Tomboulis}},\ }\href {\doibase 10.1103/PhysRevD.92.125037} {\bibfield
  {journal} {\bibinfo  {journal} {\emph {Phys. Rev. D}}\ }\textbf {\bibinfo
  {volume} {92}},\ \bibinfo {pages} {125037} (\bibinfo {year} {2015})},\
  \Eprint {http://arxiv.org/abs/1507.00981} {arXiv:1507.00981 [hep-th]}
  \BibitemShut {NoStop}%
\bibitem [{\citenamefont {Bueno}\ \emph
  {et~al.}(2024{\natexlab{a}})\citenamefont {Bueno}, \citenamefont {Cano},\
  and\ \citenamefont {Hennigar}}]{Bueno:2024dgm}%
  \BibitemOpen
  \bibfield  {author} {\bibinfo {author} {\bibfnamefont {P.}~\bibnamefont
  {Bueno}}, \bibinfo {author} {\bibfnamefont {P.~A.}\ \bibnamefont {Cano}},
  and \bibinfo {author} {\bibfnamefont {R.~A.}\ \bibnamefont {Hennigar}},\
  }\Eprint {http://arxiv.org/abs/2403.04827} {arXiv:2403.04827 [gr-qc]}
  \BibitemShut {NoStop}%
\bibitem [{\citenamefont {Bueno}\ \emph
  {et~al.}(2024{\natexlab{b}})\citenamefont {Bueno}, \citenamefont {Cano},
  \citenamefont {Hennigar},\ and\ \citenamefont {Murcia}}]{Bueno:2024zsx}%
  \BibitemOpen
  \bibfield  {author} {\bibinfo {author} {\bibfnamefont {P.}~\bibnamefont
  {Bueno}}, \bibinfo {author} {\bibfnamefont {P.~A.}\ \bibnamefont {Cano}},
  \bibinfo {author} {\bibfnamefont {R.~A.}\ \bibnamefont {Hennigar}},  and
  \bibinfo {author} {\bibfnamefont {A.~J.}\ \bibnamefont {Murcia}},\ }\Eprint
  {http://arxiv.org/abs/2412.02740} {arXiv:2412.02740 [gr-qc]} \BibitemShut
  {NoStop}%
\bibitem [{\citenamefont {Kaup}(1968)}]{Kaup:1968zz}%
  \BibitemOpen
  \bibfield  {author} {\bibinfo {author} {\bibfnamefont {D.~J.}\ \bibnamefont
  {Kaup}},\ }\href {\doibase 10.1103/PhysRev.172.1331} {\bibfield  {journal}
  {\bibinfo  {journal} {\emph {Phys. Rev.}}\ }\textbf {\bibinfo {volume}
  {172}},\ \bibinfo {pages} {1331} (\bibinfo {year} {1968})}\BibitemShut
  {NoStop}%
\bibitem [{\citenamefont {Brito}\ \emph {et~al.}(2016)\citenamefont {Brito},
  \citenamefont {Cardoso}, \citenamefont {Herdeiro},\ and\ \citenamefont
  {Radu}}]{Brito:2015pxa}%
  \BibitemOpen
  \bibfield  {author} {\bibinfo {author} {\bibfnamefont {R.}~\bibnamefont
  {Brito}}, \bibinfo {author} {\bibfnamefont {V.}~\bibnamefont {Cardoso}},
  \bibinfo {author} {\bibfnamefont {C.~A.~R.}\ \bibnamefont {Herdeiro}},  and
  \bibinfo {author} {\bibfnamefont {E.}~\bibnamefont {Radu}},\ }\href {\doibase
  10.1016/j.physletb.2015.11.051} {\bibfield  {journal} {\bibinfo  {journal}
  {\emph {Phys. Lett. B}}\ }\textbf {\bibinfo {volume} {752}},\ \bibinfo
  {pages} {291} (\bibinfo {year} {2016})},\ \Eprint
  {http://arxiv.org/abs/1508.05395} {arXiv:1508.05395 [gr-qc]} \BibitemShut
  {NoStop}%
\end{thebibliography}%

\end{document}